\documentclass[acmtog]{acmart}
\acmSubmissionID{175}

\newcommand{\revised}[1]{{\color{black} #1}}

\usepackage{booktabs} 

\usepackage[ruled]{algorithm2e} 

\SetAlFnt{\small}
\SetAlCapFnt{\small}
\SetAlCapNameFnt{\small}
\SetAlCapHSkip{0pt}

\usepackage{overpic}
\usepackage{bbold} 
\usepackage{amsfonts} 
\usepackage{amsmath} 
\usepackage{pgfplots}
\usepackage{pgf,tikz}
\usepackage{tkz-euclide}
\usepackage{bm}
\usepackage{wrapfig}
\usepackage{multirow}
\usepackage{mathtools}
\usepackage{enumerate}

\usepackage{verbatim}
\usepackage[mode=buildnew]{standalone}

\SetAlFnt{\small}
\SetAlCapFnt{\small}
\SetAlCapNameFnt{\small}
\SetAlCapHSkip{0pt}

\renewcommand{\phi}{\varphi}

\newcommand{\mynorm}[1]{\big\Vert\, {#1}\big\Vert_F^2}

\newcommand{\tran}{^{\mkern-1.5mu {T}}}	  

\usepackage{soul}

\usepackage{listings}
\usepackage{color, colortbl} 
\definecolor{mygreen}{RGB}{28,172,0} 
\definecolor{mylilas}{RGB}{170,55,241}
\definecolor{ffzzcc}{rgb}{1,0.6,0.8}
\definecolor{mytbcol}{RGB}{175,227,246}

\newtheorem{defi}{Definition}[section]
\newtheorem{remark}{Remark}[section]

\usepackage[shortlabels]{enumitem}
\usepackage{kantlipsum}

\newcommand{\R}{\mathcal{R}}
\renewcommand{\O}{\mathcal{O}}
\newcommand{\D}{\mathcal{D}}

\DeclareFontFamily{U}{mathx}{\hyphenchar\font45}
\DeclareFontShape{U}{mathx}{m}{n}{<-> mathx10}{}
\DeclareSymbolFont{mathx}{U}{mathx}{m}{n}
\DeclareMathAccent{\widebar}{0}{mathx}{"73}

\usepackage{tipa}
\UndeclareTextCommand{\!}{T3}
\DeclareTextCommand{\tipaEXCLAM}{T3}{}
\DeclareRobustCommand{\!}{%
  \ifmmode\mskip-\thinmuskip\else\expandafter\tipaEXCLAM\fi
}
\newcommand{\parallelsum}{\mathbin{\!/\mkern-5mu/\!}}

\usepackage{booktabs}       
\usepackage{mathrsfs}
\usetikzlibrary{arrows}
\usetikzlibrary{matrix}
\usetikzlibrary{positioning,calc,fadings}
\usetikzlibrary{backgrounds, fit}
\pgfplotsset{compat=1.14}
\usepackage{pgfplotstable}

\newcommand{\etal}{et al.}

\newcommand*{\Scale}[2][4]{\scalebox{#1}{$#2$}}%

\usepackage{pifont}
\newcommand{\cmark}{\ding{51}}%
\newcommand{\xmark}{{\ding{55}}}%

\renewcommand{\textcirc}{
 \begin{tikzpicture}
  \fill (0,0) circle [radius=0.25em];
 \end{tikzpicture}
}

\newcommand{\textedge}{
 \begin{tikzpicture}
 \draw [line width=1.25pt] (0em,-0.25em)--(0em,0.25em);
 \end{tikzpicture}
}

\definecolor{myblue}{RGB}{23, 195, 178}
\definecolor{myyellow}{RGB}{255, 186, 8}
\definecolor{mygray}{rgb}{0.5,0.5,0.5}
\definecolor{myred}{RGB}{254, 109, 115}
\catcode`\@ = 11
\newdimen\@InsertBoxMargin
\newcount\@numlines    
\newcount\@linesleft   
\def\ParShape{%
    \@numlines = 0
    \def\@parshapedata{ }
    \afterassignment\@beginParShape
    \@linesleft
}%
\def\@beginParShape{%
    \ifnum \@linesleft = 0
      \let\@whatnext = \@endParShape
    \else
      \let\@whatnext = \@readnextline
    \fi
    \@whatnext
}%
\def\@endParShape{%
    \global\parshape = \@numlines \@parshapedata
}%
\def\@readnextline#1 #2 #3 {
    \ifnum #1 > 0
      \bgroup  
        \dimen0 = \hsize
        \advance \dimen0 by -#2  
        \advance \dimen0 by -#3  
        \count0 = 0
        \loop
          \global\edef\@parshapedata{%
            \@parshapedata    
            #2                
            \space            
            \the\dimen0       
            \space            
          }%
          \advance \count0 by 1
          \ifnum \count0 < #1
        \repeat
      \egroup
      \advance \@numlines by #1
    \fi
    \advance \@linesleft by -1
    \@beginParShape
}%
\newbox\@boxcontent     
\newcount\@numnormal    
\newdimen\@framewidth   
\newdimen\@wherebottom  
\newif\if@byframe       
\@byframefalse
\def\InsertBoxC#1{%
  \leavevmode
  \vadjust{
    \vskip \@InsertBoxMargin
    \hbox to \hsize{\hss#1\hss}
    \vskip \@InsertBoxMargin
  }%
}%
\def\InsertBoxL#1#2{%
  \@numnormal = #1
  \setbox\@boxcontent = \hbox{#2}%
  \let\@side = 0
  \futurelet \@optionalparameter \@InsertBox
}
\def\InsertBoxR#1#2{%
  \@numnormal = #1
  \setbox\@boxcontent = \hbox{#2}%
  \let\@side = 1
  \futurelet \@optionalparameter \@InsertBox
}%
\def\@InsertBox{%
  \ifx \@optionalparameter [
    \let\@whatnext = \@@InsertBoxCorrection
  \else
    \let\@whatnext = \@@InsertBoxNoCorrection
  \fi
  \@whatnext
}%
\def\@@InsertBoxCorrection[#1]{%
  \ifx \@side 0
    \@@InsertBox{#1}{0}{{\the\@framewidth} 0cm}%
  \else
    \@@InsertBox{#1}{1}{0cm {\the\@framewidth}}%
  \fi
}%
\def\@@InsertBoxNoCorrection{%
  \@@InsertBoxCorrection[0]%
}%
\def\@@InsertBox#1#2#3{%
  \MoveBelowBox
  \@byframetrue
  \@wherebottom = \baselineskip
  \multiply \@wherebottom by \@numnormal
  \advance \@wherebottom by 2\@InsertBoxMargin
  \advance \@wherebottom by \ht\@boxcontent
  \advance \@wherebottom by \pagetotal
  \ifdim \pagetotal = 0cm
    \advance \@wherebottom by -\baselineskip  
  \fi
  \advance \@wherebottom by #1\baselineskip
  \@framewidth = \wd\@boxcontent
  \advance \@framewidth by \@InsertBoxMargin
  \bgroup  
    \ifdim \pagetotal = 0cm
      \dimen0 = \vsize
    \else
      \dimen0 = \pagegoal
    \fi
    \ifdim \@wherebottom > \dimen0
      \immediate\write16{+--------------------------------------------------------------+}%
      \immediate\write16{| The box will not fit in the page. Please, re-edit your text. |}%
      \immediate\write16{+--------------------------------------------------------------+}%
      \vrule width \overfullrule
    \fi
  \egroup
  \prevgraf = 0
  \vbox to 0cm{%
    \dimen0 = \baselineskip
    \multiply \dimen0 by \@numnormal
    \advance \dimen0 by -\baselineskip
    \setbox0 = \hbox{y}%
    \vskip \dp0
    \vskip \dimen0
    \vskip \@InsertBoxMargin
    \ifnum #2 = 1
      \vtop{\noindent \hbox to \hsize{\hss \box\@boxcontent}}%
    \else
      \vtop{\noindent \box\@boxcontent}%
    \fi
    \vss
  }%
  \vglue -\parskip
  \vskip -\baselineskip
  \everypar = {%
    \ifdim \pagetotal < \@wherebottom
      \bgroup  
        \dimen0 = \@wherebottom
        \advance \dimen0 by -\pagetotal
        \divide \dimen0 by \baselineskip
        \count1 = \dimen0
        \advance \count1 by 1
        \advance \count1 by -\@numnormal
        \ifnum #2 = 1
          \ParShape = 3
                      {\the\@numnormal}   0cm   0cm
                      {\the\count1}       0cm   {\the\@framewidth}
                      1                   0cm   0cm
        \else
          \ParShape = 3
                      {\the\@numnormal}   0cm                  0cm
                      {\the\count1}       {\the\@framewidth}   0cm
                      1                   0cm                  0cm
        \fi
      \egroup
    \else
      \@restore@    
    \fi
  }%
  \def\par{%
      \endgraf
      \global\advance \@numnormal by -\prevgraf
      \ifnum \@numnormal < 0
        \global\@numnormal = 0
      \fi
      \prevgraf = 0
  }%
}%
\def\MoveBelowBox{%
  \par
  \if@byframe
    \global\advance \@wherebottom by -\pagetotal
    \ifdim \@wherebottom > 0cm
      \vskip \@wherebottom
    \fi
    \@restore@
  \fi
}%
\def\@restore@{%
    \global\@wherebottom = 0cm
    \global\@byframefalse
    \global\everypar = {}%
    \global\let \par = \endgraf
    \global\parshape = 1 0cm \hsize
}%
\ifx \documentclass \@Dont@Know@What@It@Is@
\else
  \let \pageno = \c@page
\fi

\catcode`\@ = 12

\graphicspath{{figures/}}

\setcopyright{rightsretained}
\acmJournal{TOG}
\acmYear{2021}
\acmVolume{40}
\acmNumber{6}
\acmArticle{249}
\acmMonth{12} 
\acmDOI{10.1145/3478513.3480494}

\begin{document}
\title[Intuitive and Efficient Roof Modeling for Reconstruction and Synthesis]{Intuitive and Efficient Roof Modeling for Reconstruction and Synthesis}  

\author{Jing Ren}
\authornote{Work was done during the author's research internship at Alibaba Group}
\affiliation{\institution{KAUST and Alibaba Group}}
\email{jing.ren@kaust.edu.sa}

\author{Biao Zhang}
\affiliation{\institution{KAUST}}
\email{biao.zhang@kaust.edu.sa}

\author{Bojian Wu, Jianqiang Huang, Lubin Fan}
\affiliation{\institution{Alibaba Group}}
\email{ustcbjwu@gmail.com, jianqiang.hjq@alibaba-inc.com, lubinfan@gmail.com}

\author{Maks Ovsjanikov}
\affiliation{
  \institution{LIX, \'Ecole Polytechnique}}
\email{maks@lix.polytechnique.fr}

\author{Peter Wonka}
\affiliation{
  \institution{KAUST}}
\email{pwonka@gmail.com}
\renewcommand{\shortauthors}{Jing Ren, Biao Zhang, Bojian Wu, Jianqiang Huang, Lubin Fan, Maks Ovsjanikov, and Peter Wonka}

\ccsdesc[500]{Computing methodologies~Shape modeling; Shape mesh modeling}
\keywords{Roof modeling, Optimization, Interactive editing, Roof synthesis}

\begin{abstract} We propose a novel and flexible roof modeling approach that can be used for constructing planar 3D polygon roof meshes.  Our method uses a graph structure to encode roof topology and enforces the roof validity by optimizing a simple but effective planarity metric we propose.  This approach is significantly more efficient than using general purpose 3D modeling tools such as 3ds Max or SketchUp, and more powerful and expressive than specialized tools such as the straight skeleton.  Our optimization-based formulation is also flexible and can accommodate different styles and user preferences for roof modeling.  We showcase two applications. The first application is an interactive roof editing framework that can be used for roof design or roof reconstruction from aerial images. We highlight the efficiency and generality of our approach by constructing a mesh-image paired dataset consisting of 2539 roofs.  Our second application is a generative model to synthesize new roof meshes from scratch.  We use our novel dataset to combine machine learning and our roof optimization techniques, by using transformers and graph convolutional networks to model roof topology, and our roof optimization methods to enforce the planarity constraint.  \end{abstract}

\begin{teaserfigure}
    \centering
    \vspace{-3pt}
    \includegraphics[width=1\linewidth]{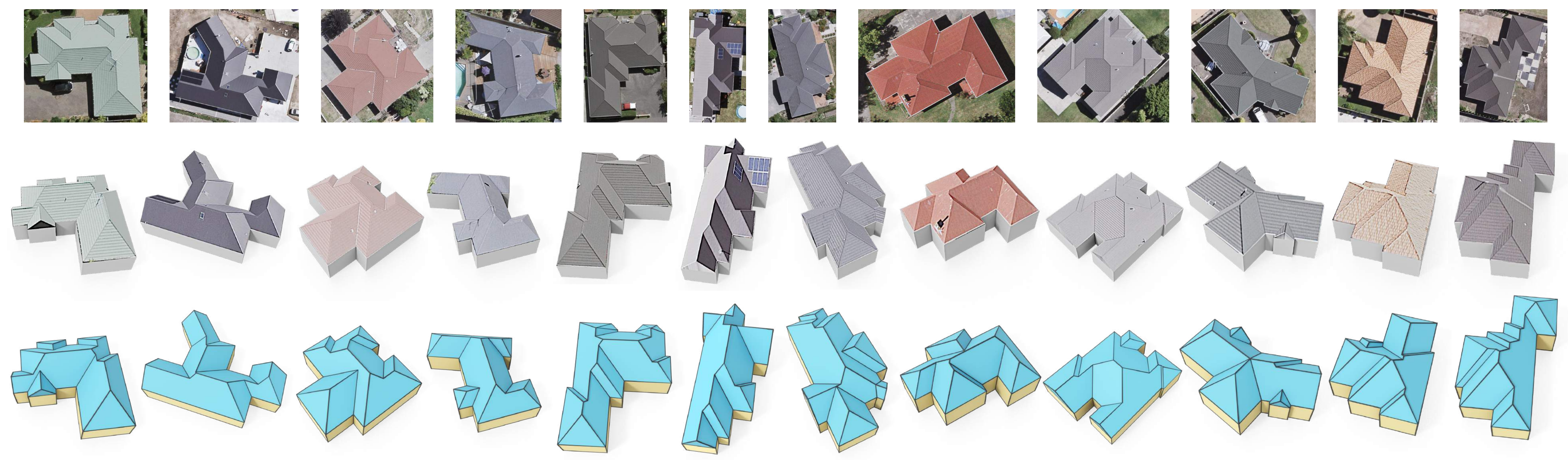}
    \vspace{-22pt}
    \caption{\revised{Our method can be used to reconstruct buildings from aerial images. \emph{Top}: aerial images, on which a user can use our UI to specify the roof topology. \emph{Middle}: the reconstructed buildings obtained by our geometric optimization (textured by the input image). \emph{Bottom}: 3D geometry of the reconstructed buildings.}}
    \label{fig:eg:roofs_with_texture}
\end{teaserfigure}

\maketitle

\begin{table}[!t]
    \caption{Different solutions for roof modeling.} \label{tb:baselines}
    \vspace{-3pt}
        \footnotesize
    {\def\arraystretch{1.3}\tabcolsep=0.5em 
    \begin{tabular}{lcccc}
    \toprule[1pt]
    \multicolumn{1}{c}{Property}  & {\def\arraystretch{0.8}\begin{tabular}[c]{@{}c@{}} \scriptsize{straight}\\ \scriptsize{skeleton}\end{tabular}}    & {\def\arraystretch{0.8}\begin{tabular}[c]{@{}c@{}}\scriptsize{weighted} \\ \scriptsize{straight}\\ \scriptsize{skeleton} \end{tabular}}    &{\def\arraystretch{0.8}\begin{tabular}[c]{@{}c@{}} \scriptsize{commercial}\\ \scriptsize{software}\end{tabular}}  &  \footnotesize{\textbf{Ours}}\\
    \bottomrule[1pt]
    \itshape\textbf{Easy} to use for \textbf{beginners}? & \cmark & \cmark & \xmark & \cmark\\ \rowcolor{mytbcol!30}
    \itshape\textbf{Efficient} for roof construction? & \cmark & \cmark & \xmark & \cmark\\  
    \itshape  \textbf{Accurately} reconstruct roofs from images? & \xmark & \xmark & \cmark &  \cmark\\    
    \rowcolor{mytbcol!30}    
    \itshape \textbf{Light} user input? & \cmark & \cmark & \xmark & \cmark\\ 
    \itshape \textbf{Allow} editing operations? & \xmark & \cmark & \cmark & \cmark\\
    \rowcolor{mytbcol!30} 
    \itshape \textbf{Intuitive} editing operations? & \xmark & \xmark & \cmark & \cmark\\
    \itshape \textbf{Insensitive} to \textbf{noisy} user input? & \xmark & \xmark & \cmark & \cmark \\
    \bottomrule[1pt] 
    \end{tabular}
    }
    \vspace{-9pt}
\end{table} 

\section{Introduction}
Roof modeling is an important topic in urban reconstruction. 
In our work, we propose a novel and simple formulation for roof modeling which is expressive enough to handle a large range of roofs. Our formulation is also suitable for two applications: interactive reconstruction of roofs from aerial images (See Fig.~\ref{fig:eg:roofs_with_texture}) and the generative modeling of new roofs. 

The main challenges of roof modeling are how to mathematically describe the roof structure and how to enforce the planarity of the roof faces. A popular tool for roof construction is to use the straight skeleton~\cite{aichholzer1996novel,aichholzer1996straight} or one of its extensions to increase its modeling expressiveness~\cite{eppstein1999raising,biedl2015weighted,kelly2011interactive}. 
These methods typically use a closed roof outline to describe the roof structure assuming that the interior topology can be determined by the roof outline. The roof face planarity is enforced during the computation of roof topology from the input outline.
Another solution is to use general purpose commercial software such as 3ds Max and SketchUp to model roofs. Though the commercial software can construct large variations of realistic roofs, the roof planarity is either ignored (e.g., 3ds Max) or implicitly enforced by triangulation (SketchUp).

\begin{figure}[!t]
    \centering
    \begin{overpic}[trim=0cm 0cm 0cm 0cm,clip,width=1\linewidth,grid=false]{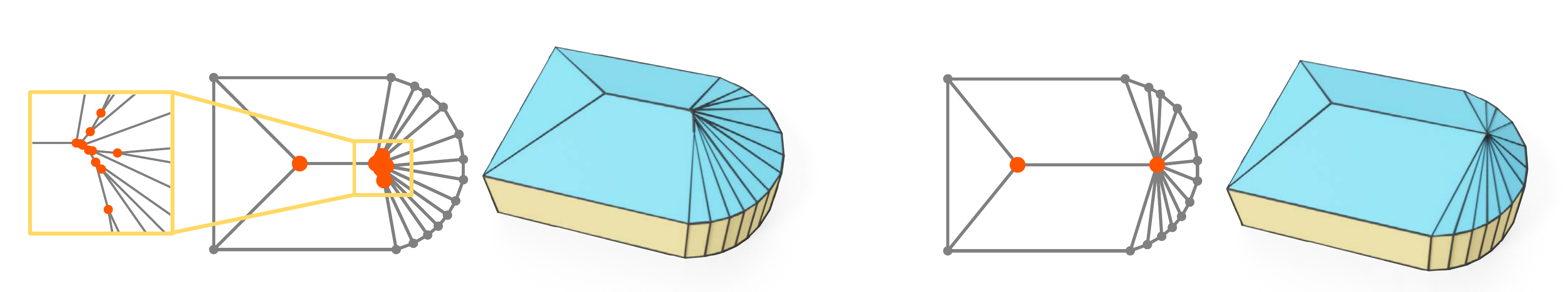}
    \put(13,17){\footnotesize (a) Straight Skeleton}
    \put(72,17){\footnotesize (b) \textbf{Ours}}
    \end{overpic}\vspace{-12pt}
    \caption{For a given outline with 16 vertices, we run the straight skeleton method to obtain the 2D roof embedding and construct the corresponding building as shown in (a). The straight skeleton computes a planar but complicated 3D roof with 12 roof vertices (colored in red). As a comparison, our method constructs a more plausible roof with 2 roof vertices.}\vspace{-3pt}
    \label{fig:intro:ss_prob_eg1_extra_vtx} 
\end{figure}

\begin{figure}[!t]
    \centering
    \begin{overpic}[trim=0cm 0cm 0cm 0cm,clip,width=1\linewidth,grid=false]{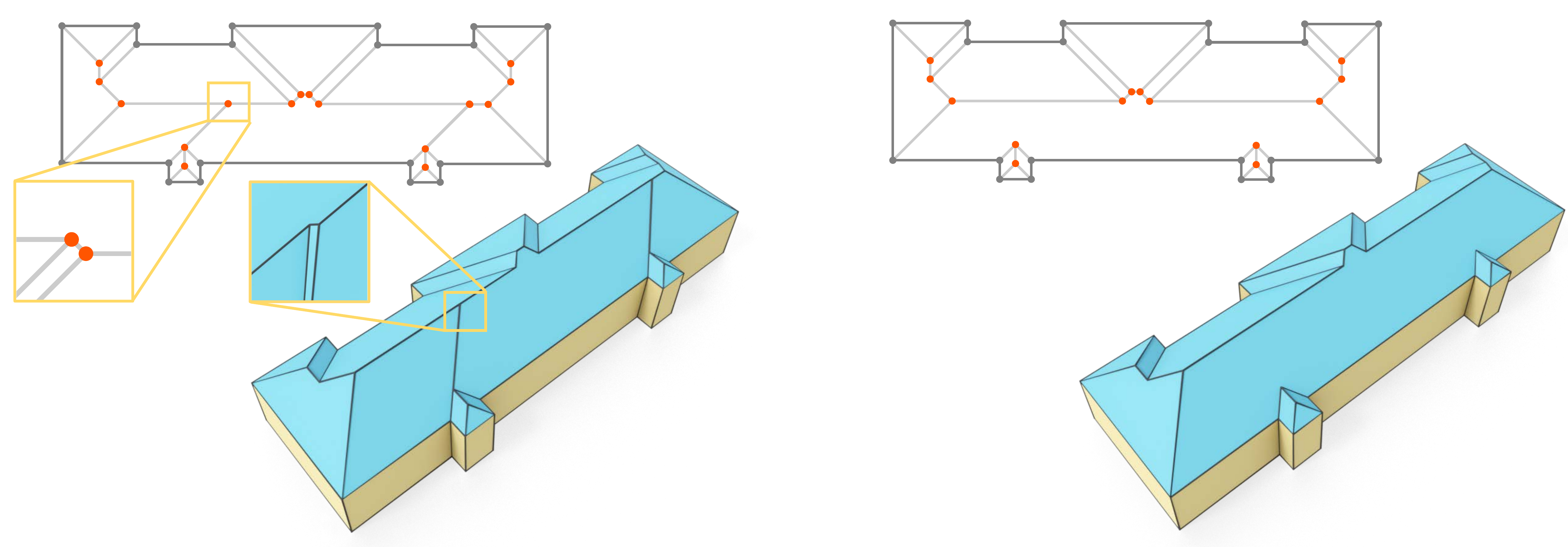}
    \put(0,10){\footnotesize (a) Straight}
    \put(3.7,7){\footnotesize Skeleton}
    \put(57,10){\footnotesize (b) \textbf{Ours}}
    \end{overpic}\vspace{-9pt}
    \caption{The straight skeleton method cannot handle the cases where there exists a face that contains multiple outline edges. \emph{Top}: we show the roof graph computed by the straight skeleton and our method. \emph{Bottom}: we show the corresponding constructed roofs.}
    \label{fig:intro:ss_prob_eg2_multi_outlineEdge}\vspace{-9pt}
\end{figure}

We observe that these solutions have limitations in different aspects (see Table~\ref{tb:baselines}).
For example, using only the outline for roof structure specification is simple but 
not sufficient for roof modeling, since different roofs can have the same outline.
Moreover, determining the roof topology from the outline can be error-prone.
For example, for some outlines, the straight skeleton based methods create spurious additional vertices close-by (Fig.~\ref{fig:intro:ss_prob_eg1_extra_vtx}), and fail to recover the correct roof topology when there exists a face corresponding to multiple outline edges (Fig.~\ref{fig:intro:ss_prob_eg2_multi_outlineEdge}).
On the other hand, commercial software provides large freedom, but modeling is significantly more complicated especially for non expert users and it is not easy to enforce geometric constraints.

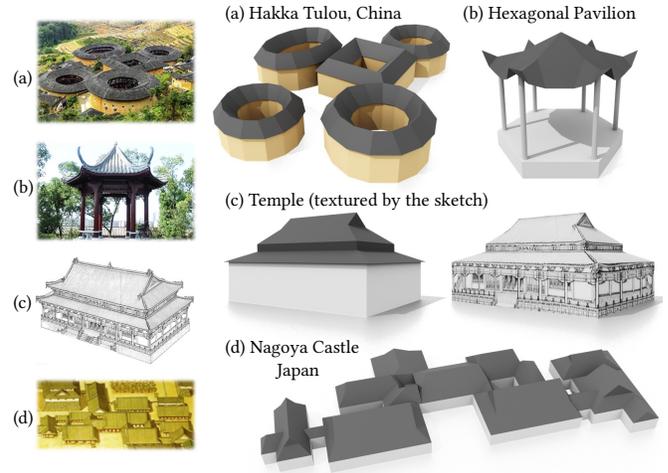
\begin{figure}[!t]
    \centering
    \vspace{-3pt}
    \begin{overpic}[trim=3cm 15cm 42cm 2cm,clip,width=1\linewidth,grid=false]{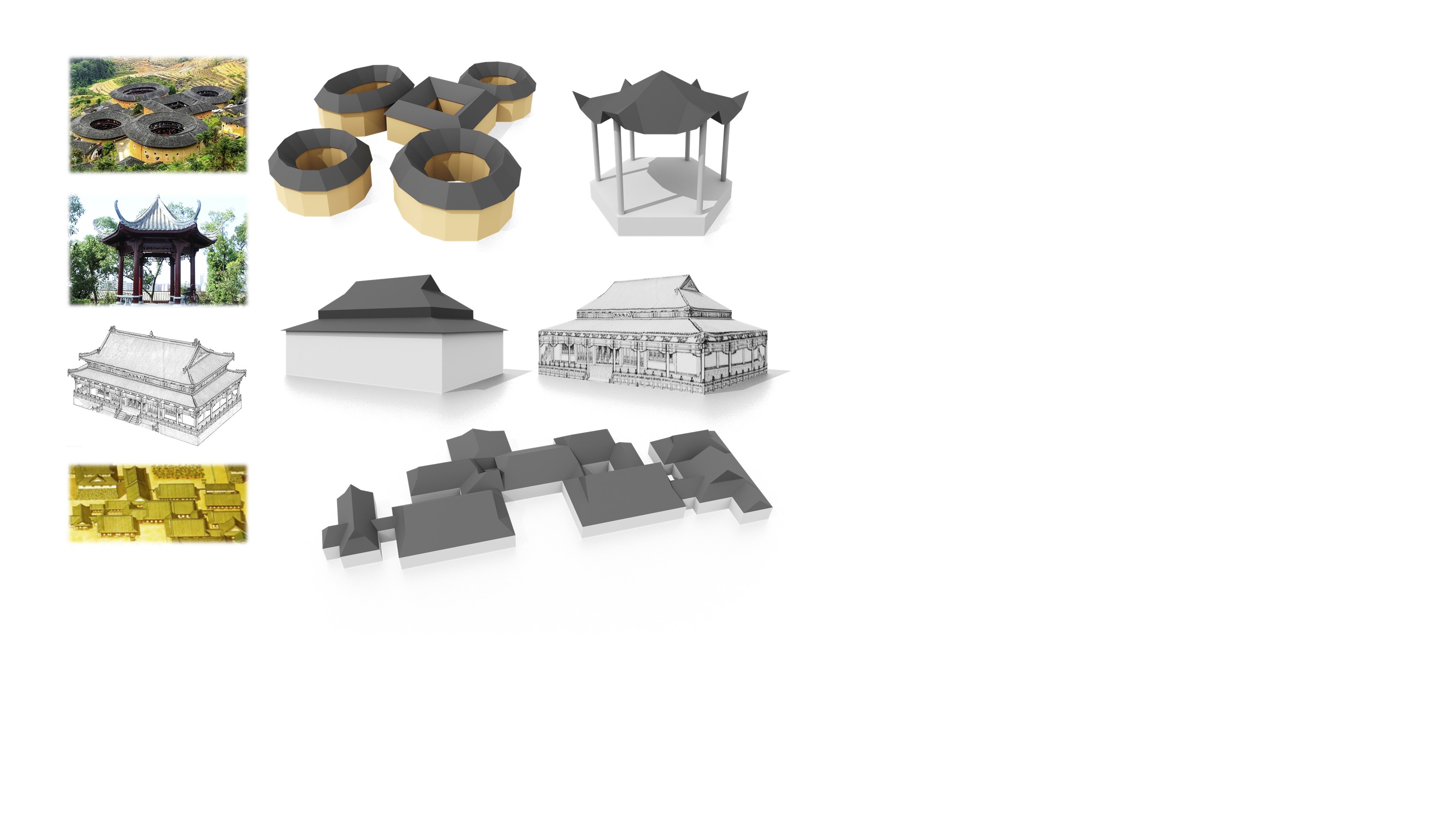}
    \put(32,72){\footnotesize (a) Hakka Tulou, China}
    \put(69,72){\footnotesize (b) Hexagonal Pavilion}
    \put(32,43){\footnotesize (c) Temple (textured by the sketch)}
    \put(32,20){\footnotesize (d) Nagoya Castle}
    \put(40,16.5){\footnotesize Japan}
    \put(-1,62){\footnotesize (a)}
    \put(-1,45){\footnotesize (b)}
    \put(-1,27){\footnotesize (c)}
    \put(-1,9){\footnotesize (d)}
    \end{overpic}\vspace{-10pt}
    \caption{Building meshes of Asian architecture created by our method. The reference images on the left are collected from internet. }\vspace{-6pt}
    \label{fig:eg_temple}
\end{figure}

In our work, we propose to use a two-step procedure where we first model the roof topology and optionally the approximate geometry and then refine the geometry using optimization.
Specifically, we propose to use a \emph{roof graph} to specify roof topology which is simple and flexible enough to represent a large range of roofs including residential buildings (Fig.~\ref{fig:eg:roofs_with_texture}) and architecture with different styles (Fig.~\ref{fig:eg_temple}).
We then propose an optimization-based method for roof modeling from an input roof graph, where we introduce a simple planarity metric. Our method is generic and can be adapted to different settings such as including user-specified regularizers. 
Compared to the straight skeleton based methods, our solution has stronger expressiveness with fewer assumptions placed on the underlying roofs. Meanwhile, our solution is more robust and can better reconstruct roofs from an image with higher accuracy. 
Compared to the general purpose 3D modeling software, our method has a more natural and systematic roof structure representation and is explicitly designed to output planar 3D \emph{polygon} roofs. Our roof modeling framework \revised{can be easily used by novices requiring light user input}.

Our method has two practical applications, interactive roof editing and roof synthesis from scratch. Specifically, our roof graph representation allows different editing operations for modifying roof topology, while our roof optimization efficiently updates the roof embedding by enforcing planarity constraints. These features allow a user to interactively model a planar roof by iteratively modifying roof topology and optimizing the roof planarity. 
Another useful and novel application is to automatically synthesize realistic roofs. 
Roof synthesis in general is a difficult problem, since it has a mixture of discrete (roof topology) and continuous (roof planarity) constraints. 
To tackle this issue, we train neural networks to generate a discrete roof graph then run our roof optimization method to enforce the continuous geometric constraints. 


In summary, our main contributions are:
\begin{itemize}
    \item A novel roof modeling method including a simple roof graph representation that encodes roof topology and a new planarity metric that can be used to enforce geometric constraints for generating 3D polygonal roof meshes.
    \item An optimization-based framework that is complementary to learning algorithms and user input for automatic roof synthesis and interactive editing.
    \item We created a dataset consisting of 2539 roof meshes paired with images without topological or geometric errors (see Fig.~\ref{fig:dataset}) using our method, which can be helpful for different visual computing tasks. \revised{ Code and data are available.\footnote{\revised{Demo Code: \href{https://github.com/llorz/SGA21_roofOptimization}{\url{https://github.com/llorz/SGA21_roofOptimization}} }}}
\end{itemize}

\begin{figure}[!t]
    \centering
    \vspace{-6pt}
    \begin{overpic}[trim=0cm 12.5cm 0cm 0cm,clip,width=1\linewidth,grid=false]{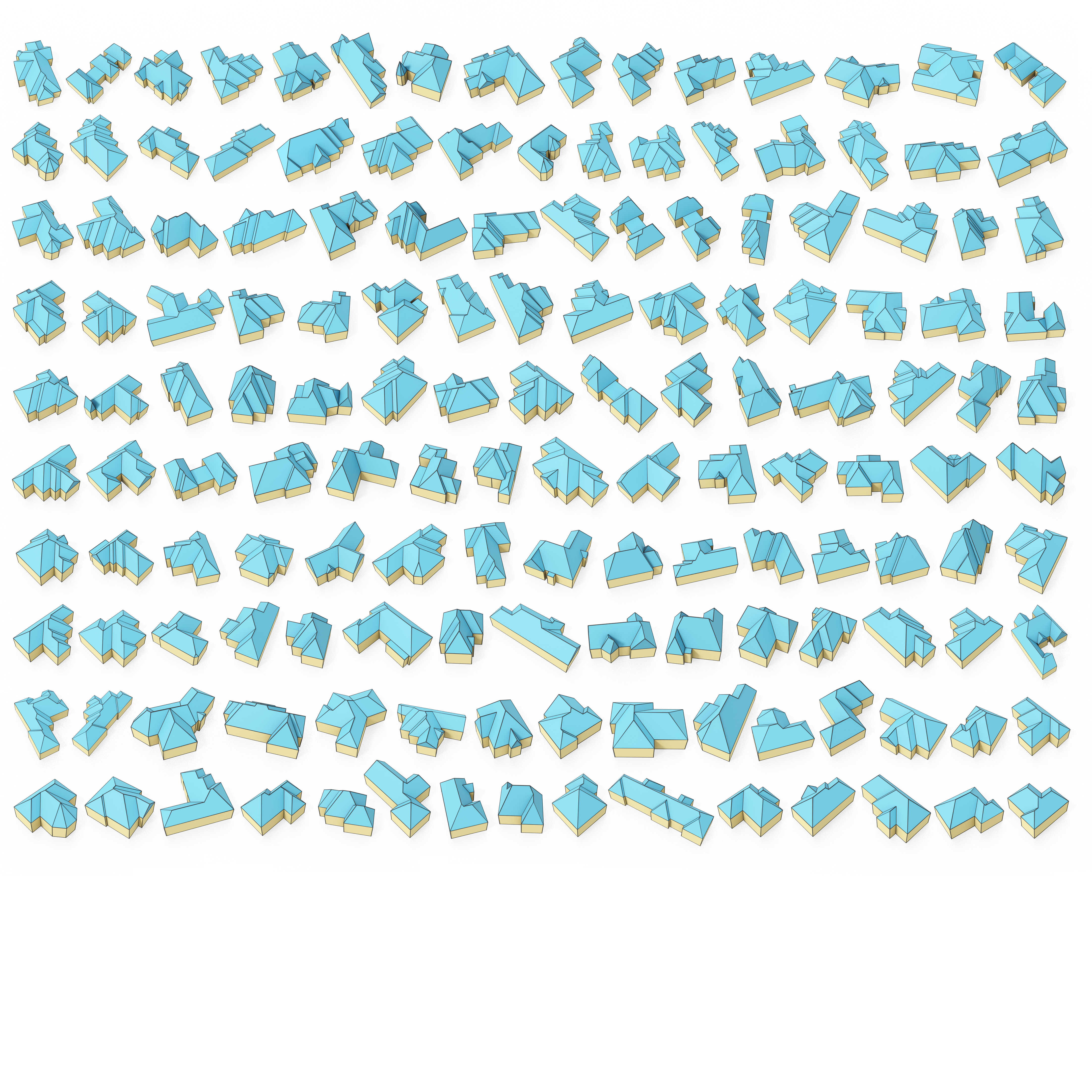}
    \end{overpic}\vspace{-10pt}
    \caption{We created a dataset of 3D planar roofs paired with the corresponding aerial images. Here we show 150 example roofs and highlight the topology. The paired images are shown in Fig. 17 in the supplementary materials.}
    \label{fig:dataset}
\end{figure}

\section{Related work}
We review related work in three categories: roof construction algorithms, roof reconstruction, and generative modeling.

\subsection{Roof Construction Algorithms}
In computer graphics, previous work proposes solutions to model various aspects of architectural models, such as the room layout and arrangement~\cite{merrell2010computer,hu2020graph2plan} and indoor scene synthesis~\cite{yu2011make,fisher2015actsynth}. Roofs are one of the components previous works attempted to model using algorithmic or procedural approaches.

The \textit{straight skeleton}
algorithm~\cite{aichholzer1996novel,aichholzer1996straight} is a popular geometric algorithm for the generation of complex roof structures \revised{from user-specified roof outlines}~\cite{laycock2003automatically}. \revised{The weighted straight skeleton~\cite{eppstein1999raising,biedl2015weighted, eder2018computing} is an extension to improve the modeling expressiveness to facilitate the modeling of roof faces at different angles specified by users.} Several large-scale urban modeling projects are inspired by these techniques. Larive and Gaildrat~\shortcite{larive2006wall} combine 3D building descriptions, including the height of footprints and roofs, with GIS information to create building models. Kelly and Wonka~\shortcite{kelly2011interactive} propose a subsequent extension to the weighted straight skeleton for the interactive modeling of complete buildings. Besides, Buron~\etal~\shortcite{buron2013gpu} consider to bring parallelism to grammar-based roof generation in order to improve the computational efficiency. Other extensions to the straight skeleton include the works of Sugihara~\shortcite{sugihara2013straight,sugihara2019straight} and Held and Palfrader~\shortcite{held2017straight}.
An alternative approach to modeling complex roofs is to create a combination of elementary roof primitives, e.g. as done in procedural modeling~\cite{muller2006procedural}. Complementary to our work in roof modeling, several magnificent real-world buildings feature smooth roof shapes. The design of such roofs~\cite{liu2006geometric,pottmann2007geometry,pottmann2008freeform} requires different and specialized tools and should be considered as a separate topic.

\subsection{Roof Reconstruction}
Urban reconstruction aims at automatically generating 3D models from real physical measurements, such as multi-view images or point clouds~\cite{musialski2013survey,demir2015procedural}. We focus our review on recent methods that have roof reconstruction as a major component. 
One category of algorithms uses optimization.
For example, Zhou and Neumann published a series of papers to reconstruct coarse building models including roofs from LiDAR point clouds~\shortcite{zhou2008fast,zhou20102,zhou20112}. 
\revised{
Lin~\etal~\shortcite{lin2013semantic} propose to find a combination of planar primitives that best explains the input LiDAR data for roof reconstruction. Dehbi~\etal~\shortcite{dehbi2021robust} propose an active sampling strategy for RANSAC to fit plane approximations to input point clouds.
}
Nan and Wonka~\shortcite{nan2017polyfit} and Kelly~\etal~\shortcite{kelly2017bigsur} employ integer programming to reconstruct coarse planar building models. Arikan~\etal~\shortcite{arikan2013snap} propose an initial automatic method to extract candidate planes and estimate coarse polygons from unorganized point clouds, and then allow users to interactively edit the model by optimization-based snapping operations. Verdie~\etal~\shortcite{verdie2015lod} and Zhu~\etal~\shortcite{zhu2018large} focus on segmenting parts belonging to the roof surface, and then fitting a collection of piece-wise planar planes~\cite{liu2018planenet} or prior defined templates for roof reconstruction.
\revised{
Habbecke and Kobbelt~\shortcite{habbecke2012linear} propose an interactive geometric modeling system that can be used for polygonal roof editing with geometric regularities or constraints. 
Salinas~\etal~\shortcite{salinas2015structure} present a mesh decimation approach that generates planar abstractions of roof meshes. 
Bauchet and Lafarge~\shortcite{bauchet2020kinetic} adopt a kinetic data structure to partition the 3D space into convex polyhedra from which the underlying surface mesh of the input point cloud can be extracted.
}
Another category of papers use deep learning~\cite{yu2021automatic,alidoost2020shaped,zeng2018neural,zhang2020conv} to directly output the reconstruction of 3D roof structures. 
\revised{However, these methods do not propose a solution for enforcing geometric constraints, while in our work, we mainly focus on enforcing geometric constraints in an optimization-based formulation during interactive roof reconstruction.}

\subsection{Generative Modeling}
\revised{
Generative models aim to generate new samples that follow a similar distribution as a collection of training samples. 
For example, generative adversarial networks (GANs)~\cite{goodfellow2014generative} can be extended to 3D by synthesizing details on surfaces, e.g., \cite{kelly2018frankengan} create textures on coarse building models. Volumetric GANs~\cite{wu2016learning} can create 3D models with a voxel grid representation.
Normalizing flows~\cite{rezende2015variational,DBLP:journals/corr/DinhKB14} (NF) have been extended to 3D by modeling the distribution of point clouds as an invertible parameterized transformation from a probability density embedded in 3D space~\cite{yang2019pointflow,kim2020softflow,stypulkowski2019conditional}.
In the following, we mainly discuss variational autoencoders (VAEs) and auto-regressive models (ARs) for 3D tasks that are closely related to our roof synthesis application.

Variational autoencoders~\cite{DBLP:journals/corr/KingmaW13} parameterize latent variable models with deep neural networks. Compared to GANs, VAEs are easier to train but they typically produce images of lower visual quality~\cite{van2017neural,razavi2019generating}. While extending to 3D geometry~\cite{brock2016generative}, they have the potential to synthesize 3D shapes, especially on domain-specific tasks that require limited topological variations, such as face~\cite{ranjan2018generating} or human body~\cite{tan2018variational} modeling. 
VAEs also have been used for generating structured models, such as furniture~\cite{mo2019structurenet,yang2020dsmnet,gaosdmnet2019}. In these instances, there are separate generative models for the individual object parts and the part arrangement.

Auto-regressive models (ARs) factorize a distribution over a sequence into several conditional densities. Each conditional density models a single element in the sequence conditioned on all previous elements. Language models are auto-regressive in nature~\cite{radford2019language, DBLP:conf/acl/DaiYYCLS19}. Some works also model images as sequences, \emph{e.g.}, PixelRNN~\cite{van2016pixel} and its follow-up works PixelCNN~\cite{van2016conditional}, PixelCNN++~\cite{DBLP:conf/iclr/SalimansK0K17}, and PixelSNAIL~\cite{chen2018pixelsnail}.
With the rise of transformers, autoregressive models have been increasingly applied to 3D data, including polygonal meshes~\cite{DBLP:conf/icml/NashGEB20}, floor plans~\cite{DBLP:journals/corr/abs-2011-13417}, and scenes~\cite{wang2020sceneformer}. We argue that the discrete nature of ARs is a unique advantage for modeling 3D data that does not consist of smoothly varying models, but needs a latent representation that can handle discontinuities. Therefore, we propose to adapt ARs to the generation of roof outlines. We will demonstrate that ARs are more suitable than VAEs (the currently most popular generative model) in the results.
}

\begin{figure}[!t]
    \centering
    \begin{overpic}[trim=0cm 0cm 0cm -1cm,clip,width=1\linewidth,grid=false]{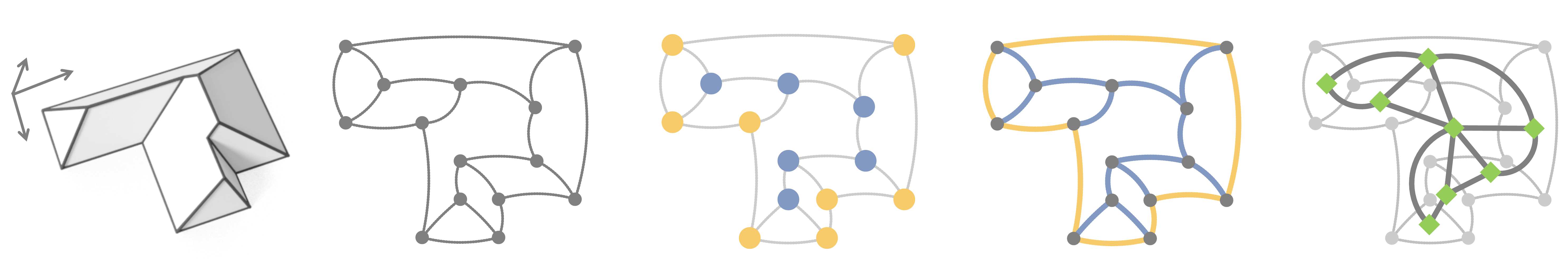}
    \put(5,18){\footnotesize 3D roof}
    \put(23,18){\footnotesize Roof Graph}
    \put(44.5,18){\footnotesize Vertex Set}
    \put(66,18){\footnotesize Edge Set}
    \put(84.5,18){\footnotesize Dual Graph}
    \put(1,0){\footnotesize\bfseries (a)}
    \put(21,0){\footnotesize\bfseries (b)}
    \put(40.5,0){\footnotesize\bfseries (c)}
    \put(62,0){\footnotesize\bfseries (d)}
    \put(82,0){\footnotesize\bfseries (e)}
    \put(1,6){\scriptsize $x$}
    \put(4.5,11.8){\scriptsize $y$}
    \put(0.8,13.5){\scriptsize $z$}
    \put(41,14){\tiny $1$}
    \put(41,8){\tiny $2$}
    \put(46.5,7){\tiny $3$}
    \put(46,0){\tiny $4$}
    \put(53.5,0){\tiny $5$}
    \put(53.5,2.8){\tiny $6$}
    \put(58.5,3){\tiny $7$}
    \put(58.5,13.8){\tiny $8$}
    \put(43.5,10.5){\tiny $9$}
    \put(48.3,12.5){\tiny ${10}$}
    \put(55,11){\tiny ${11}$}
    \put(55,7.2){\tiny ${12}$}
    \put(49,7.2){\tiny ${13}$}
    \put(49,1.9){\tiny ${14}$}
    
    \put(82.5,10.5){\tiny $f_1$}
    \put(87,7.8){\tiny $f_2$}
    \put(93,9.3){\tiny $f_3$}
    \put(89,0.5){\tiny $f_4$}
    \put(92.5,2){\tiny $f_5$}
    \put(96,3.6){\tiny $f_6$}
    \put(99,7){\tiny $f_7$}
    \put(90,14.5){\tiny $f_8$}
\end{overpic}\vspace{-15pt}
    \caption{3D roof represented as a \emph{roof graph}: \textbf{(a)} a simple 3D roof with 8 faces; \textbf{(b)} the topology of the roof represented as a graph $G = (V, E)$; we highlight the vertex set $V$ in \textbf{(c)} and the edge set $E$ in \textbf{(d)}, where the \emph{outline} vertices (edges) are colored in orange, and the \emph{roof} vertices (edges) are colored in blue. \textbf{(e)} We can also construct a dual graph of the roof graph, where each face of the roof is represented as a node, and two nodes are connected to each other if the corresponding roof faces are adjacent.}
    \label{fig:roof_graph}
\end{figure}
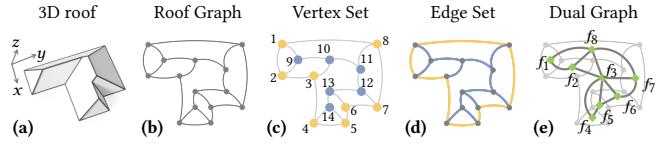

\section{Background \& Problem Formulation}

\subsection{Definitions}
For a planar 3D roof, we can use a roof graph to represent its topology (as shown in Fig.~\ref{fig:roof_graph} (b)). We first categorize the vertices $V$ in the roof graph into two sets, the outline vertices $V_{\O}$ and the roof vertices $V_{\R}$. 
The outline vertices are on the boundary of the (2D projection of the) 3D embedding of the roof graph and the remaining vertices are roof vertices.
We denote by $n_{\O} = \vert V_{\O} \vert$, the number of outline vertices, and $n_{\R} = \vert V_{\R} \vert$, the number of roof vertices.
We also use $n_v = n_{\O} + n_{\R}$ to denote the total number of roof vertices.
For example, for the roof graph shown in Fig.~\ref{fig:roof_graph}, we have $V_{\O} =\big\{v_1, \cdots,v_8\big\}, V_{\R} = \big\{v_9, \cdots, v_{14}\big\}$.
Similarly, we can also categorize the edges in the roof graph into two sets, the outline edges $E_{\O}$ (both of the two endpoints of the edge are outline vertices), and the roof edges $E_{\R}$ (at least one of the endpoints is a roof vertex). 
A region bounded by a set of edges and vertices in the embedding of the roof graph defines a \emph{roof face}. For example, we have $f_1 = \big\{v_1, v_2, v_9\big\}$ and $f_3 = \big\{ v_3, v_{10}, v_{11}, v_{12}, v_{13}, v_{14}, v_{4}\big\}$. Therefore, we can equivalently represent the roof graph $G$ by $(V, F)$, where the edge set $E$ can be easily extracted from the face set $F$. 
We denote $n_f = \vert F \vert$, the number of roof faces in the roof graph.

\begin{figure}[!t]
    \centering
    \input{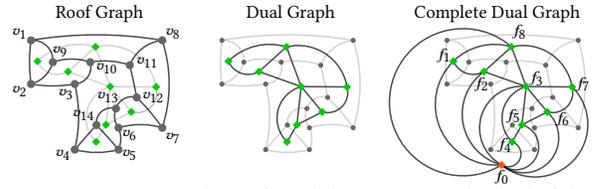}\vspace{-12pt}
    \caption{We can recover the roof graph by computing the dual of the dual graph. \emph{Left}: a given roof graph. \emph{Middle}: its corresponding dual graph. \emph{Right}: we add a node $f_0$ indicating the outside region in the dual graph. We can see that the roof graph (\emph{left}) is the dual of the complete dual graph (\emph{right}).}
    \label{fig:mtd:complete_dual}\vspace{-6pt}
\end{figure}

\paragraph{\textbf{Dual graph}}
For each roof graph $G = (V, F)$, we can construct its \emph{dual graph} $G^{\D} = (V^{\D}, E^{\D})$, where each face in the roof graph $G$ is represented as a node in the dual graph $G^{\D}$, i.e., $\vert V^{\D} \vert = \vert F \vert$. Two nodes in $V^{\D}$ are connected to each other by an edge (stored in $E^{\D}$) if and only if the corresponding two faces are adjacent (i.e., share an edge) in the roof graph $G$. We can store this connectivity information into an adjacency matrix $\Scale[0.9]{A^{\D}\in \{0,1\}^{n_{f}\times n_{f}}}$, i.e., $\Scale[0.9]{A^{\D}_{ij} = 1}$ if $f_i$ is adjacent to $f_j$.
We can therefore equivalently represent the dual graph as $G^{\D} = (V^{\D}, A^{\D})$.
In Fig.~\ref{fig:roof_graph} (e) we show the dual graph which is placed above the original roof graph. Note that it is possible not only to construct a dual graph $G^{\D}$ from the primal roof graph $G$, but also to recover the primal roof graph from a dual graph by computing the dual of the dual graph as shown in Fig.~\ref{fig:mtd:complete_dual} (see Sec.~\ref{sec:mtd:opti_dual} for more details). 

We can embed a roof graph in 3D (2D) by assigning each vertex a 3D (2D) coordinate. For a vertex set $V$, we denote its 2D embedding as $\Scale[0.9]{\widebar{X}\in \mathbf{R}^{n_v\times 2}}$ and its 3D embedding as $\Scale[0.9]{X\in\mathbf{R}^{n_v\times 3}}$, where we store the vertex positions in rows. 
See Fig.~\ref{fig:roof_graph} (a) for a 3D roof that stems from a 3D embedding of the roof graph in Fig.~\ref{fig:roof_graph} (b).

\begin{figure*}[!t]
    \centering
    \begin{overpic}[trim=0cm 0cm 0cm 0cm,clip,width=1\linewidth,grid=false]{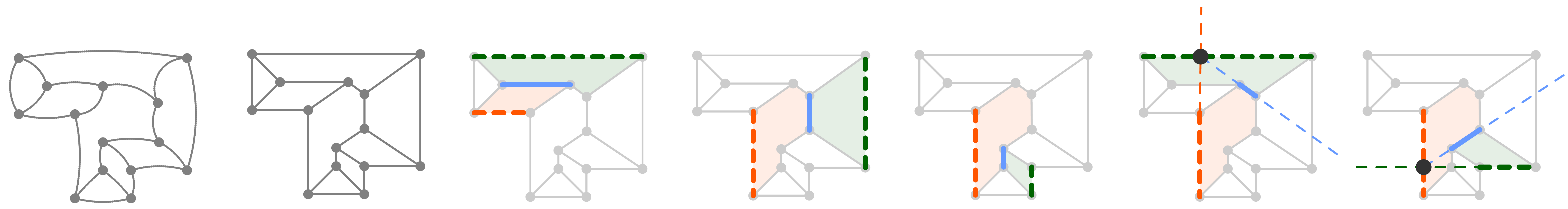}
    \put(3,12){\footnotesize Roof Graph}
    \put(20,13){\footnotesize Valid}
    \put(17.5,11){\footnotesize 2D Embedding}
    \put(0.5,0.5){\footnotesize\bfseries (a)}
    \put(15.5,0.5){\footnotesize\bfseries (b)}
    \put(29.5,0.5){\footnotesize\bfseries (c)}
    \put(43.8,0.5){\footnotesize\bfseries (d)}
    \put(57.7,0.5){\footnotesize\bfseries (e)}
    \put(72,0.5){\footnotesize\bfseries (f)}
    \put(86.3,0.5){\footnotesize\bfseries (g)}
    
    \put(-0.5,10){\tiny $v_1$}
    \put(-0.5,6){\tiny $v_2$}
    \put(3.5,5){\tiny $v_3$}
    \put(3,0.5){\tiny $v_4$}
    \put(8.8,0.5){\tiny $v_5$}
    \put(8.8,1.8){\tiny $v_6$}
    \put(12.2,1.8){\tiny $v_7$}
    \put(12.2,10){\tiny $v_8$}
    \put(3,8.5){\tiny $v_9$}
    \put(6.5,8.5){\tiny $v_{10}$}
    \put(10.2,7.2){\tiny $v_{11}$}
    \put(10.2,4.7){\tiny $v_{12}$}
    \put(6,5){\tiny $v_{13}$}
    \put(5.6,1.2){\tiny $v_{14}$}
    
    \put(34, 10.6){\tiny $e_{1,8}$}
    \put(33, 8.5){\tiny $e_{9,10}$}
    \put(31, 5){\tiny $e_{2,3}$}
    \put(31, 12){\footnotesize $e_{1,8}\parallelsum e_{2,3} \parallelsum e_{9,10}$}
    
    \put(45.5,3){\tiny $e_{3,4}$}
    \put(48.5,5.8){\tiny $e_{11,12}$}
    \put(53,6){\tiny $e_{7,8}$}
    \put(45,12){\footnotesize $e_{3,4}\parallelsum e_{7,8} \parallelsum e_{11,12}$}
    
    \put(59.8, 3){\tiny $e_{3,4}$}
    \put(62.8, 5){\tiny $e_{13,14}$}
    \put(66.2, 1.5){\tiny $e_{5,6}$}
    \put(59,12){\footnotesize $e_{3,4}\parallelsum e_{5,6} \parallelsum e_{13,14}$}
    
    \put(77, 10.6){\tiny $e_{1,8}$}
    \put(74,3){\tiny $e_{3,4}$}
    \put(79.5,8){\tiny $e_{10,11}$}
    \put(73.3,12){\footnotesize $e_{3,4}\lor e_{1,8} \lor e_{10,11}$}
    
    \put(88.2, 3.2){\tiny $e_{3,4}$}
    \put(95,1.5){\tiny $e_{6,7}$}
    \put(91.2,5.8){\tiny $e_{12,13}$}
    \put(87,12){\footnotesize $e_{3,4}\lor e_{6,7} \lor e_{12,13}$}
    
    \put(16.3,7.7){\tiny $f_1$}
    \put(18,7){\tiny $f_2$}
    \put(21,5){\tiny $f_3$}
    \put(21,1.3){\tiny $f_4$}
    \put(22,2.45){\tiny $f_5$}
    \put(23.2,3.6){\tiny $f_6$}
    \put(25,6){\tiny $f_7$}
    \put(21,8.8){\tiny $f_8$}
\end{overpic}\vspace{-18pt}
    \caption{\textbf{Valid 2D embedding of a roof graph}. For a pair of adjacent faces in the input roof graph \textbf{(a)}, we consider their outline edges and their shared edge. If for every pair of adjacent faces \textbf{(c-g)}, these three edges are either \emph{parallel to each other} or \emph{intersect at the same point}, according to Remark~\ref{rmk:valid_2D}, this 2D embedding \textbf{(b)} is \emph{valid}. Here the notation $e_1 \lor e_2 \lor e_3$ means these three edges $e_1, e_2, e_3$ intersect at the same point.} 
    \label{fig:valid_graph}\vspace{-3pt}
\end{figure*}
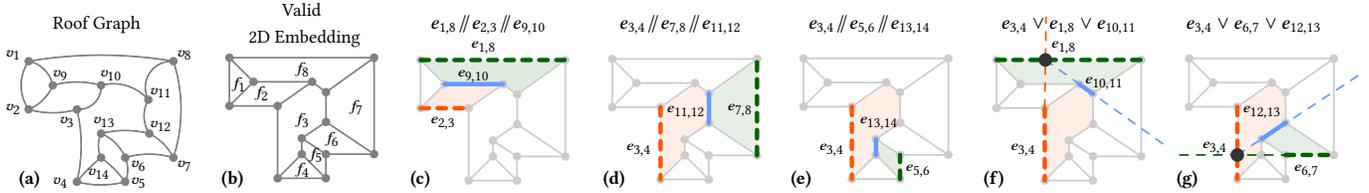

\subsection{Valid Roofs}\label{sec:validity}
An important constraint for a roof is that all faces have to be planar. We can therefore use the planarity constraint to define valid 2D and 3D embeddings of roof graphs:
\begin{defi}\label{def:valid_3D}
We call a 3D embedding of a roof graph \emph{valid} if each 3D roof face is planar and the roof has non-zero height.
\end{defi}

\begin{defi}\label{def:valid_2D}
We call a 2D embedding of a roof graph \emph{valid} if there exists a \emph{valid} 3D embedding such that the projection of the 3D embedding in the $xy$ plane is exactly the same as the 2D embedding.
\end{defi}
Therefore, we can always obtain a valid 2D embedding by projecting a valid 3D embedding to the $xy$ plane. At the same time, we can get a valid 3D embedding by lifting up a valid 2D embedding along the $z$-axis (i.e., assigning each vertex a $z$-axis value). 

\paragraph{\textbf{Verification of the validity}}
To verify the validity of a given 3D embedding, we can simply check if each 3D face is planar or not. To verify the validity of a given 2D embedding, according to the definition, we need to check if there exists a set of $z$-values that combined with the given 2D embedding forms a valid 3D embedding.  An alternative and much easier way to verify the validity of a given 2D embedding is to use basic geometry:
\begin{remark}\label{rmk:valid_2D}
The intersecting line of two adjacent 3D planar faces with fixed outline edges, is either parallel to both outline edges, or intersects the two outline edges at the same point. The same conclusion holds when we project the 3D planar faces to $xy$-plane.
\end{remark}
See Fig.~\ref{fig:remark_proof} and Appendix~\ref{appendix:remark_proof} for a simple proof. Remark~\ref{rmk:valid_2D} gives both a necessary and sufficient condition. Therefore, we can use it to check the validity of a 2D embedding. 
See Fig.~\ref{fig:valid_graph} for an example of a \emph{valid} 2D embedding. 

\begin{figure}[!t]
    \centering
    \vspace{-6pt}
    \input{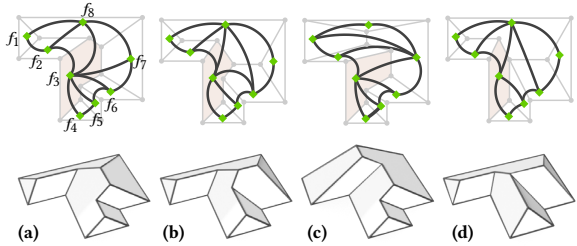}\vspace{-9pt}
    \caption{\textbf{Different roof styles with the same outline}. The roof style can be encoded into the dual graph of the roof graph. \emph{Top}: different dual graphs (with green nodes and black edges) on top of their corresponding primal roof graph (gray nodes and gray edges). \emph{Bottom}: we show the corresponding reconstructed roofs with different styles.}
    \label{fig:mtd:diff_adjacency}\vspace{-3pt}
\end{figure}

\subsection{Background: Straight Skeleton Methods}
%
\InsertBoxR{2}{\parbox{0.35\linewidth}{
\centering
\begin{overpic}[trim=3.2cm 8.5cm 5cm 9cm,clip,width=1\linewidth,grid=false]{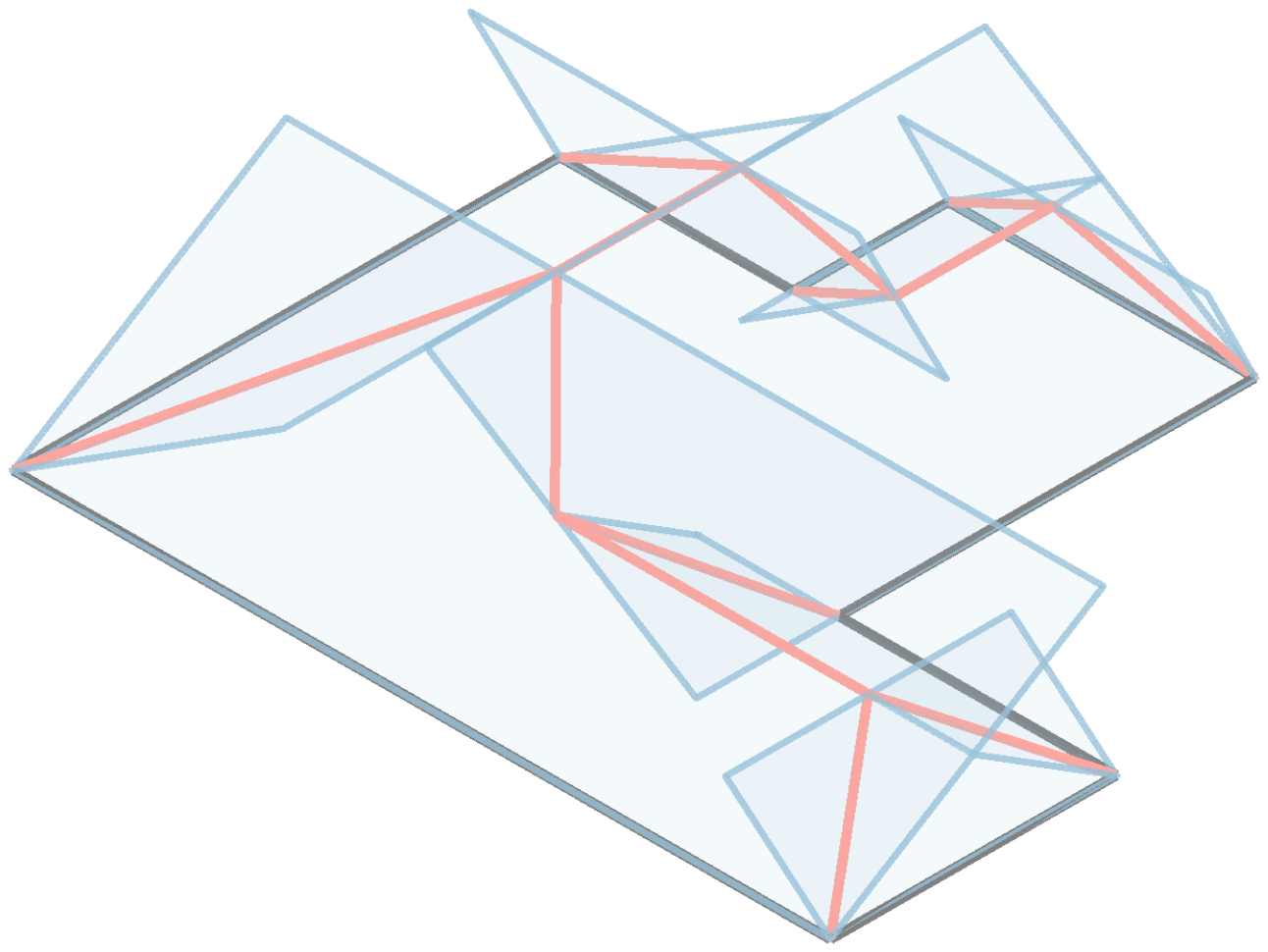}
\end{overpic}
}}[2]
The straight skeleton~\cite{aichholzer1996novel,aichholzer1996straight} or its extension the weighted straight skeleton~\cite{eppstein1999raising,biedl2015weighted,kelly2011interactive} are popular tools for roof construction. The straight skeleton based methods take a 2D roof outline as input, and output a valid 2D/3D roof embedding by solving for the roof \emph{topology} and roof \emph{embedding} at the same time.
Specifically, the straight skeleton methods formulate the roof construction problem as determining how roof planes with given slope from a given roof outline intersect with each other. See the inset figure for an example, where blue planes stemming from the outline intersect and form the roof structure colored in red. 
This can be equivalently formulated as shrinking the input outline edges with a constant rate and determining how the resulting interior polygon changes. Once there is a change, an interior roof vertex or roof edge can be detected accordingly. See~\cite{felkel1998straight} for a more detailed description.

\subsection{Observations \& Challenges}
\begin{figure}[!t]
    \centering
    \includegraphics[width=1\linewidth]{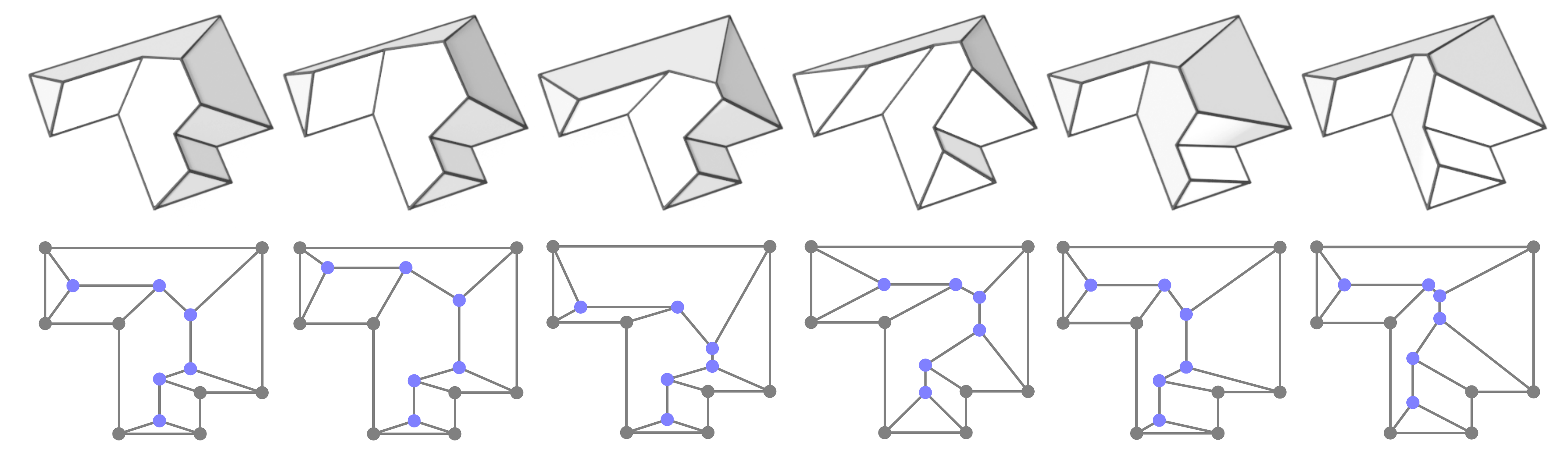}\vspace{-12pt}
    \caption{\textbf{Multiple valid embeddings for the same roof graph}. \emph{Top}: a set of 3D planar roofs with the same outline and topology. \emph{Bottom}: $xy$ projection of the corresponding roofs.}
    \label{fig:mtd:local_minima}\vspace{-6pt}
\end{figure}

The straight skeleton based methods can construct planar roofs from given roof outline efficiently. However, they still have some limitations. For example, the roof topology is determined at the same time as the roof embedding, with an implicit assumption that a single roof topology corresponds to the input roof outline, which is not the case in practice. 
For example, we can observe that:
\begin{itemize}[leftmargin=*]
\item \textbf{Roofs with the same outline can have different styles}. In Fig.~\ref{fig:mtd:diff_adjacency} we show four different roof constructions for the \emph{same} outline. We can observe that these four roofs have different style and structure.
\item \textbf{Roofs with the same outline and topology can have different embeddings}. In Fig.~\ref{fig:mtd:local_minima} we show a set of 3D planar roofs with exactly the same outline and the same adjacency between faces. All of the shown roofs are valid but the 3D roof embeddings are different, i.e., the roof vertices have different locations. 
\end{itemize}
These observations suggest that it is \emph{not} enough to only use the roof outline for roof construction, as the straight skeleton based methods do. We also need to specify the \emph{roof topology} and \emph{geometry} in some way. As a result, to solve the roof construction problem, we need to tackle the following challenges:
\begin{itemize}[leftmargin=*]
    \item How to specify the \emph{roof topology} or \emph{roof style}, i.e., the geometry of the interior vertices of a roof?
    \item How to formulate or enforce \emph{roof planarity}?
    \item How to \emph{avoid undesirable but valid} roof embeddings? 
\end{itemize}
\revised{
In the following, we will discuss how the proposed primal-dual roof graph representation can help to tackle these challenges and solve for desirable planar roofs.
}

\section{Methodology: Roof Optimization}\label{sec:mtd:roof_optimization}
\begin{figure}[!t]
    \begin{overpic}[trim=0cm 36cm 16cm 0cm,clip,width=1\linewidth,grid=false]{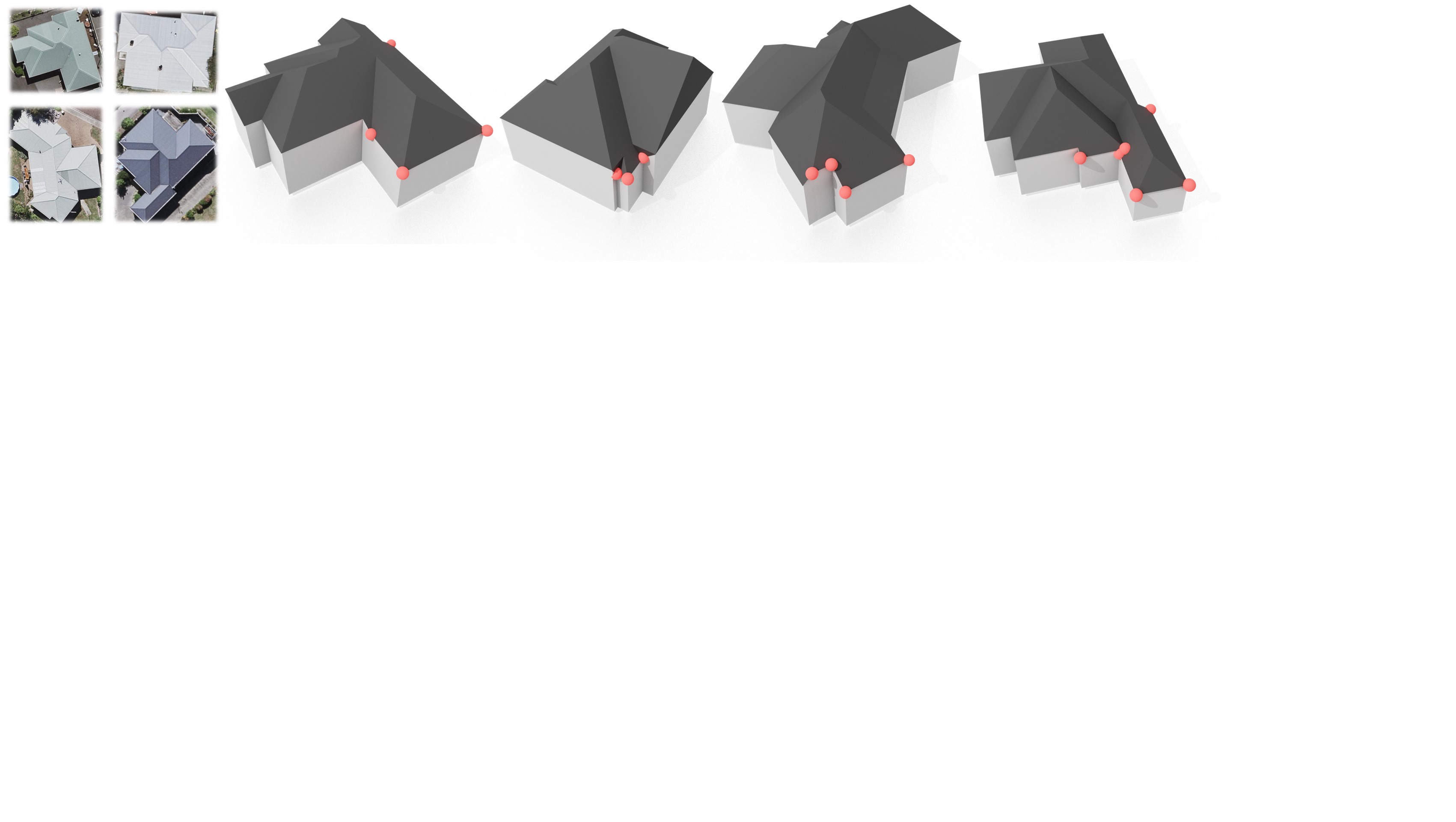}
    \end{overpic}\vspace{-10pt}
    \caption{Roofs with outline vertices in different height. We highlight the outline vertices with non-zero height in red.}
    \label{fig:mtd:eg_building_outline_diff_height}
\end{figure}

In this section, we introduce an optimization-based method to construct a 3D planar roof where the roof structure/style is encoded into a primal or dual roof graph.
We first discuss how to formulate the roof planarity, where we introduce a planarity metric to measure the validity of an arbitrary 3D embedding of a roof graph (Sec.~\ref{sec:mtd:def_planarity}).
We then discuss how to reconstruct a planar roof from its primal roof graph (Sec.~\ref{sec:mtd:opti_primal}) or its dual graph (Sec.~\ref{sec:mtd:opti_dual}) respectively. For simplicity of method description, we make the same assumptions as the straight skeleton method: (1) the outline vertices of a roof are in the same height, and (2) each roof face stems from one of the outline edges. We then discuss in Sec.~\ref{sec:mtd:relax_assumption} how to relax these assumptions to deal with roofs with outline vertices in different height (e.g., Fig.~\ref{fig:mtd:eg_building_outline_diff_height}) and roofs with faces having multiple outline edges (e.g., Fig.~\ref{fig:intro:ss_prob_eg2_multi_outlineEdge}), which are not supported by straight skeleton based methods.

\subsection{Roof Planarity Formulation}\label{sec:mtd:def_planarity}
Assume we have a 3D embedding $X$ for the roof graph $G = (V, F)$, i.e., $X_i$ is a 3D position for the vertex $v_i$ in the roof graph, how can we evaluate the validity of the embedding $X$? 
\paragraph{\textbf{Planarity metric on a point set}}
We first propose to use the following metric to measure the planarity of a set of 3D points $Z$: $g(Z) = \sigma_1 \big(\text{Cov}(Z)\big)$, where $\sigma_1(M)$ is the \emph{smallest eigenvalue} of the square matrix $M$, $\text{Cov(Z)}$ gives the \emph{covariance} matrix of the set of points $Z$.
We know that  $g(Z) \ge 0$ since the covariance matrix is positive semi-definite.
If the 3D points in $Z$ are coplanar, the rank of the $Z$ is at most 2. Then, the smallest eigenvalue of the covariance matrix is 0 and we have $g(Z) = 0$. 
Therefore, for an arbitrary set of 3D points $Z$, the smaller $g(Z)$ is, the more coplanar the points $Z$ are. 
We therefore use $g(Z)$ to measure the planarity of a set of points $Z$. 
Note that $g(Z)$ is \textit{differentiable} with respect to $Z$.
\paragraph{\textbf{Polygonal roof planarity}}
With the planarity metric in hand, we can easily measure the validity (planarity) of a roof embedding $X$.
Specifically, we denote as $X_{f_i}$ the corresponding 3D embedding of the face $f_i$ in $F$. As discussed above, $g(X_{f_i})$ measures the planarity of the embedding of face $f_i$. We can sum over the planarity metric on each face $f_i$ as a measure of the roof planarity:
\begin{equation}\label{eq:mtd:planarity}
\Scale[0.9]{ \mathllap{\mathbf{E}_{\text{planarity}}}\big(X\big)= \hspace{2pt}\sum_{i=1}^{n_f}  \sigma_1\Big(\text{Cov}\big(X_{f_i}\big)\Big)}
\end{equation}
Further, we can construct a valid roof by solving for an embedding $X$ that has zero planarity error as defined above, which can be formulated as an optimization problem. In the following, we will discuss in details how to achieve a roof construction from a primal or a dual roof graph, respectively.

\subsection{Roof Construction from Primal Graph}\label{sec:mtd:opti_primal}
We assume that we are given a primal roof graph $G = (V, F)$. For example, a user can draw a roof graph similar to Fig.~\ref{fig:roof_graph} (b). In this case, a 2D embedding $\widebar{X}^{\text{user}}$ is also provided by the user. 
Recall that we use $\widebar{X}$ to denote a 2D embedding and use $X$ to denote a 3D embedding of the vertex set $V$. 
Note that this 2D embedding $\widebar{X}^{\text{user}}$ is \emph{unlikely} to be valid due to the noise in the user annotations, but it provides a strong prior of the expected positions of the roof vertices from the user. Then our goal is to solve for a \emph{valid} 3D embedding $X$ from the user input.

\begin{figure}[!t]
    \centering
    \input{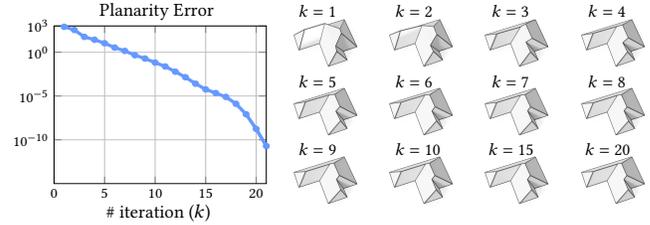}\vspace{-9pt}
    \caption{Planarity error over iterations. We visualize the 3D embedding updated over iterations by minimizing the planarity measure.}
    \label{fig:mtd:energy_per_iteration}\vspace{-6pt}
\end{figure}

\paragraph{\textbf{Preprocessing}}
We first lightly regularize the 2D positions of the outline vertices from user input, i.e. $\widebar{X}^{\text{user}}_{\O}$,  to promote the accuracy of the parallel edges. Specifically, the outline edges labeled or drawn by users can be inaccurate: some outline edges that should be parallel are only approximately parallel. Therefore, for a pair of edges that has a smaller angle than the threshold $\theta$, we modify the outline vertex positions a bit to make them parallel to each other numerically, and this leads to new outline vertex positions $\widebar{X}_{\O}$.

\paragraph{\textbf{Problem formulation}}
To find a valid 3D embedding, without loss of generality $X$, we can fix the outline vertices $V_{\O}$ to $X_{\O} = [\widebar{X}_{\O},\mathbf{0}]$ (i.e., with 0 $z$-axis value). Our goal is to find a 3D embedding $X_{\R}$ for the roof vertices $V_{\R}$. We propose to solve the following problem:
\begin{equation}\label{eq:mtd:prob_primal}
    \Scale[0.9]{\min_{X_{\R}}\,\,\mathbf{E}_{\text{planarity}}\big(X\big) + \lambda \mynorm{\widebar{X}_{\R} - \widebar{X}^{\text{user}}_{\R}} \quad \text{s.t.}\,\, x_z^* = h}
\end{equation}
where $x^*$ is a randomly selected roof vertex in $X_{\R}$, $h$ is a pre-defined roof height parameter. 
We can optimize the above problem with initialization $X_{\R} = [\widebar{X}^{\text{user}}_{\R}, \mathbf{h}]$, i.e., set the $z$-axis value of all the roof vertices to $h$.
Our objective function promotes the planarity of the embedding $X$ and at the same time promotes its corresponding 2D embedding $\widebar{X}$ to be close to the user input. 
See Fig.~\ref{fig:mtd:energy_per_iteration} for an example where we show the intermediate roofs over iterations.
We also include a hard constraint that enforces one of the roof vertices to have $z$-value (height) as $h$. This design choice has two advantages: (1) it helps to avoid degenerate global minimizers. Without this hard constraint, we can see that any arbitrary 2D embedding of the roof graph with zero $z$-axis values leads to a 3D embedding of the graph where the planarity of each face is satisfied. To avoid this type of degenerate solutions, we can force that at least one roof vertex has non-zero height. (2) this hard constraint also provides the user a way to control the overall height of the constructed 3D roof.

\subsection{Roof Construction from Dual Graph}\label{sec:mtd:opti_dual}
Another scenario is that we are given a dual roof graph $G^{\D} = (V^{\D}, A^{\D})$. 
Recall that each roof face $f_i$ is represented as a node in $V^{\D}$, and $A^{\D}$ stores the face adjacency.
In practice, such a dual graph can be described by the roof outline $V_{\O}$ and $A^{\D}$, 
Specifically, the 2D roof outline is given as a list of consecutive 2D points. Once stored in a matrix we have $\widebar{X}_{\O}\in\mathbf{R}^{n_{\O}\times 2}$. I.e., we have $n_{\O}$ outline vertices $V_{\O}$ embedded in 2D with vertex positions $\widebar{X}_{\O}$.  Then we can obtain $n_{\O}$ outline edges, $E_{\O} = \big\{e_{1,2},\cdots, e_{i, i+1}, \cdots, e_{n_{\O},1}\big\}$, where the edge $e_{i,i+1}$ connects the outline vertices $v_{i}$ and $v_{i+1}$. We assume each roof face stems from one of the outline edges, and we denote $f_i$ as the face that is associated with the outline edge $e_{i, i+1}$. Then the face adjacency is specified in the matrix $A^{\D}\in\{0, 1\}^{n_{\O}\times n_{\O}}$.

The roof outline $\widebar{X}_{\O}$ can either be drawn by a user or generated by a transformer. Similarly the face adjacency $A^{\D}$ can either be specified by a user or predicted by a trained network.

\begin{figure}[!t]
    \centering\vspace{2pt}
    \input{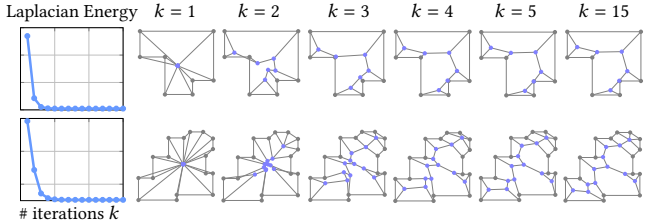}\vspace{-15pt}
    \caption{2D Spectral Embedding. Here we show two examples of embedding a roof graph into 2D with fixed outline by minimizing the Laplacian energy. We initialize all the roof vertices at the center of the roof outline.}
    \label{fig:mtd:2D_spectral}
\end{figure}

\paragraph{\textbf{Recovering primal from dual}}
Since the roof planarity is defined on the primal roof graph representation, we need to first recover the primal graph from its dual graph.
Specifically, we first add an outside node to the dual graph, and connect all the node in $G^{\D}$ to the outside node to obtain a complete dual graph. Then the primal graph can be recovered by \revised{computing} the dual of the complete dual graph (see Fig.~\ref{fig:mtd:complete_dual}).

\paragraph{\textbf{Problem formulation}} With the recovered primal roof graph, we can solve for a roof embedding $X$ by optimizing the roof planarity as before. However, there is a big difference from the previous case, where we do not have the user-specified roof interior structure $\widebar{X}^{\text{user}}_{\R}$ for initialization and for guiding the roof optimization to a preferred structure. We therefore propose a new energy:
\begin{equation}\begin{split}\label{eq:energy:planarity}
 \Scale[0.9]{\min_{X_{\R}}\,\, \mathbf{E}_{\text{planarity}}\big(X\big) + \gamma \mathbf{E}_{\text{aesthetic}}\big(X\big) \quad \text{s.t.}\,\, x_z^* = h}
\end{split}\end{equation}
where $\mathbf{E}_{\text{\revised{aesthetic}}}\big(X\big)$ encodes some additional aesthetic constraints which can help to solve for a \emph{planar} roof with \emph{preferred} properties. 
\begingroup
\setlength{\columnsep}{0pt}%
\setlength{\intextsep}{0pt}%
\begin{wrapfigure}{r}{0.45\linewidth}
\centering
\begin{overpic}[trim=0.2cm 0.2cm 0.18cm 0.2cm,clip,width=1\linewidth,grid=false]{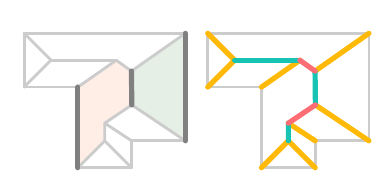}
\put(24.5,25){\scriptsize $e_p$}
\put(7.5,15){\scriptsize $e_{p_1}$}
\put(37.5,26){\scriptsize $e_{p_2}$}
\put(4,4){\scriptsize\bfseries (a)}
\put(55,4){\scriptsize\bfseries (b)}
\end{overpic}
\end{wrapfigure}
For example, we can categorize the \emph{roof edges} into three categories according to Remark~\ref{rmk:valid_2D}: the roof edges parallel to the corresponding outline edges are colored green ($\,\color{myblue}{\textedge}\,$), the roof edges that connect to an outline vertex are colored yellow ($\,\color{myyellow}{\textedge}\,$), and the other roof edges are colored red ($\,\color{myred}{\textedge}\,$), as illustrated in (b) of the inset figure. In practice, we would like to have (1) the green roof edge has equal distance to the corresponding outline edges, i.e., in the medial axis; (2) the yellow roof edge is an angle bisector that equally splits the angle formed by the two corresponding outline edges. We therefore have the following aesthetic constraints:
\begin{equation*}
\Scale[0.9]{
    \mathbf{E}_{\text{aes.}}=
     \sum\limits_{p\in \{\,{\color{myyellow}{\textedge}}\,\}  } \big\Vert\langle \vec{e}_{p}, \vec{e}_{p_1} \rangle - \langle\vec{e}_{p}, \vec{e}_{p_2}\rangle\big\Vert_F^2 + \sum\limits_{q\in\{\,{\color{myblue}{\textedge}}\,\}}\big\Vert  \text{dist}(\vec{e}_{q}, \vec{e}_{q_1}) - \text{dist}(\vec{e}_{q}, \vec{e}_{q_2})\big\Vert_F^2} 
\end{equation*}
where $\vec{e}$ is an unit vector on edge $e$; for a roof edge $e$, we can find the outline edge $e_{p_1}, e_{p_2}$ in its neighboring faces as illustrated in (a) of the inset figure; $\Scale[0.9]{\text{dist}(\vec{a}, \vec{b})}$ gives the distance between two parallel unit vectors $\vec{a}$ and $\vec{b}$.

\endgroup

\begin{figure}[!t]
    \centering
    \input{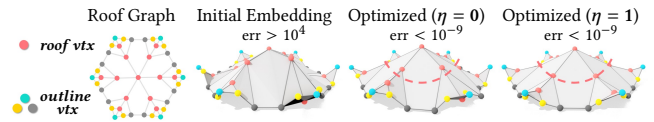}
    \vspace{-22pt}
    \caption{Constructing the roof of the hexagonal pavilion shown in Fig.~\ref{fig:eg_temple}, where the outline vertices have different height. Starting from the initial embedding (\emph{left}), we optimize the roof planarity with (\emph{right}) and without (\emph{middle}) the variance energy defined on vertex height, i.e., with $\eta = 1$ and $\eta = 0$ respectively. The red dashed and curved lines highlights the fact that the roof embedding is more symmetric with the variance energy. We also report the planarity error ("err") of the three embeddings.}
    \label{fig:mtd:pavilion_outline_vtx}
\end{figure}

\paragraph{\textbf{Initialization: 2D embedding by spectral drawing}}
Without user inputs, we propose to use spectral graph drawing in 2D as initial embedding for planarity optimization as opposed to random initialization, which can help to avoid self-intersections. 
Specifically, we first find a 2D embedding $\widebar{X}_{\R}$ for the roof vertices by minimizing the Dirichlet energy using the graph Laplacian~\cite{ren2017graph}, then we initialize the 3D embedding by $X_{\R} = [\widebar{X}_{\R}, \textbf{h}]$. We can then update the 3D embedding by minimizing the roof planarity as discussed above. 

To embed the roof graph in 2D by spectral embedding, we first construct the adjacency matrix $A_V\in\mathbf{R}^{n\times n}$ between the \emph{vertices} in $V$, i.e., $A_V(p,q)$ equals to 1 if $(v_p, v_q)$ is an edge in some face $f_i$, and equals to $0$ otherwise. We can then construct the graph Laplacian $L_V = \text{diag}\big(\mathbf{1}_{n} \tran A_V\big) - A_V$. Then we can embed the roof graph with fixed outline by minimizing the Laplacian energy:
\begin{equation}\label{eq:energy:laplacian}
    \Scale[0.9]{
    \min_{\widebar{X}_{\R}} \quad
    \Bigg\Vert
    \begin{pmatrix}
    \widebar{X}_{\O}\\
    \widebar{X}_{\R}
    \end{pmatrix}\tran L_V
    \begin{pmatrix}
    \widebar{X}_{\O}\\
    \widebar{X}_{\R}
    \end{pmatrix}
    \Bigg\Vert_F^2
    }
\end{equation}
Note that, the spectral energy is considered at the complete roof graph while we only solve for the roof vertices $\widebar{X}_{\R}$ with fixed outline $\widebar{X}_{\O}$. In this case, we can obtain a planar 2D embedding without self-intersections (see Fig.~\ref{fig:mtd:2D_spectral} for some examples).

\subsection{Relaxing the Assumptions on Roof Graphs}\label{sec:mtd:relax_assumption}
Here we discuss how to use our optimization-based formulation to handle roofs with outline vertices at different height and roofs with faces containing multiple outline edges.

\paragraph{\textbf{Roof with outline vertices at different heights}} 
We can simply extend our method to handle a roof with outline vertices at different heights by setting the outline vertices as free variables for the optimization.
Our method can handle common cases such as two roof outline edges of different heights emanating from the same vertex or sloped roof outline edges.
See Fig.~\ref{fig:mtd:pavilion_outline_vtx} for an example, where we label the outline vertices in three categories colored in green ($\textcolor{myblue}{\textcirc}$), yellow ($\textcolor{myyellow}{\textcirc}$) , and gray ($\textcolor{gray}{\textcirc}$) respectively. 
To construct a realistic pavilion, the green and yellow outline vertices are expected to be higher than the gray outline vertices. Therefore, we propose to solve the following problem:
\begin{equation*}
\Scale[0.9]{
    \min\limits_{x^{\textcolor{myred}{\textcirc}}_{xyz},\, x_{z}^{\textcolor{myblue}{\textcirc}, \textcolor{myyellow}{\textcirc}}} \mathbf{E}_{\text{planarity}}(X) + \lambda\mynorm{\widebar{X} - \widebar{X}^{\text{user}}} + \eta \text{Var}(x_{z}^{\textcolor{myblue}{\textcirc}}) + 
    \eta \text{Var}(x_{z}^{\textcolor{myyellow}{\textcirc}})}
\end{equation*}
where $x^{\textcolor{myred}{\textcirc}}_{xyz}$ means that the $xyz$-axis values of the red vertices are variables for optimization,  and $x_{z}^{\textcolor{myblue}{\textcirc}, \textcolor{myyellow}{\textcirc}}$ means that only the $z$-axis value of the green/yellow vertices are variables while their $xy$-axis values are fixed. 
There are two modifications made to Eq.~\eqref{eq:mtd:prob_primal}: (1) we set the $z$-axis value of the green and yellow outline vertices as variables for the optimization besides the positions of the roof vertices. (2) we add extra energy terms to regularize the \emph{variance} of the height of the green/yellow vertices, $\text{Var}(x_{z}^{\textcolor{myblue}{\textcirc}})$ and $\text{Var}(x_{z}^{\textcolor{myyellow}{\textcirc}})$. The additional regularizers can help to construct a more symmetric and realistic pavilion (see Fig.~\ref{fig:mtd:pavilion_outline_vtx} with $\eta=0$ and $\eta=1$).
In summary, our optimization-based formulation is flexible to address user preferences by adding variables to the optimization and including different types of regularizers for different types of roofs.
Fig.~\ref{fig:mtd:eg_building_outline_diff_height} shows more examples of roofs with outline vertices with different heights.

\begin{figure}[!t]
    \centering
    \vspace{-6pt}
    \input{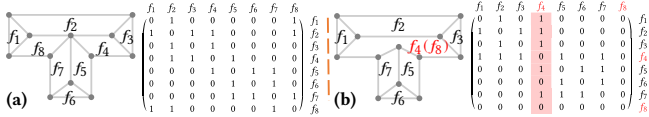}
    \vspace{-22pt}
    \caption{\textbf{Roof face containing multiple outline edges}. We show two examples (a-b) of using dual graph (\emph{right}) to encode the roof topology (\emph{left}). In \textbf{(b)}: we merge the face $f_4$ and $f_8$ in (a) into a single face, and we highlight the changes in the face adjacency matrix on the right.}
    \label{fig:appendix:adj_multi_outline}\vspace{-6pt}
\end{figure}

\paragraph{\textbf{Roof face containing multiple outline edges}}
Our roof optimization from the primal graph can handle the case where a face contains multiple outline edges directly. Here we only discuss how to handle this with the dual graph as input.
The main issue is how to represent such a roof in a dual graph. This can be easily handled by modifying the face adjacency matrix $A^{\D}$.
Specifically, each outline edge corresponds to a roof face; and for the set of outline edges that correspond to the same face, the corresponding rows in $A^{\D}$ are merged to the first outline edge, while the rest outline edges are ignored by setting all the entries to 0.
%
See Fig.~\ref{fig:appendix:adj_multi_outline} for an example of how $A^{D}$ is constructed.
In this way, we can use a dual graph to represent the roof topology where faces can have multiple outline edges. Note that the straight skeleton methods do not support this feature (e.g., see Fig.~\ref{fig:intro:ss_prob_eg2_multi_outlineEdge}, Fig.~\ref{fig:intro:edit_straight_skeleton}, and Fig.~\ref{fig:res2:ss_wss_ours}).

\begin{figure}[!t]
    \centering
    \includegraphics[width=1\linewidth]{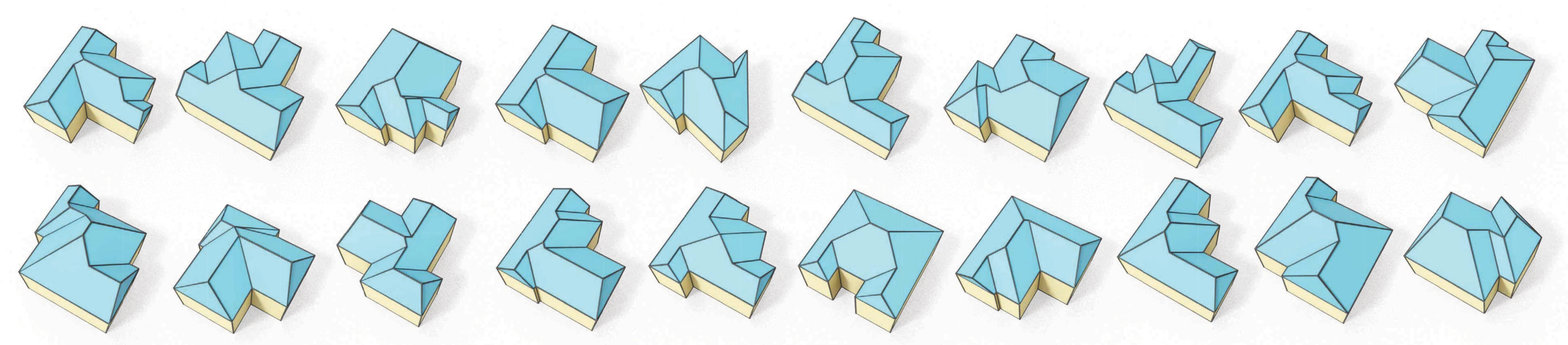}\vspace{-12pt}
    \caption{\textbf{Our synthesized buildings from scratch}. Our method can automatically generate realistic roof outlines and correctly predict face adjacency. We then run our roof optimization method to construct roofs from the learned components.}
    \label{fig:res:synthesized_roofs}
\end{figure}

\begin{figure}[!t]
    \centering
    \definecolor{wwccqq}{rgb}{0.4,0.8,0}
\definecolor{wwccff}{rgb}{0.4,0.8,1}
\definecolor{ffzzcc}{rgb}{1,0.6,0.8}
\definecolor{orange}{RGB}{255,204,51}
\definecolor{violet}{RGB}{125,125,255}
\definecolor{olive}{RGB}{255,127,0}
\begin{tikzpicture}[remember picture]
    \tikzstyle{transformer} = [inner sep=0pt, rectangle, rounded corners, minimum width=2cm, minimum height=0.5cm, text centered, draw=ffzzcc!80, fill=ffzzcc!10]
    \tikzstyle{embedding} = [inner sep=0pt, rectangle, rounded corners, minimum width=1.5cm, minimum height=0.4cm, text centered, draw=orange!80, fill=orange!10]
    \tikzstyle{block} = [inner sep=0pt, rectangle, rounded corners, minimum width=1.5cm, minimum height=0.4cm, text centered, draw=violet!80, fill=violet!10]
    \tikzstyle{attention} = [inner sep=0pt, rectangle, rounded corners, minimum width=1.3cm, minimum height=0.4cm, text centered, draw=wwccff!80, fill=wwccff!10]
    \tikzstyle{mlp} = [inner sep=0pt, rectangle, rounded corners, minimum width=1.3cm, minimum height=0.4cm, text centered, draw=olive!80, fill=olive!10]
    \matrix[matrix of nodes, column sep=1mm, nodes={rectangle, minimum width=0.5cm, minimum height=0.4cm, text centered, draw=wwccqq!80, fill=wwccqq!10, anchor=center}] (token) {
     \small\phantom{0} & \small$\nu_1$ & \small$\nu_2$ & |[draw=gray!40, fill=gray!5]| \small$\nu_3$ & |[draw=gray!40, fill=gray!5]| \tiny$\cdots$ \\
    };
    \node[below = 0.4cm of token, transformer] (trans) {\footnotesize Transformer};
    \begin{scope}[on background layer]
    \draw[line width=0.8pt, dashed,draw=wwccff!80] (token-1-1.south west) -- (trans.north);
    \draw[line width=0.8pt, dashed,draw=wwccff!80] (token-1-3.south east) -- (trans.north);
    \fill [opacity=0.1,wwccff!80] (token-1-1.south west) -- (trans.north) -- (token-1-3.south east) -- cycle;
    \end{scope}
    
    \node[below = 0.4cm of trans] (prob) {$p(\nu_3|\nu_{<3})$};
    \draw[line width=0.8pt, -stealth, draw=gray!90] (trans) -- (prob);
    
    \node[right = 1.0cm of trans, transformer, minimum width=2.0cm, minimum height=2.4cm, fill=ffzzcc!3] (trans2) {
        \begin{tikzpicture}[remember picture]
            \node[embedding] (embeddings) {\footnotesize Embeddings};
            \node[below = 0.2cm of embeddings, block] (block1) {\footnotesize Block 1};
            \draw[line width=0.8pt, -stealth, draw=gray!90] (embeddings) -- (block1);
            \node[below = 0.2cm of block1, inner sep=0pt, minimum width=1cm, minimum height=0.2cm] (vdots) {$\,\cdots\,$};
            \draw[line width=0.8pt, -stealth, draw=gray!90] (block1) -- (vdots);
            \node[below = 0.2cm of vdots, block] (blockl) {\footnotesize Block $L$};
            \draw[line width=0.8pt, -stealth, draw=gray!90] (vdots) -- (blockl);
        \end{tikzpicture}
    };
    \draw[line width=0.8pt, -, dashed, draw=ffzzcc!80] (trans.north east) -- (trans2.north west);
    \draw[line width=0.8pt, -, dashed, draw=ffzzcc!80] (trans.south east) -- (trans2.south west);
    
    \node[right = 0.3cm of trans2, block, minimum width=2cm, minimum height=2.4cm, fill=violet!3] (blocki) {
        \begin{tikzpicture}[remember picture]
            \node[minimum width=0.2cm, minimum height=0.1cm, inner sep=0pt] (blockinput) {};
            \node[below = 0.2cm of blockinput, attention] (atten) {\footnotesize Attention};
            \node[below = 0.2cm of atten, minimum width=0.3cm, minimum height=0.2cm, inner sep=0pt] (res) {\textcolor{gray!90}{$\oplus$}};
            \node[below = 0.2cm of res, mlp] (mlp) {\footnotesize MLP};
            \node[below = 0.2cm of mlp, minimum width=0.3cm, minimum height=0.2cm, inner sep=0pt] (res2) {\textcolor{gray!90}{$\oplus$}};
            \node[below = 0.2cm of res2, minimum width=0.3cm, minimum height=0.01cm, inner sep=0pt] (blockoutput) {};
            
            \draw[line width=0.8pt, -stealth, draw=gray!90] (blockinput.north) -- (atten);
            \draw[line width=0.8pt, -, draw=gray!90] (atten) -- (res);
            \draw[line width=0.8pt, -stealth, draw=gray!90] (res.north) -- (mlp);
            \draw[line width=0.8pt, -, draw=gray!90] (mlp) -- (res2);
            \draw[line width=0.8pt, -stealth, draw=gray!90]  (blockinput.south) -- ++(6ex,0)coordinate (a) -- ($(a|-res.north)$) -- (res.north);
            \draw[line width=0.8pt, -stealth, draw=gray!90]  (res.south) -- ++(6ex,0)coordinate (a) -- ($(a|-res2.north)$) -- (res2.north);
            \draw[line width=0.8pt, -stealth, draw=gray!90] (res2.north) -- (blockoutput);
        \end{tikzpicture}
    };

    \draw[line width=0.8pt, -, dashed, draw=violet!80] (block1.north east) -- (blocki.north west);
    \draw[line width=0.8pt, -, dashed, draw=violet!80] (block1.south east) -- (blocki.south west);
\end{tikzpicture}
    \vspace{-9pt}
    \caption{Our auto-regressive transformer with input of a flattened vertex sequence, and output of the probability distribution of the next token.}
    \label{fig:transformer}
\end{figure}
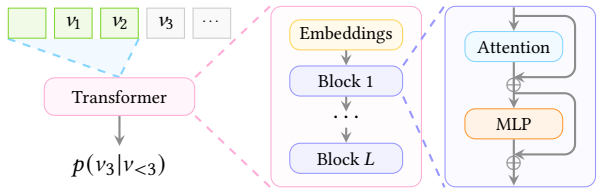

\revised{
\paragraph{\textbf{Alternative solutions}}
In Appendix~\ref{sec:diff_planarity} we discuss three alternative planarity metrics that can be used for planar roof modeling as well. 
In Appendix~\ref{sec:opti_2D} we introduce another solution to optimize for a valid 2D embedding directly based on Remark~\ref{rmk:valid_2D}.
}

\begin{figure}[!t]
    \centering
    \vspace{-9pt}
    \input{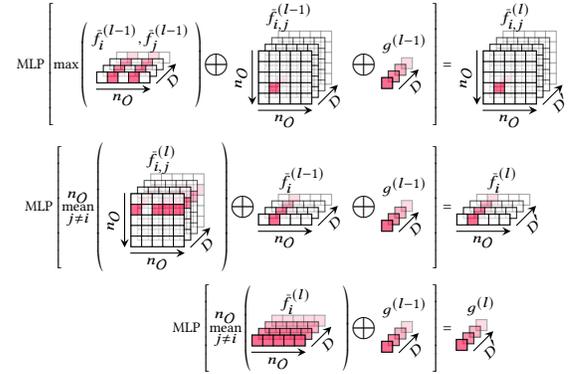}\vspace{-15pt}
    \caption{\textbf{Face adjacency prediction building block}. $D$ is the feature dimensions. $\bigoplus$ is the concatenation operator. From top to bottom, we show the adjacency model, the edge model, and the global model}
    \label{fig:adj-model}\vspace{-6pt}
\end{figure}

\begin{figure}[!t]
    \centering
    \includegraphics[width=1\linewidth]{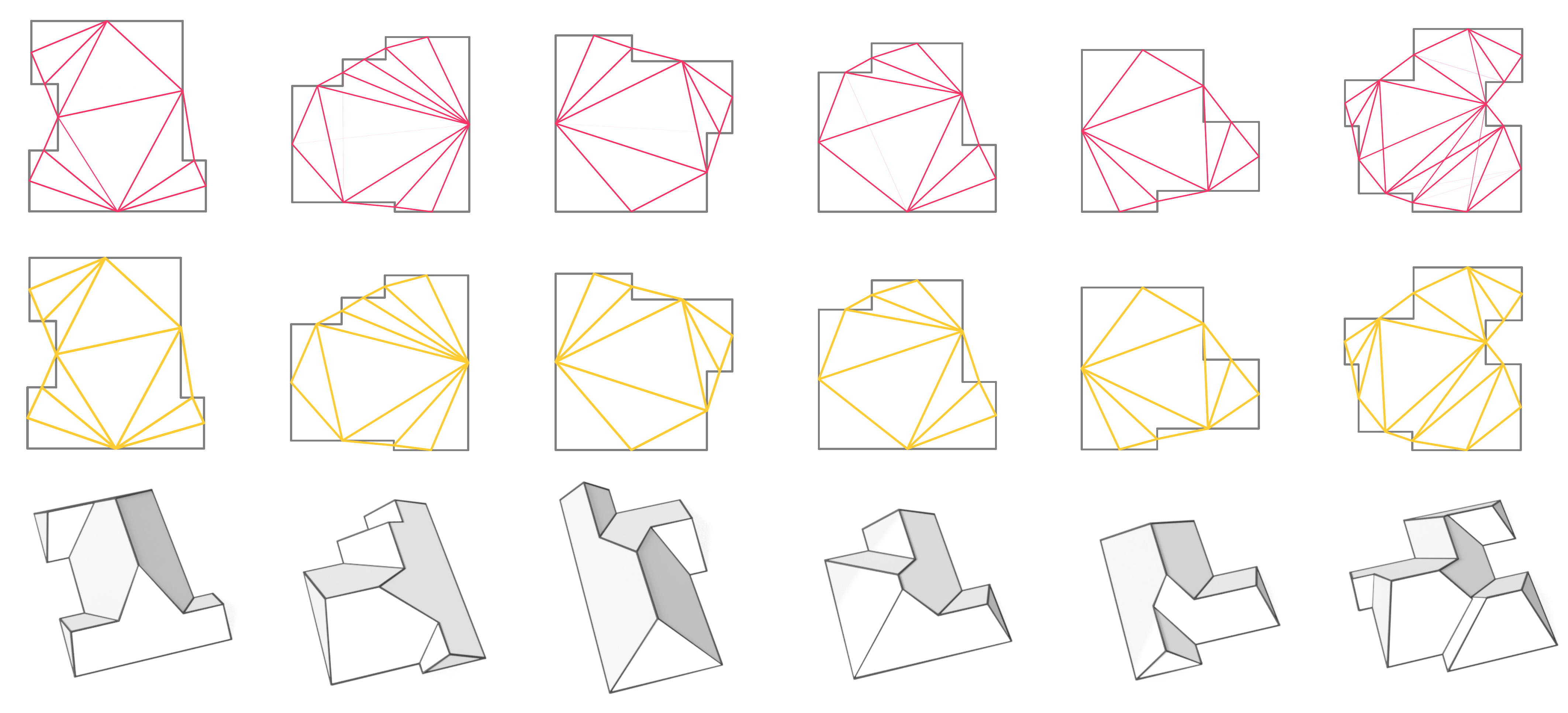}
    \vspace{-22pt}
    \caption{\emph{Top}: the predicted adjacency with probability using our transformer. \emph{Middle}: post-processed adjacency that forms a valid dual graph. \emph{Bottom}: the corresponding constructed 3D roofs using our method.\label{fig:res:learned_adj}}
    \vspace{-3pt}
\end{figure}

\begin{figure*}[!t]
    \centering
    \begin{overpic}[trim=0cm 2cm 0cm 0cm,clip,width=1\linewidth,grid=false]{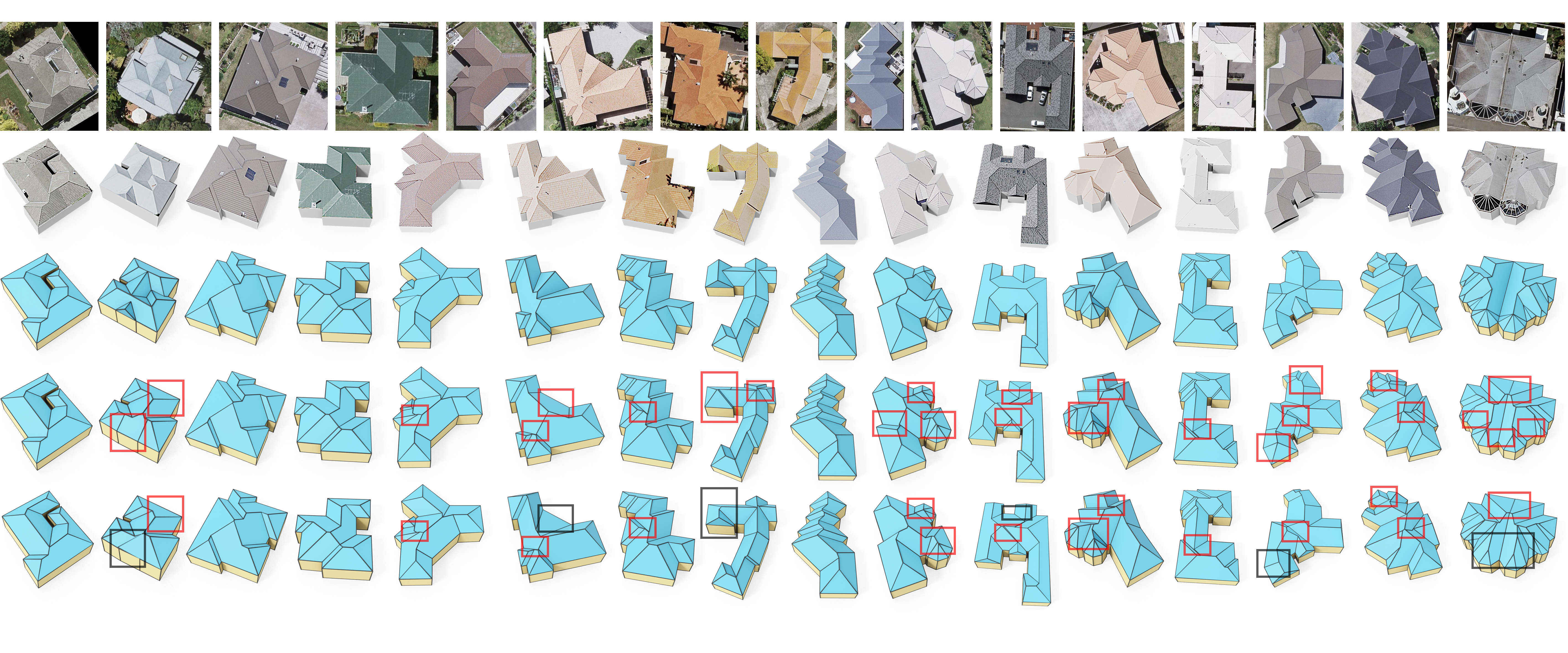}
    \put(-2,33){\footnotesize\bfseries (a)}
    \put(-2,24.5){\footnotesize\bfseries (b)}
    \put(-2,17){\footnotesize\bfseries (c)}
    \put(-2,9.5){\footnotesize\bfseries (d)}
    \put(-2,2){\footnotesize\bfseries (e)}
    \end{overpic}
    \vspace{-20pt}
    \caption{\textbf{Roof reconstruction from aerial images}. \textbf{(a)} 16 input images of roofs with different complexity. \textbf{(b)} our reconstructed roofs with texture. \textbf{(c)} the topology of our reconstructed roofs. \textbf{(d)} the results of the straight skeleton where the visually erroneous regions are colored in red. \textbf{(e)} we use the weighted straight skeleton method to refine the results in (d) aiming to make them more consistent with the input image. The weighted straight skeleton successfully resolved 5 inconsistencies. However it also introduced new inconsistencies (colored in black) and failed to resolve several inconsistencies (colored in red).}
    \label{fig:res2:img_recon}
\end{figure*}
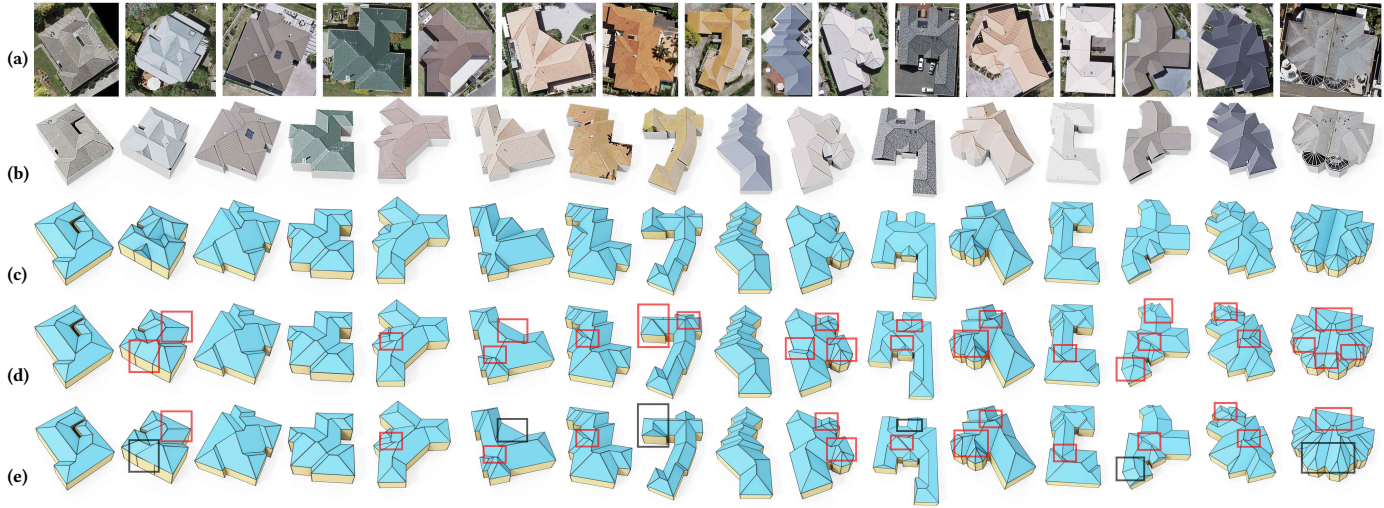

\revised{
\section{Generative Models for Roof Synthesis}\label{sec:mtd:roof_synthesis}

One direct application of our method is roof synthesis from scratch (see Fig.~\ref{fig:res:synthesized_roofs}). We believe that synthesizing a valid roof directly can be hard since the model needs to take care of the discrete constraints (roof topology) and the continuous constraints (roof embedding) at the same time.
We propose to tackle the roof synthesis problem by combining generative models for roof topology generation (dealing with discrete constraints only) with roof optimization (dealing with continuous constraints only).
Specifically, we propose a \emph{transformer} for roof outline generation and a \emph{graph neural network} for face adjacency prediction. The architectures for both networks can be found in the supplementary materials.
\paragraph{\textbf{Outline generation}}
Our goal is to model a distribution of the 2D roof outline $X_{\O}\in\mathbf{R}^{n_{\O}\times 2}$, where we assume the outline vertices are in counter-clockwise order and the first vertex is the one closest to the lower left corner.
We flatten the coordinate matrix to $N^{seq} = \left\{\nu_1, \nu_2, \cdots, \nu_{2n_{\O}}\right\}$. 
The vertex values are first normalized to range $[0, 1]$ and then quantized to $b$-bits, i.e., $\nu_i$ belongs to the set $\{1, 2, \cdots, 2^b\}$ for any $i$. We also append the sequence $N^{seq}$ with a stopping token $s$. Consequently, the sequence has the length of $2n_{\O}+1$ and each entry of the sequence has $2^b+1$ kinds of tokens.
We train a transformer~\cite{vaswani2017attention} to convert the input tokens $N^{seq}$ to roof outline embeddings. 
The probability of $N^{seq}$ can be factorized into a chain of conditional probabilities:
$\Scale[0.9]{p\left(N^{seq};\phi\right) = \prod^{2n_{\O}}_{i=1}p\left( \nu_i|\nu_{<i};\phi \right)}$,
where $\phi$ are the parameters of the model. The model is an auto-regressive network implemented with a transformer. The network outputs a probability $p$ at time step $i$ based on $\nu_{<i}=\left\{ \nu_1, \nu_2, \cdots, \nu_{i-1} \right\}$.
See Fig.~\ref{fig:transformer} for the structure of our transformer. We train this model by minimizing the negative log-likelihood over all training sequences.
}

\revised{
\paragraph{\textbf{Face adjacency prediction}}
With the generated roof outline, we propose a GCN~\cite{kipf2016semi} to predict the face adjacency, i.e., $p_{i,j}$, the \emph{probability} of having the face stemming from the edge $e_{i,i+1}$ being adjacent to the face stemming from the edge $e_{j,j+1}$, for all $1\le i, j \le n_{\O}$.
The network is built by $L$ basic building blocks. The $l$-th block updates 3 types of representations: (1) an edge model updates the feature representation $\Scale[0.9]{\bar{f}_i^{(l)}}$ for the edge $e_{i,i+1}$; (2) an adjacency model updates the feature representation $\Scale[0.9]{\bar{f}_{i,j}^{(l)}}$ for the adjacency $(e_{i,i+1},e_{j,j+1})$; (3) a global model updates the global feature representation $g^{(l)}$. See Fig.~\ref{fig:adj-model} for an illustration of these three building blocks.
Specifically, the input roof outline is transformed by $L$ blocks and we obtain the final adjacency representation $\Scale[0.9]{\bar{f}_{i,j}^{(L)}}$. We project the representation through a fully-connected layer which outputs the adjacency probability: $\Scale[0.9]{p_{i,j} =\mathrm{Sigmoid}\left(\mathrm{FC}\left(\bar{f}_{i,j}^{(L)}\right)\right) \in [0,1]}$.
The loss function of our GCN is the binary cross entropy between the predicted probability $p_{i,j}$ and the ground-truth adjacency $A_F$. 
See Fig.~\ref{fig:res:learned_adj} for some qualitative examples, where we visualize the probability $p_{i,j}$ via opacity (top row). We can extract the dual graph with the highest predicted probability (second row) and construct planar roofs by using our roof optimization method (bottom row).
See the supplementary materials for more details and discussions about our generative models.
}

\section{Results}
In this section, we show results of our roof optimization method from the primal graph and dual graph. We demonstrate the advantages of our method over the straight skeleton based methods and commercial software.

\subsection{Roof Reconstruction from Aerial Images}

\subsubsection{Comparison to Straight Skeleton based methods}
In Fig.~\ref{fig:res2:img_recon} we compare to the straight skeleton based methods on roof reconstruction, \revised{which take user-specified roof outlines as input}~\cite{aichholzer1996straight,eppstein1999raising}. 
We test on 16 aerial images containing roofs with different structure and complexity. Then the primal roof graph of the input image is specified by a user. We run our roof optimization method to reconstruct the roofs, and report the runtime including user labeling and optimization in Table~\ref{tb:res:img_recon} and show our reconstructed roofs in Fig.~\ref{fig:res2:img_recon} (b-c).
Note that, for the most complicated roof with more than 50 of roof vertices, it only takes less than 5 minutes to label and reconstruct a realistic roof from the aerial image. 

We then compare to the straight skeleton and the weighted straight skeleton using the same roof outline as ours. In row (d) of Fig.~\ref{fig:res2:img_recon}, we show the results of the straight skeleton. Though globally, the obtained results appear reasonable, note that the reconstructed roofs using straight skeleton contain a lot of structural inconsistencies w.r.t. the input images and unrealistic errors (highlighted in red). We then use the weighted straight skeleton to fine-tune the results in order to fix the errors by tuning the edge weights. We use the GUI provided by~\cite{kelly2011interactive} for the weighted straight skeleton where the user is allowed to change the weight for each outline edge. We asked a well-trained user to tune the edge weights until the structure of the reconstructed roof is as consistent as possible with the one shown in the image. We can see that the weighted straight skeleton can fix some inconsistencies and errors in the roofs constructed using the straight skeleton (for those successful edits, we changed the highlighting color from red to black). However, there are still many structural errors that cannot be fixed by changing the weights (highlighted in red). 

\begin{table}[!t]
\caption{\textbf{Comparison to straight skeleton based methods}.
We report the complexity of the roofs shown in Fig.~\ref{fig:res2:img_recon}, including the number of vertices ($n_v$) and faces $n_f$. For each image shown in Fig.~\ref{fig:res2:img_recon} (with underlying roof having $n_v$ vertices and $n_f$ faces), 
We report the number of visual inconsistencies (\#err) between the constructed roof and the image, the number of vertices ($\widebar{n_v}$) and faces ($\widebar{n_f}$) on the reconstructed roof, and the construction time ($t$) of different methods.}
\label{tb:res:img_recon}
\vspace{-6pt}

\footnotesize
{\def\arraystretch{1.15}\tabcolsep=0.22em
\begin{tabular}{cccccccccccccccccc}
\toprule[1pt]
\multicolumn{2}{c}{No.} & 1 & 2 & 3 & 4 & 5 & 6 & 7 & 8 & 9 & 10 & 11 & 12 & 13 & 14 & 15 & 16 \\\rowcolor{mytbcol!20}
\multicolumn{2}{c}{$n_v$} & 22   & 24   & 25   & 25   & 27   & 27   & 32   & 33   & 33   & 33   & 34   & 35   & 36   & 39   & 39   & 51   \\
\multicolumn{2}{c}{$n_f$} & 12   & 17   & 14   & 14   & 15   & 16   & 17   & 20   & 18   & 20   & 18   & 21   & 19   & 22   & 21   & 38   \\\bottomrule[1pt]
\rowcolor{mytbcol!30} 
\cellcolor{mytbcol!30} &\textbf{Ours} & 0 & 0 & 0 & 0 & 0 & 0  & 0 & 0 & 0 & 0 & 0 & 0 & 0 & 0 & 0 & 0  \\
\cellcolor{mytbcol!30} & \cellcolor{mytbcol!30} ss  & 0 &   \bfseries 2 & 0 & 0 &  \bfseries 1 & \bfseries 2  & \bfseries  1 &  \bfseries 2 & 0 & \bfseries 3 & \bfseries  2 & \bfseries  2 &  \bfseries 1 &  \bfseries 3 &  \bfseries 2 & \bfseries  4\\
\rowcolor{mytbcol!30} 
\multirow{-3}{*}{\cellcolor{mytbcol!30}\#err} & wss & 0 &  \bfseries 2 & 0 & 0 &  \bfseries 1 &  \bfseries 2  &  \bfseries 1 &  \bfseries 1 & 0 & \bfseries  2 & \bfseries   2 &\bfseries  2 & \bfseries 1 &  \bfseries 2 & \bfseries  2 &  \bfseries 2\\ \hline

\rowcolor{gray!10} 
\cellcolor{gray!10} & \textbf{Ours} & 22 & 24 & 25 & 25 & 27 & 27 & 32 & 33 & 33 & 33 & 34 & 35 & 36 & 39 & 39 & 51 \\
\cellcolor{gray!10} & \cellcolor{gray!10} ss & 22 & {22} & {26} & {26} & {30} & {28} & {34} & {40} & {34} & {38} & {38} & {40} & {38} & {40} & {44} & {50} \\
\rowcolor{gray!10} 
\multirow{-3}{*}{\cellcolor{gray!10}$\widebar{n_v}$} & wss & 22 & {22} & {26} & {26} & {30} & {28} & {34} & {40} & {34} & {38} & {38} & {40} & {38} & {40} & {44} & {50} \\ \hline

\rowcolor{mytbcol!30} 
\cellcolor{mytbcol!30} & \textbf{Ours} & 12 & 17 & 14 & 14 & 15 & 16 & 17 & 20 & 18 & 20 & 18 & 21 & 19 & 22 & 21 & 38 \\
\cellcolor{mytbcol!30} & \cellcolor{mytbcol!30} ss & 12 & {12} & 14 & 14 & {16} & {15} & {18} & {21} & 18 & 20 & {20} & 21 & {20} & {21} & {23} & {26} \\
\rowcolor{mytbcol!30} 
\multirow{-3}{*}{\cellcolor{mytbcol!30}$\widebar{n_f}$} & \cellcolor{mytbcol!30} wss & 12 & {12} & 14 & 14 & {16} & {15} & {18} & {21} & 18 & 20 & {20} & 21 & {20} & {21} & {23} & {26}\\ \hline

\rowcolor{gray!10} 
\cellcolor{gray!10} &\textbf{Ours} & 89.4 & 115 & 153 & 97.9 & 114 & 92.9 & 151 & 145 & 155 & 167 & 148 & 158 & 179 & 199 & 178 & 284  \\
\cellcolor{gray!10} & \cellcolor{gray!10} ss  & 16 & 20 & 24 & 18 & 20 & 16 & 28 & 29 & 27 & 33 & 32 & 35 & 31 & 33 & 38 & 38 \\
\rowcolor{gray!10} 
\multirow{-3}{*}{\cellcolor{gray!10}\begin{tabular}[c]{@{}c@{}}$t$\\  (s)\end{tabular}} & wss & - & 300 & - & - & 60 & 180 & 180 & 300 & - & 120 & 180 & 60 & 60 & 480 & 180 & 360\\ 

\bottomrule[1pt]
\end{tabular}
}

\end{table}

\begin{figure}[!t]
    \centering
    \vspace{-6pt}
    \begin{overpic}[trim=0cm 0cm 0cm 0cm,clip,width=1\linewidth,grid=false]{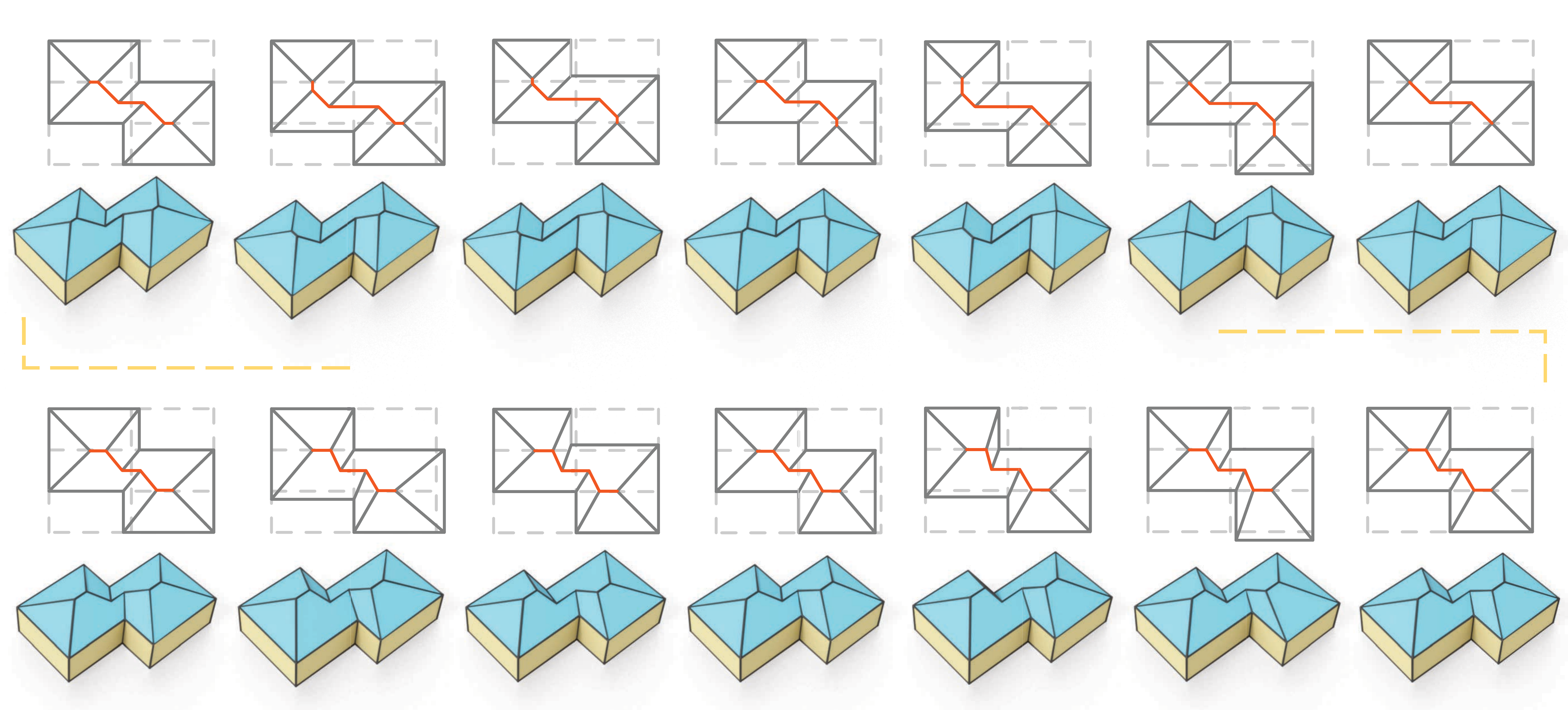}
    \put(2,22.8){\footnotesize (a) Straight Skeleton}
    \put(88,21){\footnotesize (b) \textbf{Ours}}
    \end{overpic}\vspace{-12pt}
    \caption{\emph{Top}: the straight skeleton algorithm is sensitive to the input and can lead to different roof structure for very similar roof outlines. \emph{Bottom}: as a comparison, our model allows to fix the roof structure with different outlines. We overlay the roof graph with a fixed-scale grid colored in gray to better visualize the difference between the outlines.}
    \label{fig:intro:ss_prob_eg3_robustness}\vspace{-5pt}
\end{figure}

In Table~\ref{tb:res:img_recon} we report the topology of the reconstructed roofs from different methods, including the number of vertices ($\widebar{n_v}$) and faces ($\widebar{n_f}$). We can see that the roofs reconstructed by the standard or the weighted straight skeleton method always have the same number of vertices and faces. It suggests that although the weighted straight skeleton can fix some visual inconsistencies from the standard method, it cannot change the roof topology. 
The structural errors of the straight skeleton based methods in Fig.~\ref{fig:res2:img_recon} show that these methods have much less expressiveness power in roof topology representation than our method. 
As suggested in Fig.~\ref{fig:intro:ss_prob_eg3_robustness}, the straight skeleton method is very sensitive to the input condition and can lead to different topology for similar building outlines. 

\begin{table}[!t]
\caption{\textbf{Comparison to commercial software.} We compare to the roofs constructed in 3ds Max (3D) and SketchUP (SU) and report if the constructed roof is planar (“valid”), the number of topological errors ("err"), the ratio of the polygon faces in the roof ("poly\%"), the number of vertices ($\widebar{n_v}$) and faces ($\widebar{n_f}$) of the constructed roof, and the modeling time in minutes ($t$).}
\label{tb:tb_res_softwares}
\vspace{-3pt}
\footnotesize
{\def\arraystretch{1.2}\tabcolsep=0.34em
\begin{tabular}{cccccccccccccccccc}
\toprule[1pt]
\multicolumn{2}{c}{No.} & 1 & 2 & 3 & 4 & 5 & 6 & 7 & 8 & 9 & 10 & 11 & 12 & 13 & 14 & 15 & 16 \\
\bottomrule[1pt]



\rowcolor{mytbcol!30} 
\cellcolor{mytbcol!30} &\textbf{Ours} & \cmark & \cmark & \cmark & \cmark & \cmark & \cmark & \cmark & \cmark & \cmark & \cmark & \cmark & \cmark & \cmark & \cmark & \cmark & \cmark \\
\cellcolor{mytbcol!30} & \cellcolor{mytbcol!30} 3D & \xmark & \xmark & \xmark & \xmark & \xmark & \xmark & \xmark & \xmark & \xmark & \xmark & \xmark & \xmark & \xmark & \xmark & \xmark & \xmark  \\
\rowcolor{mytbcol!30} 
\multirow{-3}{*}{\cellcolor{mytbcol!30} valid} & SU & \cmark & \cmark & \cmark & \cmark & \cmark & \cmark & \cmark & \cmark & \cmark & \cmark & \cmark & \cmark & \cmark & \cmark & \cmark & \cmark \\\hline

\rowcolor{gray!10} 
\cellcolor{gray!10} &\textbf{Ours} & 0 & 0 & 0 & 0 & 0 & 0  & 0 & 0 & 0 & 0 & 0 & 0 & 0 & 0 & 0 & 0\\
\cellcolor{gray!10} & \cellcolor{gray!10} 3D &  9 & 12 & 10 & 11 & 12 & 12 & 16 & 15 & 15 & 11 & 16 & 15 & 16 & 18 & 17 & 20\\
\rowcolor{gray!10} 
\multirow{-3}{*}{\cellcolor{gray!10} \#err} & SU & 20 & 16 & 21 & 19 & 16 & 27 & 33 & 21 & 25 & 38 & 28 & 30 & 33 & 33 & 29 & 42 \\\hline

\rowcolor{mytbcol!30} 
\cellcolor{mytbcol!30} &\textbf{Ours} &   75 &   65 &   71 &   79  & 67 &  69 &  82 &  75 &  83 &  60 &  72 & 57 & 79 &  73 &  71 &  42\\
\cellcolor{mytbcol!30} & \cellcolor{mytbcol!30} 3D &    75 &  62  & 71 &  79 &  75  &  67 &   87 &   75 &   83 & 52 &  76 &  60 &  80 &  72 & 74  & 56  \\
\rowcolor{mytbcol!30} 
\multirow{-3}{*}{\cellcolor{mytbcol!30} \begin{tabular}[c]{@{}c@{}}poly\\  \scriptsize{(\%)}\end{tabular}} & SU &  0  & 24  &  2.6  & 3.3  &  20 & 2.3  &   2.0  &  22 &   7.0 & 10 &  11 &  9.8 & 3.8 & 15 & 14 & 0 \\ \hline

\rowcolor{gray!10} 
\cellcolor{gray!10} & \textbf{Ours} & 22 & 24 & 25 & 25 & 27 & 27 & 32 & 33 & 33 & 33 & 34 & 35 & 36 & 39 & 39 & 51 \\
\cellcolor{gray!10} & \cellcolor{gray!10} 3D &  22 & 31 & 37 & 26    & 28 & 28 & 44 & 34 & 34 & 35 & 44 & 35 & 36 & 40 & 39 & 53 \\
\rowcolor{gray!10} 
\multirow{-3}{*}{\cellcolor{gray!10}$\widebar{n_v}$} & SU  & 22  & 29 & 26 & 25 & 28 & 30 & 36 & 45 & 33 & 52 & 39 & 40 & 37 & 46 & 42 & 58 \\ \hline

\rowcolor{mytbcol!30} 
\cellcolor{mytbcol!30} & \textbf{Ours} & 12 & 17 & 14 & 14 & 15 & 16 & 17 & 20 & 18 & 20 & 18 & 21 & 19 & 22 & 21 & 38 \\
\cellcolor{mytbcol!30} & \cellcolor{mytbcol!30} 3D & 12 & 21 & 14 & 14 & 16 & 18 & 18 & 20 & 18 & 21 & 21 & 25 & 20 & 25 & 23 & 36\\
\rowcolor{mytbcol!30} 
\multirow{-3}{*}{\cellcolor{mytbcol!30}$\widebar{n_f}$} & \cellcolor{mytbcol!30} SU & 32 & 33 & 35 & 33 & 31 & 43 & 50 & 41 & 43 & 58 & 46 & 51 & 52 & 55 & 50 & 82 \\ \hline

\rowcolor{gray!10} 
\cellcolor{gray!10} &\textbf{Ours} & 1.5 & 1.9 & 2.6 & 1.6 & 1.9 & 1.6  & 2.5 & 2.4 & 2.6 & 2.8 & 2.5 & 2.6 & 3.0 & 3.3 & 3.0 & 4.7\\
\cellcolor{gray!10} & \cellcolor{gray!10} 3D  & 6 & 7 & 7 & 6 & 6  & 12 & 12  & 11 & 7  & 14 & 6 & 9 & 10 & 8 & 10 &  23\\
\rowcolor{gray!10}
\multirow{-3}{*}{\cellcolor{gray!10}\begin{tabular}[c]{@{}c@{}}$t$\\  \scriptsize{(min)}\end{tabular}} & SU & 12  & 16  & 15 & 12 & 22 & 20 & 21  & 22  & 21 & 25 & 19 & 32 & 22 & 14 & 25 & 36 \\ 
\bottomrule[1pt]
\end{tabular}
}
\end{table}

\begin{figure}[!t]
    \centering
    \begin{overpic}[trim=0cm 18cm 11.2cm 0.2cm,clip,width=1\linewidth,grid=false]{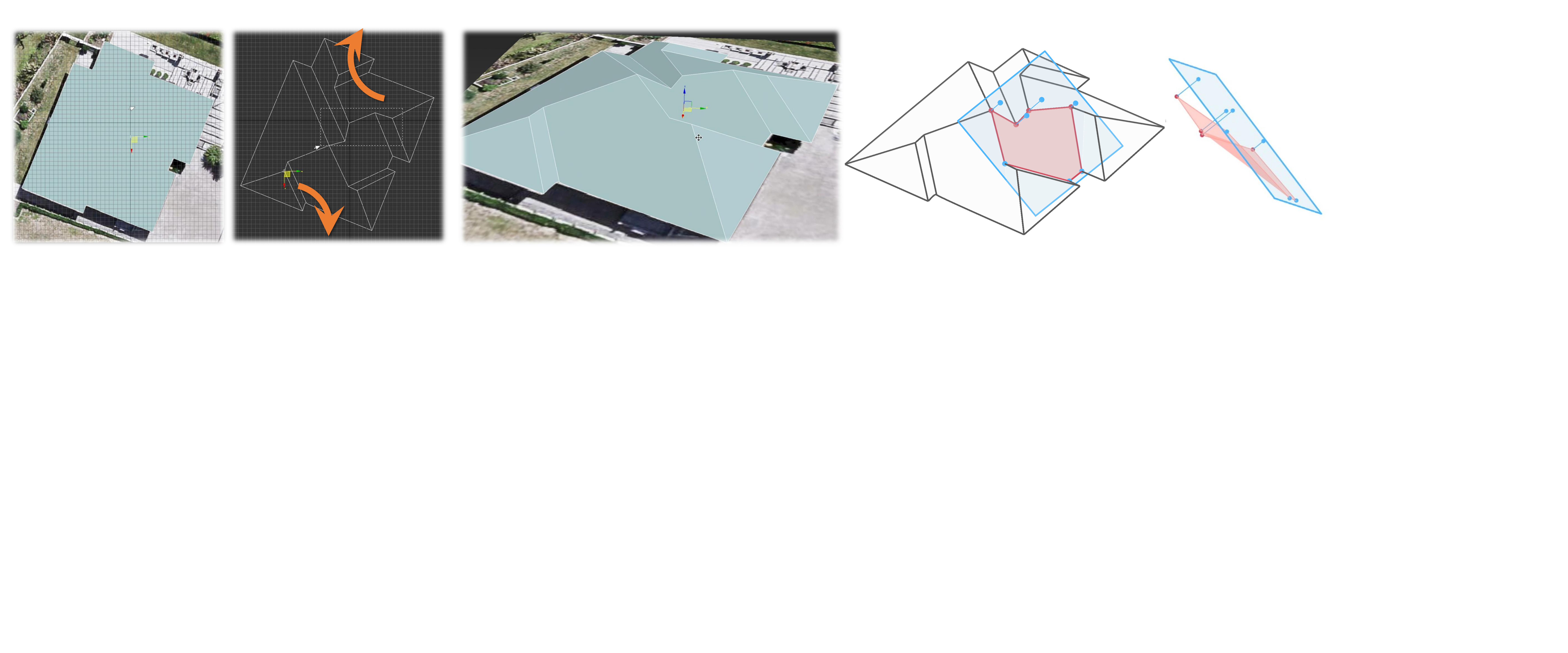}
    \put(1.5,19){\scriptsize\itshape\bfseries draw outline}
    \put(20,19){\scriptsize\itshape\bfseries cut faces}
    \put(22,-0.5){\scriptsize\itshape\bfseries move vtx}
    \put(37,19){\scriptsize\itshape\bfseries move vtx along $z$-axis}
    \put(75,19){\scriptsize\itshape\bfseries constructed roof} 
    \put(75,0){\scriptsize\itshape\bfseries (red face: non-planar)}
    \end{overpic}\vspace{-6pt}
    \caption{Workflow of using 3ds Max for roof modeling (No.3 roof in Fig.~\ref{fig:res2:img_recon}).}
    \label{fig:res:3dmax_workflow}
\end{figure}

\begin{figure}[!t]
    \centering
    \begin{overpic}[trim=0cm 20cm 25.8cm 0.2cm,clip,width=1\linewidth,grid=false]{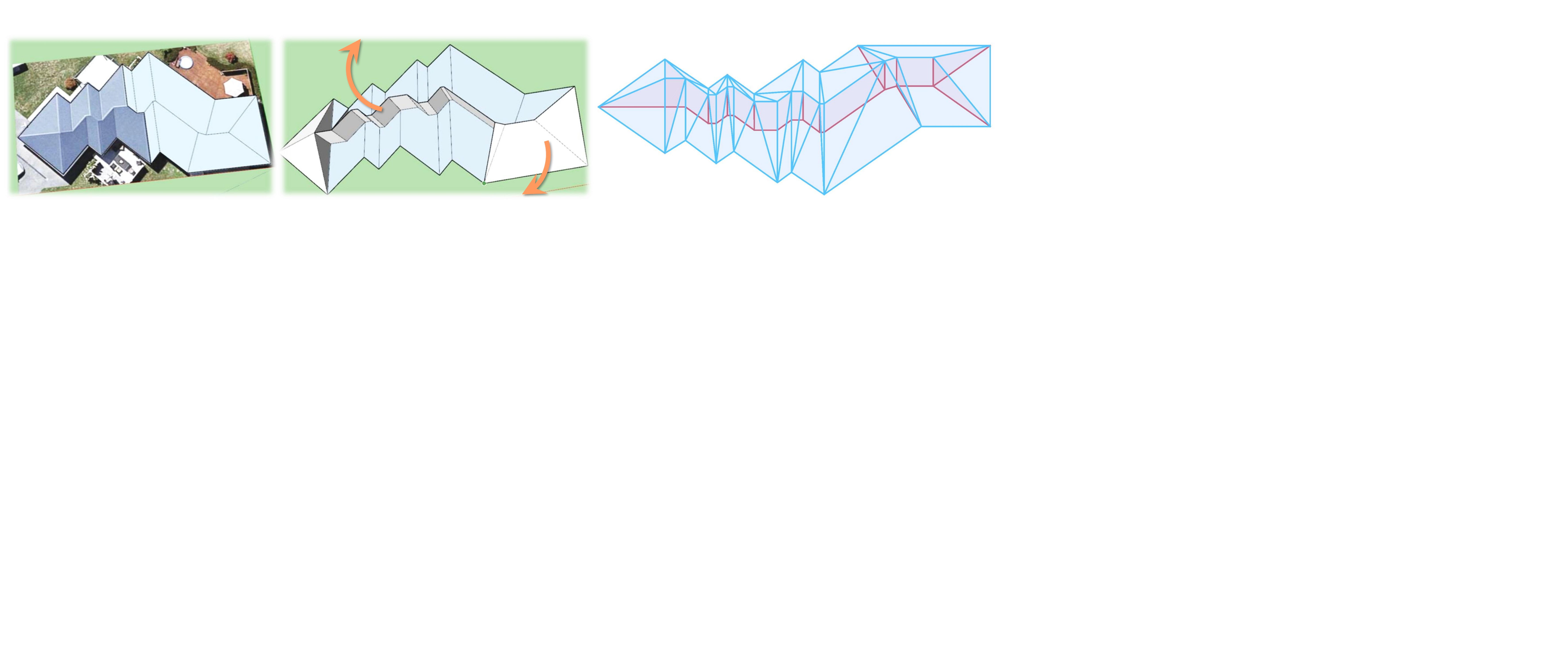}
    \put(2,19){\scriptsize\itshape\bfseries specify roof topology}
    \put(30,19){\scriptsize\itshape\bfseries build roof beams}
    \put(35,0){\scriptsize\itshape\bfseries add planar roof tops}
    \put(70,19){\scriptsize\itshape\bfseries constructed roof} 
    \put(70, 0){\scriptsize\itshape\bfseries (red: interior structure)}
    \end{overpic}\vspace{-6pt}
    \caption{Workflow of using SketchUp for roof modeling (No.9 roof in Fig.~\ref{fig:res2:img_recon})}
    \label{fig:res:su_workflow}
\end{figure}
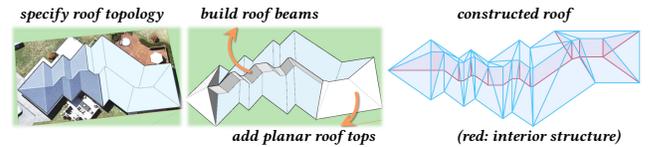

\subsubsection{Comparison to Commercial Software}
We also compare to two commercial software frameworks for roof reconstruction. 
We asked two experts, one with 5-year experience of modeling in 3ds Max, and the other with 3-year experience of modeling in SketchUp, to reconstruct the roofs shown in Fig.~\ref{fig:res2:img_recon}. 
The only instruction we gave to the experts was to model a polygonal roof that is as consistent as possible with the given image and as simple as possible. 
The two experts that worked independently followed a similar logic:
they first specified the roof topology on top of the imported image by drawing the outline and then by adding/cutting faces; then they constructed a 3D roof based on the roof topology. Specifically, the 3ds Max expert chose to move vertices mainly along the z-axis and checked in different views until satisfied with the reconstructed roofs, as shown in Fig.~\ref{fig:res:3dmax_workflow}. The SketchUp expert chose to first build the roof beams, i.e., vertical planes such that the rooftop planes can be placed on top of it. See Fig.~\ref{fig:res:su_workflow} for an example. We show some quantitative comparison in Table~\ref{tb:tb_res_softwares}. For the roofs that are constructed in SketchUp, we ignore the constructed interior structures and only evaluate the rooftops.

\begin{figure}[!t]
\centering
\vspace{12pt}
\begin{overpic}[trim=0.2cm 16cm 42cm 1cm,clip,width=1\linewidth,grid=false]{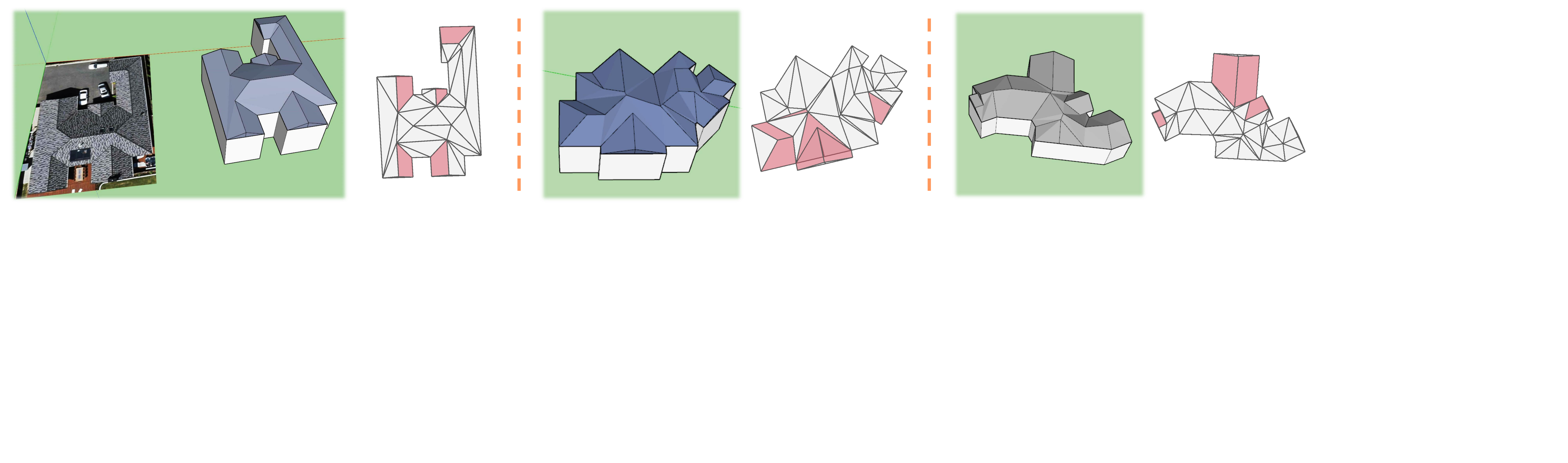}
\put(15,23){\footnotesize\bfseries ERR = 28}
\put(73,23){\footnotesize\bfseries ERR = 29}
\put(2,20){\tiny\itshape\bfseries \underline{SketchUp}}
\put(37,20){\tiny\itshape\bfseries \underline{Real Model}}
\put(60,20){\tiny\itshape\bfseries \underline{SketchUp}}
\put(80.5,20){\tiny\itshape\bfseries \underline{Real Model}}
\put(25,2.5){\tiny $(n_1 = 18)$ }
\put(42,2.5){\tiny $(n_2 = 46)$ }
\put(65,2.5){\tiny $(n_1 = 21)$}
\put(87,2.5){\tiny $(n_2 = 50)$}
\end{overpic}\vspace{-12pt}
\caption{\textbf{Evaluating roofs constructed in SketchUp.} The expert working in SketchUp hid some edges to make the constructed roofs visually consistent with the input image (\emph{left}). However, the real models shown on the \emph{right} are more complicated. We therefore measure the \emph{unwanted complexity}, i.e., the difference between the number of faces in the real model ($n_2$) and the number of faces in the visually expected model ($n_1$), to evaluate SketchUp. We highlight the faces in the real models that are not triangles in red.}
\label{fig:res:su_eval}\vspace{-9pt}
\end{figure}
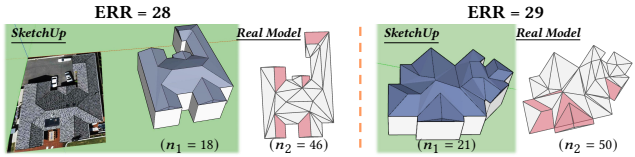

In general, commercial software provides more modeling tools and can model a larger variety of polygonal meshes. There are still some limitations for roof modeling. First, commercial software needs domain knowledge to be used efficiently, while our user input is light and friendly for novice users.
For example, the 3ds Max expert used different types of operations including creating polygons by adding edges, moving vertex positions in a 2D plane, cutting faces, extruding faces, translating grouped vertices and so on. During the modeling process, the expert had to frequently change the views or even switch to the four-view editing mode to operate and check the constructed 3D roof. 
As a comparison, our method allows the user to specify the roof topology in 2D with one simple operation, i.e. clicking on the image to construct a primal or dual roof graph. 
More importantly, it is hard to explicitly enforce the planarity of the 3D roof faces using 3ds Max or SketchUp. For example, 3ds Max allows to create a polygon from a set of non-coplanar 3D points. Therefore, it relies on the user to adjust the vertex positions to make the polygonal roof faces planar, which is extremely hard to achieve even visually. See Fig.~\ref{fig:res:3dmax_workflow} for an example where we highlight a non-planar polygon (colored red) w.r.t. a reference plane (colored blue). Also as reported in Table~\ref{tb:tb_res_softwares}, all the roofs constructed in 3ds Max are not planar numerically.
On the other hand, SketchUp does not allow to construct a non-planar polygon. Therefore, to form a rooftop from a set of non-coplanar vertices, a user need to manually triangulate the rooftop. 
We show the constructed roofs by 3ds Max and SketchUp in Fig.~\ref{fig:res:3dmax} and Fig.~\ref{fig:res:su} in Appendix.  
In Table~\ref{tb:tb_res_softwares} we report the ratio of polygonal faces in the constructed roof. We can see that this ratio of the roofs constructed using SketchUp is significantly lower than our method or using 3ds Max due to the implicit planarity constraint in SketchUp.

\begingroup
\setlength{\columnsep}{1pt}%
\setlength{\intextsep}{1pt}%
\begin{wrapfigure}{r}{0.5\linewidth}
\centering
\vspace{2pt}
\begin{overpic}[trim=0.2cm 19cm 38cm 1cm,clip,width=1\linewidth,grid=false]{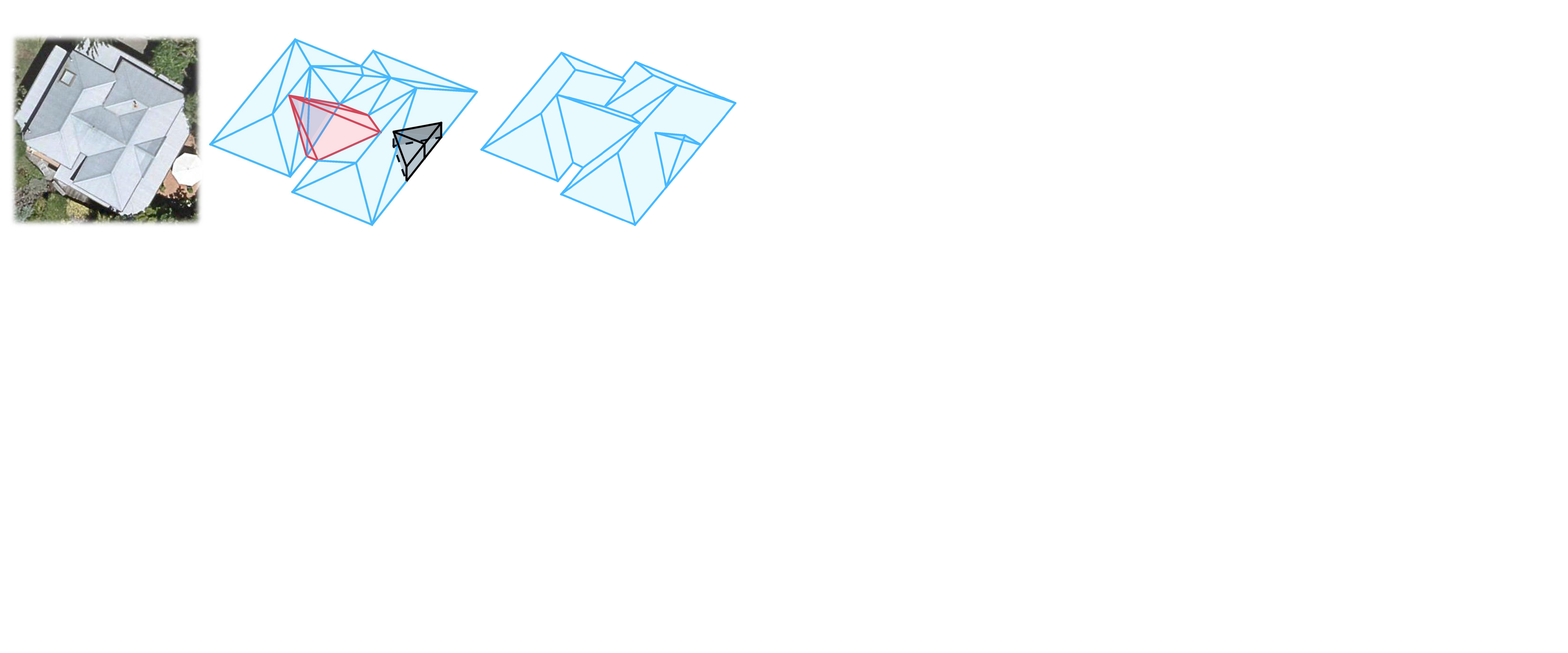}
\put(36,29){\scriptsize\itshape\bfseries SketchUp}
\put(79,29){\scriptsize\itshape\bfseries Ours}
\put(30,21){\tiny\itshape not connected}
\put(45,5.5){\tiny\itshape self-intersection}
\end{overpic}
\end{wrapfigure}
Another issue of roof modeling using the commercial software is that it heavily depends on the user's preference of how to specify roof topology. See the inset figure for an example, the roof constructed using SketchUp is visually consistent with the input image. However, it would be more plausible to have the roof faces colored in red being connected to the main roof body colored in blue. In practice, it is typically preferable to avoid self-intersections as highlighted by the gray part. As a comparison, our proposal of using a roof graph to describe roof topology is general and can result in roofs with simpler topology. For example, as shown in Fig.~\ref{tb:tb_res_softwares}, the roofs constructed by our method have a smaller number of vertices and faces. Please see the supplementary video for more examples.

\endgroup

We quantitatively measure the topological errors ("\#err") of the constructed roofs and report them in Table~\ref{tb:tb_res_softwares}. Specifically, we measure how many \emph{non-planar} faces in each roof constructed in 3ds Max. In SketchUp, the expert hid some edges of the constructed roofs as shown in Fig.~\ref{fig:res:su_eval}. We can then use the ``unwanted face complexity'' as a measure to evaluate the topological errors in SketchUp.
In summary, compared to the commercial software for roof modeling, our method is more efficient and simpler for novice users. The roofs constructed using our method have a simpler topology while the roof planarity and the visual consistency to the input image are enforced.

\begin{figure}[!t]
    \centering
    \definecolor{mycolor1}{rgb}{0.00000,0.44700,0.74100}%
\definecolor{myblue}{RGB}{147,227,247}
\definecolor{myyellow}{RGB}{255,230,153}
\definecolor{myyellow_dark}{RGB}{254,216,111}
\definecolor{myyellow_dark}{rgb}{0.5,0.5,0.5}
\pgfplotsset{
axis line style={myyellow_dark},
compat=1.11,
legend image code/.code={
\draw[mark repeat=2,mark phase=2]
plot coordinates {
(0cm,0cm)
(0.0cm,0cm)        
(0.3cm,0cm)         
};%
}
}
\begin{tikzpicture}

\begin{axis}[%
width=0.42\linewidth,
height=0.3\linewidth,
at={(0in,0in)},
scale only axis,
xmin=0,
xmax=20,
every x tick/.style={myyellow_dark},
every y tick/.style={myyellow_dark},
grid style={myyellow_dark!15}, 
xlabel style={font=\color{white!15!black}},
ymin=0,
ymax=1000,
yticklabel style={font=\footnotesize},
xticklabel style={font=\footnotesize},
ylabel style={font=\color{white!15!black}},
axis background/.style={fill=white},
title style={font=\bfseries,at={(0.5,0.95)}},
title={\# roof faces $n_f$},
xmajorgrids,
ymajorgrids,
legend style={legend cell align=left, align=left, draw=white!15!black}
]
\addplot[ybar interval, fill=myblue, fill opacity=1, draw=myblue, area legend] table[row sep=crcr] {%
x	y\\
4	433\\
5	1\\
6	992\\
7	27\\
8	616\\
9	14\\
10	300\\
11	11\\
12	94\\
13	5\\
14	24\\
15	6\\
16	9\\
17	1\\
18	4\\
19	2\\
20	2\\
};

\end{axis}


\begin{axis}[%
width=0.42\linewidth,
height=0.3\linewidth,
at={(0.48\linewidth,0)},
scale only axis,
xmin=0,
xmax=40,
every x tick/.style={myyellow_dark},
every y tick/.style={myyellow_dark},
grid style={myyellow_dark!15},
xlabel style={font=\color{white!15!black}\footnotesize},
ymin=0,
ymax=1000,
yticklabels={},
xticklabel style={font=\footnotesize},
ylabel style={font=\color{white!15!black}},
axis background/.style={fill=white},
title style={font=\bfseries,at={(0.5,0.95)}},
title={\# roof vertices $n_v$},
xmajorgrids,
ymajorgrids,
legend style={legend cell align=left, align=left, draw=white!15!black}
]
\addplot[ybar interval, fill=myblue, fill opacity=1, draw=myblue, area legend] table[row sep=crcr] {%
x	y\\
5	2\\
6	431\\
7	2\\
8	0\\
9	109\\
10	884\\
11	16\\
12	32\\
13	87\\
14	509\\
15	9\\
16	13\\
17	61\\
18	230\\
19	4\\
20	22\\
21	22\\
22	59\\
23	1\\
24	4\\
25	9\\
26	13\\
27	3\\
28	5\\
29	3\\
30	4\\
31	0\\
32	0\\
33	5\\
34	5\\
};

\end{axis}
\end{tikzpicture}%
\vspace{-13pt}
    \caption{Variations of our constructed dataset: we report the number of faces and vertices on each of the 2539 constructed buildings via a histogram. The number of faces ranges from 4 to 20, and the number of vertices on the roofs ranges from 5 to 34.}
    \label{fig:res:dt_stats}\vspace{-6pt}
\end{figure}

\subsection{Image-Building Paired Dataset}\label{sec:res:dataset}
\begingroup
\setlength{\columnsep}{1pt}%
\setlength{\intextsep}{1pt}%
\begin{wrapfigure}{r}{0.46\linewidth}
\centering
\begin{overpic}[trim=0.5cm 12cm 15cm 0.5cm,clip,width=1\linewidth,grid=false]{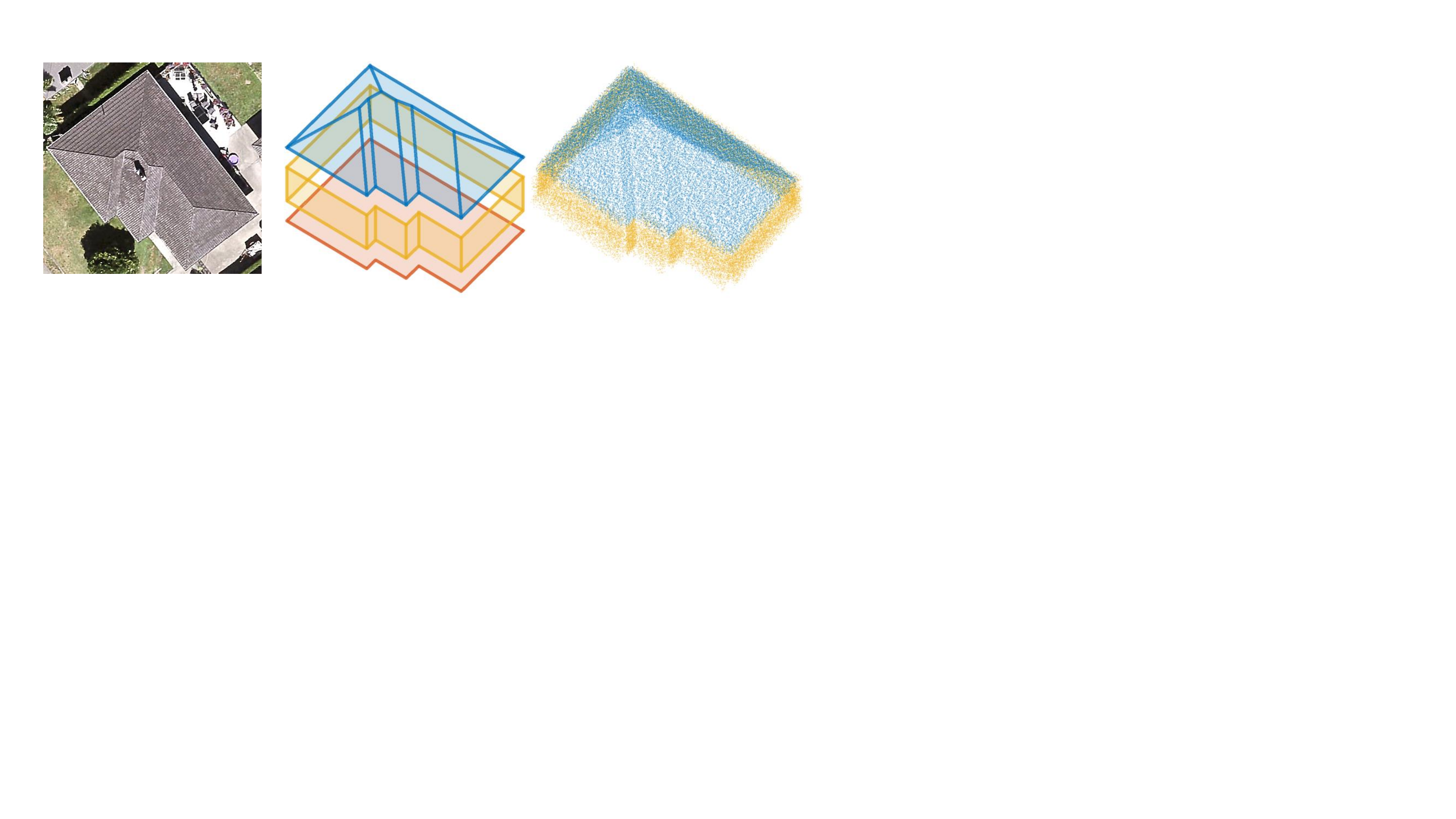}
\put(8,33){\scriptsize\itshape\bfseries image}
\put(32,33){\scriptsize\itshape\bfseries polygon mesh}
\put(70,33){\scriptsize\itshape\bfseries point cloud}
\end{overpic}
\end{wrapfigure}
We created an image-building paired dataset using our roof optimization method, where a complete building is constructed by adding facade planes along the roof outline and a base plane at the bottom. Specifically, we created a dataset consisting of 2539 buildings paired with the input aerial image and face labels (including roof faces, facade faces, and base face). \revised{The annotations are cleaned automatically by merging close-by vertices, removing duplicate or redundant edges/vertices, and our method is then used for roof reconstruction.} See Fig.~\ref{fig:dataset} for some example buildings and Fig.~\ref{fig:res:dt_stats} for a summary of the roof complexity including the number of roof vertices and roof faces in this dataset.
We believe this dataset can be helpful for different visual computing tasks. Specifically, this is a mesh-image paired dataset, which can be used for learning-based roof mesh detection. We also assign each polygon face in the building mesh a label from the set of roof face, body face, or base face. We can sample from the polygon mesh to obtain a point cloud as well (see inset figure).
This dataset also contains roofs with a larger range of complexity.
For example, in the following we discuss how to use this dataset for roof synthesis from scratch.

\endgroup

\begin{figure}[!t]
    \centering
    \begin{overpic}[trim=0cm 12.7cm 0cm 13cm,clip,width=1\linewidth,grid=false]{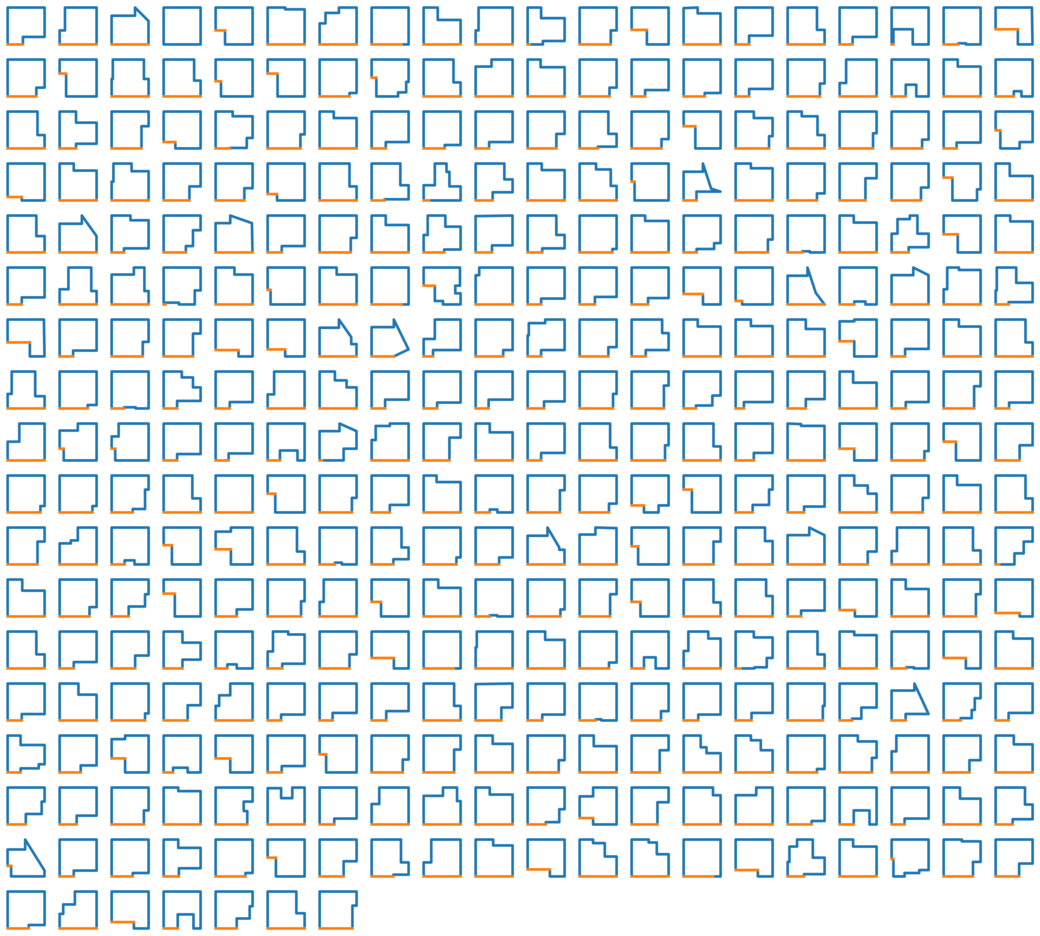}
    \end{overpic}\vspace{-10pt}
    \caption{Generated roof outlines with our auto-regressive model. We use our model to generate a sequence of 2D vertices and connect the tail vertex to the head by an orange line.}
    \label{fig:res-transformer-qual}
\end{figure}

\begin{figure}[!t]
    \centering
    \begin{overpic}[trim=0cm 0cm 0cm 0cm,clip,width=1\linewidth,grid=false]{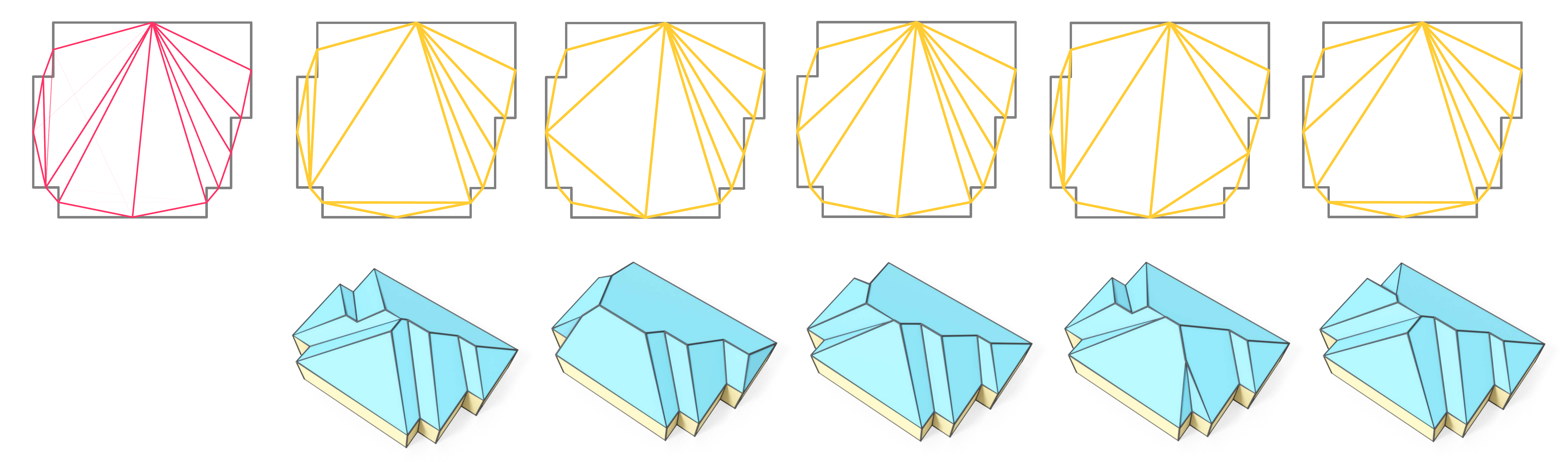}
    \put(4.5,12){\footnotesize Learned}
    \put(3,9){\footnotesize Adjacency}
    \end{overpic}
    \vspace{-22pt}
    \caption{We can extract multiple valid dual graphs from the learned adjacency (\emph{top}). We show the corresponding constructed roof on the \emph{bottom}.}
    \label{fig:res:multi_adj}\vspace{-3pt}
\end{figure}

\subsection{Application 1: Roof Synthesis from Scratch}\label{sec:app:synthesis}
\revised{
As discussed in Sec.~\ref{sec:mtd:roof_synthesis}, our roof modeling formulation can simplify the roof synthesis problem. Specifically, we propose learning-based techniques for roof topology synthesis, and then use our roof optimization method to enforce geometric planarity constraints. We use the constructed dataset in Sec.~\ref{sec:res:dataset} to train models on roof graph generation, including a transformer for roof outline generation and a graph neural network for face adjacency prediction. 
See Fig.~\ref{fig:res-transformer-qual} for some example roof outlines generated by our transformer.
Our graph neural network can predict the probability of adjacency for face pairs, from which we can extract either a single dual graph with highest probability (see Fig.~\ref{fig:res:learned_adj}) or \emph{multiple} valid dual graphs for roof construction as shown in Fig.~\ref{fig:res:multi_adj}. See the supplementary materials and videos for more discussions and results. 
}

We compare to a Variational Auto-Encoder (VAE) based generative model \cite{DBLP:journals/corr/KingmaW13} for roof graph generation. 
We used the trained VAE to synthesize 360 roof graphs, and only 119 of them are fully connected graphs while the remaining graphs have up to 19 disconnected components. We then only focus on fully connected cases for potentially valid roof graphs. Fig.~\ref{fig:res:vae_roofgraph} shows some example roof graphs synthesized by the VAE-based model. Even most of the fully connected roof graphs do not have a reasonable topology. Among the few synthesized roofs that do have a reasonable topology (e.g., the first and the last one) the geometry is not valid and violates aesthetic constraints. We therefore conclude that the task of constraint geometry generation is very difficult for a VAE. 
This shows that our strategy of separating the continuous constraints from the discrete constraints can simplify the problem and make it easier for training a generative model to learn roof topology. 

\begin{figure}[!t]
    \centering
    \includegraphics[width=1\linewidth]{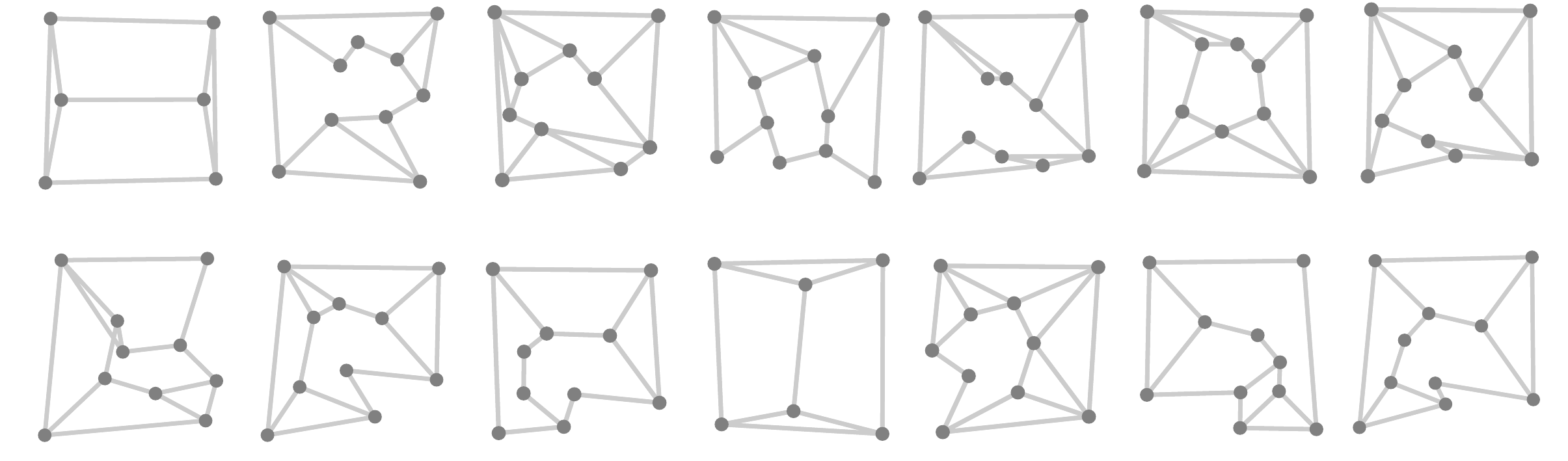}\vspace{-12pt}
    \caption{\textbf{Synthesized roof graph via VAE}. Synthesizing a valid roof directly can be hard since the model needs to take care of the discrete constraints and the continuous constraints at the same time.}
    \label{fig:res:vae_roofgraph}\vspace{-12pt}
\end{figure}

\begin{figure*}[!t]
    \centering
    \input{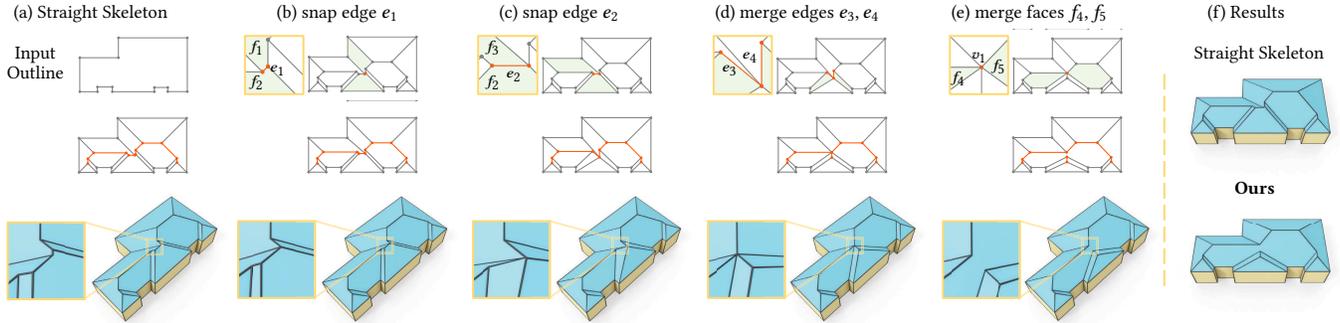}
    \vspace{-22pt}
    \caption{The roof constructed from the straight skeleton can be undesirable or unrealistic as shown in (a). Moreover, the straight skeleton formulation only supports a limited set of edits. As a comparison, our optimization-based formulation supports different types of edits (b-e). In (f), we show that after a set of edits to the result from the straight skeleton, we can obtain a more realistic roof. For (b-e), \emph{Top}: we visualize the edits made to the roof graph and we show a zoom-in version on the left to highlight the region of interest. \emph{Middle}: we show the valid 2D embedding after editing using our method and we color the roof edges in red to highlight the roof structure. \emph{Bottom}: we show the corresponding constructed planar roofs with a zoomed-in version of the regions with changes.}
    \label{fig:intro:edit_straight_skeleton}\vspace{-6pt}
\end{figure*}

\begin{figure*}[!t]
    \centering
    \vspace{6pt}
    \input{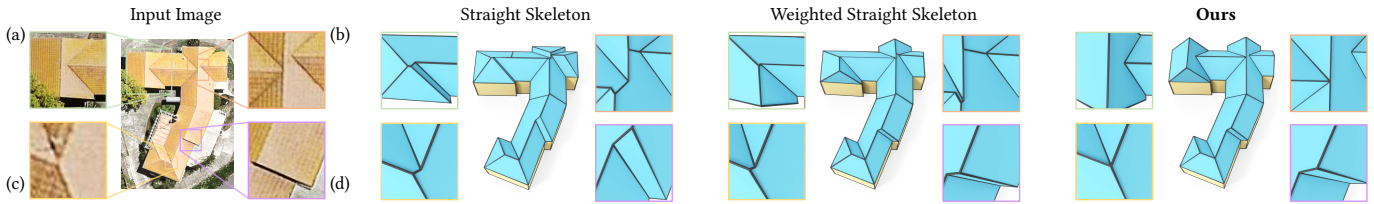}\vspace{-22pt}
    \caption{ The roofs constructed using straight skeleton can be different from the input image. For example, the highlighted region with a zoom-in view of (a) and (b) contains errors; region (b) and (c) contains extremely short edges which are unlikely to exist in reality; region (d) has inconsistent vertex positions.
    We can consider using the weighted straight skeleton to fix these issues by changing the weight for each outline edge. However, changing the edge weights is not trivial and the user could only solve the vertex inconsistency in (d), while the other structural errors in (a,b,c) are mediated, but not fully resolved. As a comparison, our operations for interactive editing are explicitly defined on the roof graph and are much easier to apply. Our method successfully fixes all the errors and obtains a consistent roof. }
    \label{fig:res2:ss_wss_ours}\vspace{-6pt}
\end{figure*}

\subsection{Application 2: Interactive Roof Editing \& Optimization}
One of the biggest advantages of our roof construction method is its flexibility. Specifically, the optimization based planarity formulation makes it possible to incorporate different regularizers. Moreover, the primal-dual roof graph representation can support different editing operations.
Therefore, our method can be used for interactive roof editing and optimization. Specifically, a user can (1) modify/edit a (valid) roof graph (either the primal or dual graph) (2) starting from the modified roof graph, run our optimization method to obtain a \emph{valid} roof graph. The user can then go back to step (1) and edit again. In this way, one can edit the roof graph until satisfied. 

Our primal-dual roof graph representation can naturally support different types of operations including moving a vertex or an edge, snapping an edge, merging two faces, splitting a face, forcing two faces to be adjacent, and so on. See  Fig.~\ref{fig:intro:edit_straight_skeleton} for an example. 
We can see that our editing operations are expressive and our optimization-based formulation is well suited for interactive editing since after applying different operations the updated roof embedding does not change too much from the previous embedding while staying valid. 

As a comparison, the weighted straight skeleton, which allows a user to change edge weights for roof editing, is not trivial or efficient enough for interactive editing, since the change of the roof structure is not continuous or easily predictable w.r.t. the change of edge weights. See Fig.~\ref{fig:res2:ss_wss_ours} for an example, where we compare our interactive editing power to the weighted straight skeleton in correcting topological errors. Our method can easily fix all the errors while the weighted straight skeleton can only fix one out of four topological errors. 
See Appendix~\ref{sec:interactive} for more discussions.

\subsection{Additional Justification of Our Formulation}

\paragraph{\textbf{Adjustable roof height}}
\begingroup
\setlength{\columnsep}{1pt}%
\setlength{\intextsep}{1pt}%
\begin{wrapfigure}{r}{0.46\linewidth}
\centering
\begin{overpic}[trim=1cm 12cm 16cm 2cm,clip,width=1\linewidth,grid=false]{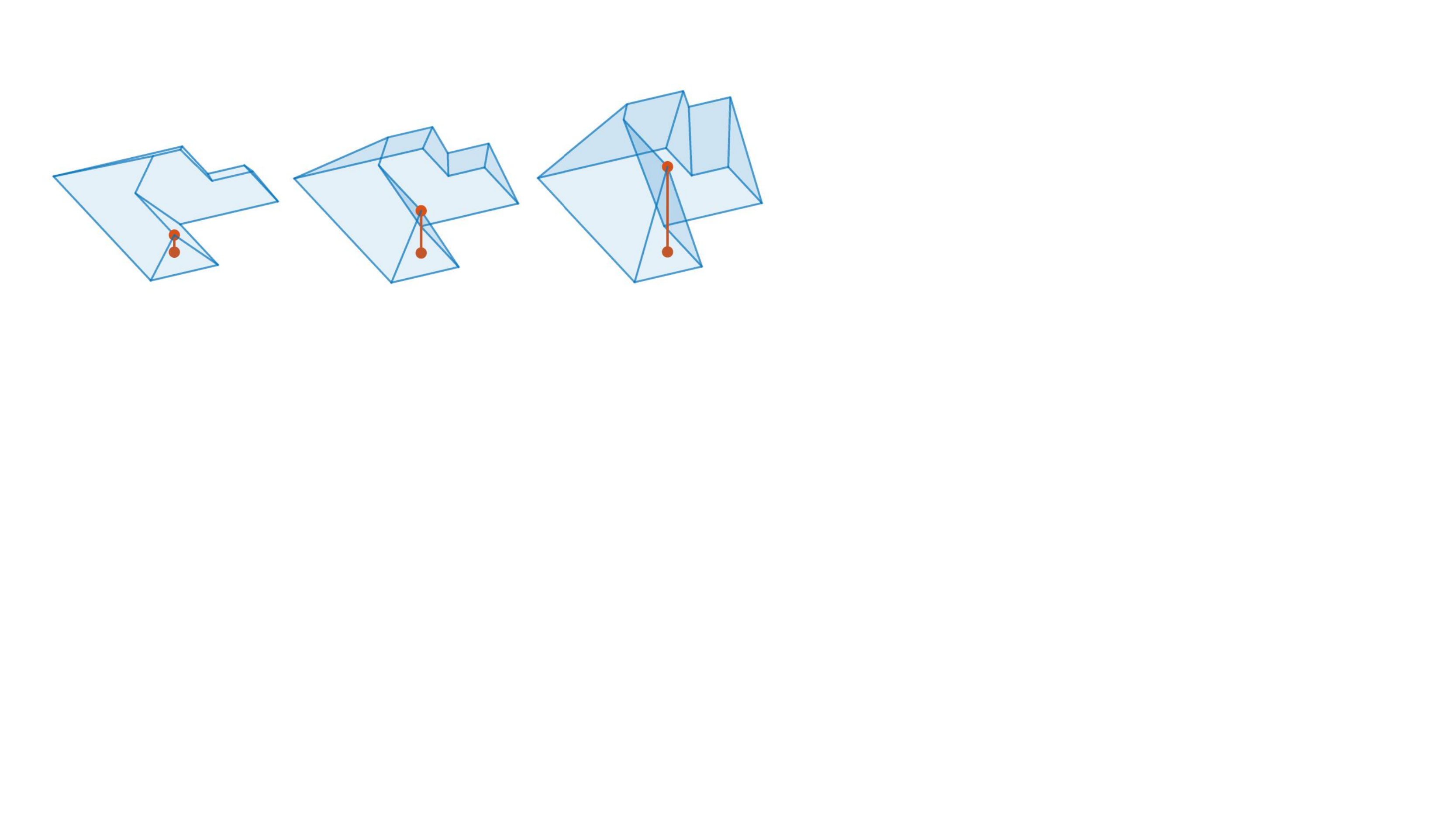}
\put(16,0){\scriptsize\itshape $h=20$}
\put(50,0){\scriptsize\itshape $h=50$}
\put(84,0){\scriptsize\itshape $h=100$}
\end{overpic}
\end{wrapfigure}
As discussed in Eq.~\eqref{eq:energy:planarity}, we optimize for a valid 3D roof embedding by minimizing the planarity energy w.r.t. a hard constraint such that a randomly selected roof vertex should have fixed height ($z$-axis value) $h$. This can help to avoid degenerate solutions where all the roof vertices have zero height, which leads to valid roofs with zero planarity error. In the inset figure we show that this parameter $h$ is not critical and we can set it to an arbitrary value to obtain valid 3D roofs. 
Additionally, this design choice can also benefit the interactive roof editing where a user can tune the roof height $h$ during the construction.
In our experiments, we usually set $h = \sqrt{S}/2$, where $S$ is the roof area.

\endgroup

\paragraph{\textbf{Usefulness of spectral embedding}}
\begingroup
\setlength{\columnsep}{1pt}%
\setlength{\intextsep}{1pt}%
\begin{wrapfigure}{r}{0.42\linewidth}
\centering
\begin{overpic}[trim=0cm 7cm 17cm 1cm,clip,width=1\linewidth,grid=false]{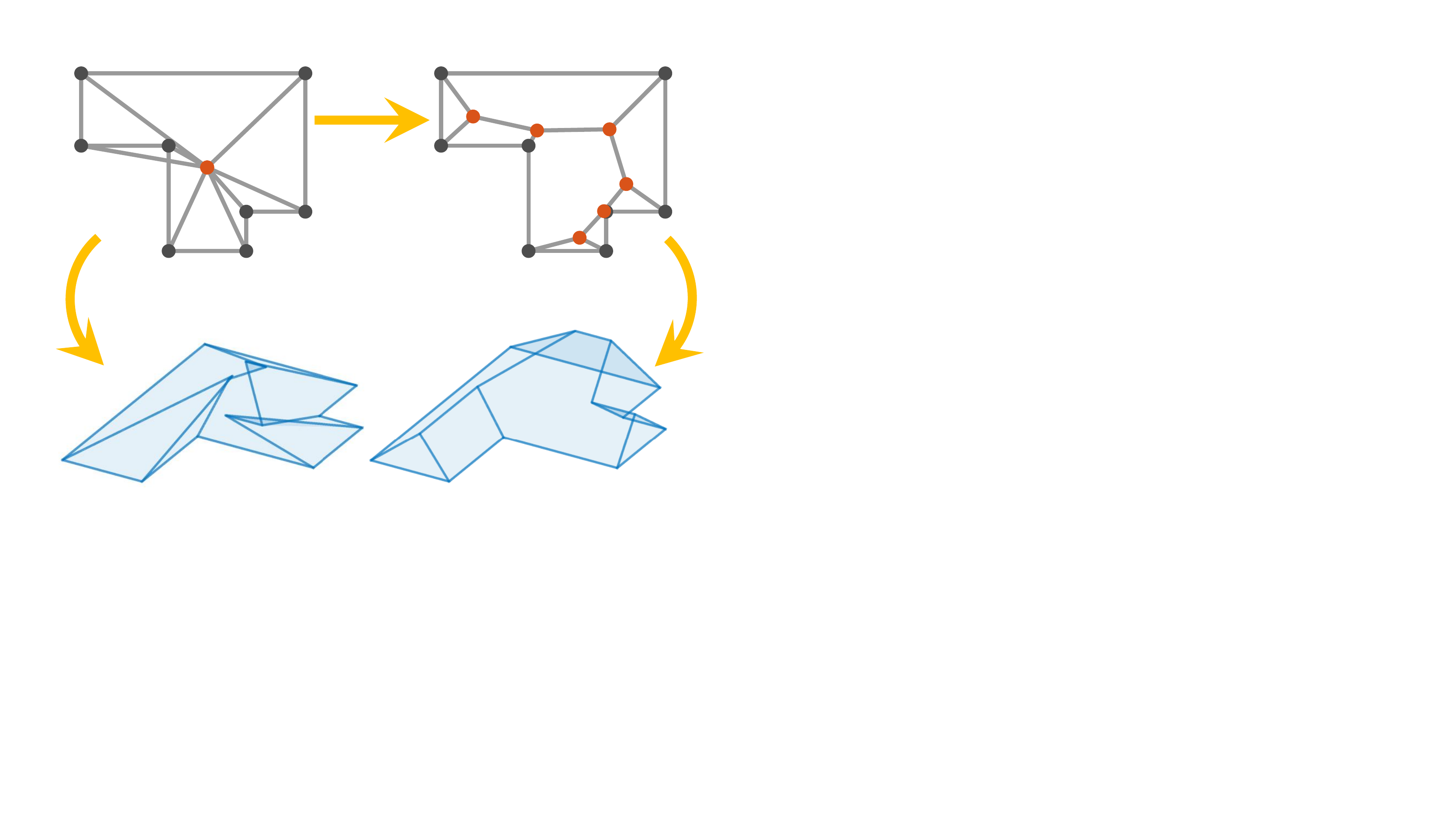}
\put(20,1.5){\scriptsize\itshape\bfseries err $< 10^{-9}$}
\put(65,1.5){\scriptsize\itshape\bfseries err $< 10^{-9}$}
\end{overpic}
\end{wrapfigure}
Taking the 2D spectral embedding as initialization can help to avoid self-intersections for roof constructions from a dual graph.
In the inset figure, we show the roofs optimized from zero embedding (left) and our spectral  
embedding (right). Both roofs are valid with planarity error ("err") smaller than $10^{-9}$. However, the roof shown on the right is more plausible with no self-intersections.
In Fig.~\ref{fig:res:roof_construction}, we show more results of our roof optimization algorithm from a dual graph. On top we show the 2D spectral embedding as the initialization, in the middle we show the optimized valid 2D embedding, and at the bottom we show the corresponding reconstructed roofs (buildings).

\endgroup

\subsection{Implementation \& Runtime}
We implemented the roof optimization methods (including spectral embedding and planarity optimizations) in MATLAB and used the build-in function "fmincon" with Quasi-Newton solver for optimization. We designed a web-based GUI with two modes for collecting user inputs of primal-dual roof graph specification for the application of roof reconstruction from images. The roof synthesis application is implemented using Pytorch~\cite{NEURIPS2019_9015}.

\begin{figure*}[!t]
    \centering
    \includegraphics[width=1\linewidth]{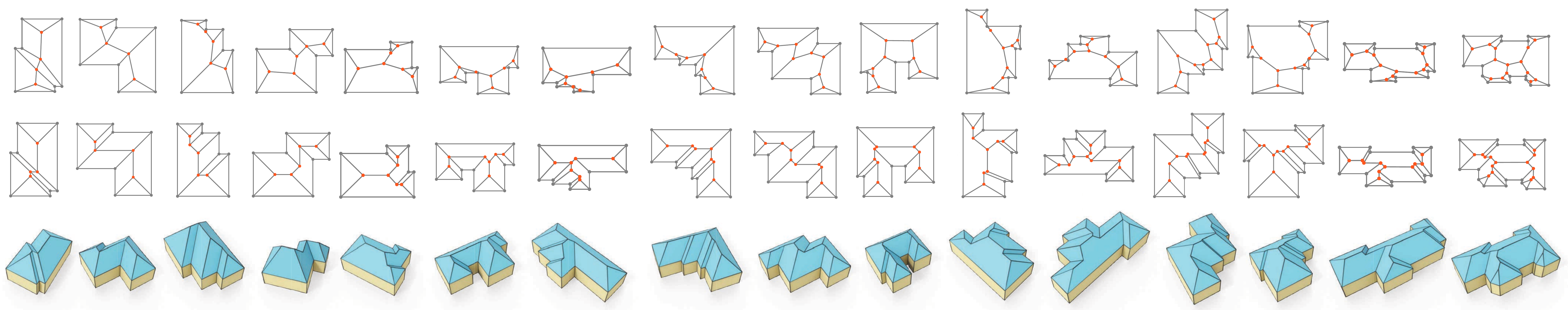}
    \vspace{-22pt}
    \caption{\textbf{Roof construction from dual graph}. \emph{Top}: the initial 2D spectral embedding. \emph{Middle}: the valid 2D embedding of each roof obtained by minimizing the planarity. \emph{Bottom}: the constructed planar roof meshes.}\vspace{-3pt}
    \label{fig:res:roof_construction}
\end{figure*}

In Table~\ref{tb:res:label_time} and Table~\ref{tb:res:roof_construction}, we report  the roof complexity and computation time of our method for each roof in Fig.~\ref{fig:res2:img_recon} and Fig.~\ref{fig:res:roof_construction} respectively.
We can see that our algorithms, both 2D spectral embedding and planarity optimization, are efficient and robust w.r.t. various roof outlines with different complexity.

\setlength{\tabcolsep}{0.16em}
\begin{table}[!t]
\caption{\textbf{Runtime of roof construction from primal graph}. For each of the examples in Fig.~\ref{fig:res2:img_recon}, 
we report the time for annotating the vertices $t_v$ and faces $t_f$ in the image for roof reconstruction by the user. $t_o$ reports the runtime \revised{in seconds} of our roof optimization algorithm. $t_{\text{ours}} = t_v + t_f + t_o$ shows the total amount of time for the roof reconstruction of our method.}
\label{tb:res:label_time}
\vspace{-6pt}
\footnotesize
{\def\arraystretch{1.2}\tabcolsep=0.2em
\begin{tabular}{ccccccccccccccccc}
\toprule[1pt]
No.   & 1    & 2    & 3    & 4    & 5    & 6    & 7    & 8    & 9    & 10   & 11   & 12   & 13   & 14   & 15   & 16   \\
\bottomrule[1pt]
$n_v$ & 22   & 24   & 25   & 25   & 27   & 27   & 32   & 33   & 33   & 33   & 34   & 35   & 36   & 39   & 39   & 51   \\\rowcolor{mytbcol!30}
$n_f$ & 12   & 17   & 14   & 14   & 15   & 16   & 17   & 20   & 18   & 20   & 18   & 21   & 19   & 22   & 21   & 38   \\
$t_v$ & 27.6 & 36.6 & 39.3 & 29.4 & 31.4 & 25.5 & 43.9 & 40.8 & 43.5 & 49.8 & 49.9 & 52.5 & 50.9 & 55.5 & 58.3 & 66.6 \\\rowcolor{mytbcol!30}
$t_f$ & 61.4 & 77.7 & 112  & 67.9 & 82.2 & 66.5 & 106  & 102  & 109  & 116  & 97.3 & 104  & 126  & 143  & 118  & 214  \\
$t_o$ & 0.42 & 1.20 & 0.73 & 0.55 & 0.61 & 0.89 & 0.94 & 2.23 & 1.47 & 1.05 & 0.72 & 1.58 & 1.24 & 1.34 & 2.52 & 3.59 \\\rowcolor{mytbcol!30}
$t_{\text{\textbf{ours}}}$  & 89.4 & 115 & 153 & 97.9 & 114 & 92.9 & 151 & 145 & 155 & 167 & 148 & 158 & 179 & 199 & 178 & 284 \\
\bottomrule[1pt]
\end{tabular}
}\vspace{-6pt}
\end{table}

\section{Conclusion, Limitation \& Future work}
We proposed an optimization-based roof construction method that first designs a primal or dual roof graph as input and then optimizes the geometry to output a planar 3D polygonal roof. Our formulation is flexible and can be adapted easily to different settings such as incorporating user-specified regularizers. Our method has two practical applications, interactive roof editing and roof synthesis from scratch. 
Our method of roof reconstruction is more expressive than the straight-skeleton based methods, and is much easier for \revised{novice users} to use than commercial software.
We also use our method to construct a image-building paired dataset with 2539 roof meshes, that can be helpful for different visual computing tasks.
\revised{
Although our method can be used to model roofs with different styles, including buildings with inner courtyards or vertical facades inside the roof, and buildings with outline edges in different height as shown in Fig.~\ref{fig:eg_temple}, it cannot directly handle curved roofs including stadiums and skyscrapers. One possible solution is to approximate curved roofs with planar sub-faces, which can be reconstructed via our method as shown in Fig.~\ref{fig:eg_temple} (a).
}
Another limitation of our work is that we do not model roof textures. While it would be interesting to research a GAN for the generation of roof textures in our framework, image synthesis is mainly orthogonal to the core topics of our paper.
Further, the interactive roof modeling is still fairly slow so that it is hard to scale up the dataset construction by another order of magnitude. For practical reasons, it might be very important to investigate efficient combinations of automatic and interactive reconstruction that would be time-saving without increasing the error rate compared to a human only baseline.
Finally, we did not touch on automatic reconstruction in this paper. 
\revised{
In future work, we would like to investigate transformers using images for cross attention following to the work of Dosovitskiy~\etal~\shortcite{dosovitskiy2021an}, and investigate how to predict roof graphs from images directly by using the image-mesh dataset we constructed. It would also be interesting to study practical constraints for roof fabricability using our optimization-based formulation, such as incorporating slope requirements of roof faces during the construction, which can be addressed by either hard or soft constraints.
}

\begin{table}[!t]
\caption{\textbf{Runtime of roof construction from dual graph}. For the 16 example roofs shown in Fig.~\ref{fig:res:roof_construction}, we report the complexity of each roof, including the number of vertices ($n_v$) and faces ($n_f$) in the roof. We also report the runtime in seconds for the 2D embedding via the spectral method ($t_1$) and 3D embedding via optimization w.r.t. the planarity and aesthetic constraints ($t_2$).}\label{tb:res:roof_construction}
\vspace{-6pt}
\footnotesize
{\def\arraystretch{1.2}\tabcolsep=0.22em 
\begin{tabular}{ccccccccccccccccc}
\toprule[1pt]
No.   & 1    & 2    & 3    & 4    & 5    & 6    & 7    & 8    & 9    & 10   & 11   & 12   & 13   & 14   & 15   & 16   \\ 
\bottomrule[1pt]
$n_v$ & 12   & 12   & 13   & 13   & 14   & 17   & 17   & 17   & 18   & 18   & 18   & 21   & 22   & 22   & 26   & 30\\ \rowcolor{mytbcol!30} 
$n_f$ & 8    & 8    & 8    & 8    & 8    & 10   & 10   & 10   & 10   & 10   & 10   & 12   & 12   & 12   & 14   & 16   \\
$t_1$ & 0.04 & 0.03 & 0.03 & 0.04 & 0.04 & 0.05 & 0.05 & 0.04 & 0.05 & 0.05 & 0.06 & 0.06 & 0.08 & 0.06 & 0.08 & 0.09 \\\rowcolor{mytbcol!30} 
$t_2$ & 0.20 & 0.17 & 0.26 & 0.24 & 0.29 & 0.41 & 0.39 & 0.37 & 0.43 & 0.44 & 0.45 & 0.60 & 0.80 & 0.69 & 1.90 & 1.15 \\ 
\bottomrule[1pt]
\end{tabular}
}\vspace{-9pt}
\end{table}

\begin{acks}
The authors thank the anonymous reviewers for their valuable comments. Parts of this work were supported by the KAUST OSR Award No. CRG-2017-3426, the ERC Starting Grant No. 758800 (EXPROTEA), the ANR AI Chair AIGRETTE, and Alibaba Innovative Research (AIR) Program. We would like to thank \textit{Guangfan Pan} and \textit{Jiacheng Ren} for helping modeling roofs in 3ds Max and SketchUp, \textit{Jialin Zhu} for helping designing the web-based roof annotation UIs,  \textit{Jianhua Guo} and \textit{Tom Kelly} for helping with the comparison to the weighted straight skeleton. We thank \textit{Muxingzi Li} for helping editing the supplementary video. We also thank \textit{Chuyi Qiu}, \textit{Tianyu He}, and \textit{Ran Yi} for their valuable suggestions and comments.
\end{acks}

\bibliographystyle{ACM-Reference-Format}
\bibliography{bibliography}


\begin{thebibliography}{77}


\ifx \showCODEN    \undefined \def \showCODEN     #1{\unskip}     \fi
\ifx \showDOI      \undefined \def \showDOI       #1{#1}\fi
\ifx \showISBNx    \undefined \def \showISBNx     #1{\unskip}     \fi
\ifx \showISBNxiii \undefined \def \showISBNxiii  #1{\unskip}     \fi
\ifx \showISSN     \undefined \def \showISSN      #1{\unskip}     \fi
\ifx \showLCCN     \undefined \def \showLCCN      #1{\unskip}     \fi
\ifx \shownote     \undefined \def \shownote      #1{#1}          \fi
\ifx \showarticletitle \undefined \def \showarticletitle #1{#1}   \fi
\ifx \showURL      \undefined \def \showURL       {\relax}        \fi
\providecommand\bibfield[2]{#2}
\providecommand\bibinfo[2]{#2}
\providecommand\natexlab[1]{#1}
\providecommand\showeprint[2][]{arXiv:#2}

\bibitem[\protect\citeauthoryear{Aichholzer and Aurenhammer}{Aichholzer and
  Aurenhammer}{1996}]%
        {aichholzer1996straight}
\bibfield{author}{\bibinfo{person}{Oswin Aichholzer} {and}
  \bibinfo{person}{Franz Aurenhammer}.} \bibinfo{year}{1996}\natexlab{}.
\newblock \showarticletitle{Straight skeletons for general polygonal figures in
  the plane}. In \bibinfo{booktitle}{\emph{International Computing and
  Combinatorics Conference}}. Springer, \bibinfo{pages}{117--126}.
\newblock


\bibitem[\protect\citeauthoryear{Aichholzer, Aurenhammer, Alberts, and
  G{\"a}rtner}{Aichholzer et~al\mbox{.}}{1996}]%
        {aichholzer1996novel}
\bibfield{author}{\bibinfo{person}{Oswin Aichholzer}, \bibinfo{person}{Franz
  Aurenhammer}, \bibinfo{person}{David Alberts}, {and} \bibinfo{person}{Bernd
  G{\"a}rtner}.} \bibinfo{year}{1996}\natexlab{}.
\newblock \showarticletitle{A novel type of skeleton for polygons}.
\newblock  (\bibinfo{year}{1996}), \bibinfo{pages}{752--761}.
\newblock


\bibitem[\protect\citeauthoryear{Alidoost, Arefi, and Hahn}{Alidoost
  et~al\mbox{.}}{2020}]%
        {alidoost2020shaped}
\bibfield{author}{\bibinfo{person}{F Alidoost}, \bibinfo{person}{H Arefi},
  {and} \bibinfo{person}{M Hahn}.} \bibinfo{year}{2020}\natexlab{}.
\newblock \showarticletitle{Y-Shaped convolutional neural network for 3D roof
  elements extraction to reconstruct building models from a single aerial
  image}.
\newblock \bibinfo{journal}{\emph{ISPRS Annals of Photogrammetry, Remote
  Sensing \& Spatial Information Sciences}} \bibinfo{volume}{5},
  \bibinfo{number}{2} (\bibinfo{year}{2020}).
\newblock


\bibitem[\protect\citeauthoryear{Arikan, Schw{\"a}rzler, Fl{\"o}ry, Wimmer, and
  Maierhofer}{Arikan et~al\mbox{.}}{2013}]%
        {arikan2013snap}
\bibfield{author}{\bibinfo{person}{Murat Arikan}, \bibinfo{person}{Michael
  Schw{\"a}rzler}, \bibinfo{person}{Simon Fl{\"o}ry}, \bibinfo{person}{Michael
  Wimmer}, {and} \bibinfo{person}{Stefan Maierhofer}.}
  \bibinfo{year}{2013}\natexlab{}.
\newblock \showarticletitle{O-snap: Optimization-based snapping for modeling
  architecture}.
\newblock \bibinfo{journal}{\emph{ACM Transactions on Graphics (TOG)}}
  \bibinfo{volume}{32}, \bibinfo{number}{1} (\bibinfo{year}{2013}),
  \bibinfo{pages}{1--15}.
\newblock


\bibitem[\protect\citeauthoryear{Battaglia, Hamrick, Bapst, Sanchez{-}Gonzalez,
  Zambaldi, Malinowski, Tacchetti, Raposo, Santoro, Faulkner,
  G{\"{u}}l{\c{c}}ehre, Song, Ballard, Gilmer, Dahl, Vaswani, Allen, Nash,
  Langston, Dyer, Heess, Wierstra, Kohli, Botvinick, Vinyals, Li, and
  Pascanu}{Battaglia et~al\mbox{.}}{2018}]%
        {DBLP:journals/corr/abs-1806-01261}
\bibfield{author}{\bibinfo{person}{Peter~W. Battaglia},
  \bibinfo{person}{Jessica~B. Hamrick}, \bibinfo{person}{Victor Bapst},
  \bibinfo{person}{Alvaro Sanchez{-}Gonzalez},
  \bibinfo{person}{Vin{\'{\i}}cius~Flores Zambaldi}, \bibinfo{person}{Mateusz
  Malinowski}, \bibinfo{person}{Andrea Tacchetti}, \bibinfo{person}{David
  Raposo}, \bibinfo{person}{Adam Santoro}, \bibinfo{person}{Ryan Faulkner},
  \bibinfo{person}{{\c{C}}aglar G{\"{u}}l{\c{c}}ehre},
  \bibinfo{person}{H.~Francis Song}, \bibinfo{person}{Andrew~J. Ballard},
  \bibinfo{person}{Justin Gilmer}, \bibinfo{person}{George~E. Dahl},
  \bibinfo{person}{Ashish Vaswani}, \bibinfo{person}{Kelsey~R. Allen},
  \bibinfo{person}{Charles Nash}, \bibinfo{person}{Victoria Langston},
  \bibinfo{person}{Chris Dyer}, \bibinfo{person}{Nicolas Heess},
  \bibinfo{person}{Daan Wierstra}, \bibinfo{person}{Pushmeet Kohli},
  \bibinfo{person}{Matthew Botvinick}, \bibinfo{person}{Oriol Vinyals},
  \bibinfo{person}{Yujia Li}, {and} \bibinfo{person}{Razvan Pascanu}.}
  \bibinfo{year}{2018}\natexlab{}.
\newblock \showarticletitle{Relational inductive biases, deep learning, and
  graph networks}.
\newblock \bibinfo{journal}{\emph{CoRR}}  \bibinfo{volume}{abs/1806.01261}
  (\bibinfo{year}{2018}).
\newblock


\bibitem[\protect\citeauthoryear{Bauchet and Lafarge}{Bauchet and
  Lafarge}{2020}]%
        {bauchet2020kinetic}
\bibfield{author}{\bibinfo{person}{Jean-Philippe Bauchet} {and}
  \bibinfo{person}{Florent Lafarge}.} \bibinfo{year}{2020}\natexlab{}.
\newblock \showarticletitle{Kinetic shape reconstruction}.
\newblock \bibinfo{journal}{\emph{ACM Transactions on Graphics (TOG)}}
  \bibinfo{volume}{39}, \bibinfo{number}{5} (\bibinfo{year}{2020}),
  \bibinfo{pages}{1--14}.
\newblock


\bibitem[\protect\citeauthoryear{Biedl, Held, Huber, Kaaser, and
  Palfrader}{Biedl et~al\mbox{.}}{2015}]%
        {biedl2015weighted}
\bibfield{author}{\bibinfo{person}{Therese Biedl}, \bibinfo{person}{Martin
  Held}, \bibinfo{person}{Stefan Huber}, \bibinfo{person}{Dominik Kaaser},
  {and} \bibinfo{person}{Peter Palfrader}.} \bibinfo{year}{2015}\natexlab{}.
\newblock \showarticletitle{Weighted straight skeletons in the plane}.
\newblock \bibinfo{journal}{\emph{Computational Geometry}}
  \bibinfo{volume}{48}, \bibinfo{number}{2} (\bibinfo{year}{2015}),
  \bibinfo{pages}{120--133}.
\newblock


\bibitem[\protect\citeauthoryear{Brock, Lim, Ritchie, and Weston}{Brock
  et~al\mbox{.}}{2016}]%
        {brock2016generative}
\bibfield{author}{\bibinfo{person}{Andrew Brock}, \bibinfo{person}{Theodore
  Lim}, \bibinfo{person}{James~Millar Ritchie}, {and}
  \bibinfo{person}{Nicholas~J Weston}.} \bibinfo{year}{2016}\natexlab{}.
\newblock \showarticletitle{Generative and Discriminative Voxel Modeling with
  Convolutional Neural Networks}. In \bibinfo{booktitle}{\emph{Neural
  Inofrmation Processing Conference: 3D Deep Learning}}.
\newblock


\bibitem[\protect\citeauthoryear{Buron, Marvie, and Gautron}{Buron
  et~al\mbox{.}}{2013}]%
        {buron2013gpu}
\bibfield{author}{\bibinfo{person}{Cyprien Buron}, \bibinfo{person}{Jean-Eudes
  Marvie}, {and} \bibinfo{person}{Pascal Gautron}.}
  \bibinfo{year}{2013}\natexlab{}.
\newblock \showarticletitle{GPU Roof Grammars}. In
  \bibinfo{booktitle}{\emph{Eurographics (Short Papers)}}.
  \bibinfo{pages}{85--88}.
\newblock


\bibitem[\protect\citeauthoryear{Chen, Mishra, Rohaninejad, and Abbeel}{Chen
  et~al\mbox{.}}{2018}]%
        {chen2018pixelsnail}
\bibfield{author}{\bibinfo{person}{Xi Chen}, \bibinfo{person}{Nikhil Mishra},
  \bibinfo{person}{Mostafa Rohaninejad}, {and} \bibinfo{person}{Pieter
  Abbeel}.} \bibinfo{year}{2018}\natexlab{}.
\newblock \showarticletitle{Pixelsnail: An improved autoregressive generative
  model}. In \bibinfo{booktitle}{\emph{International Conference on Machine
  Learning}}. PMLR, \bibinfo{pages}{864--872}.
\newblock


\bibitem[\protect\citeauthoryear{Dai, Yang, Yang, Carbonell, Le, and
  Salakhutdinov}{Dai et~al\mbox{.}}{2019}]%
        {DBLP:conf/acl/DaiYYCLS19}
\bibfield{author}{\bibinfo{person}{Zihang Dai}, \bibinfo{person}{Zhilin Yang},
  \bibinfo{person}{Yiming Yang}, \bibinfo{person}{Jaime~G. Carbonell},
  \bibinfo{person}{Quoc~Viet Le}, {and} \bibinfo{person}{Ruslan
  Salakhutdinov}.} \bibinfo{year}{2019}\natexlab{}.
\newblock \showarticletitle{Transformer-XL: Attentive Language Models beyond a
  Fixed-Length Context}. In \bibinfo{booktitle}{\emph{Proceedings of the 57th
  Conference of the Association for Computational Linguistics, {ACL} 2019,
  Florence, Italy, July 28- August 2, 2019, Volume 1: Long Papers}},
  \bibfield{editor}{\bibinfo{person}{Anna Korhonen}, \bibinfo{person}{David~R.
  Traum}, {and} \bibinfo{person}{Llu{\'{\i}}s M{\`{a}}rquez}} (Eds.).
  \bibinfo{publisher}{Association for Computational Linguistics},
  \bibinfo{pages}{2978--2988}.
\newblock


\bibitem[\protect\citeauthoryear{Dehbi, Henn, Gr{\"o}ger, Stroh, and
  Pl{\"u}mer}{Dehbi et~al\mbox{.}}{2021}]%
        {dehbi2021robust}
\bibfield{author}{\bibinfo{person}{Youness Dehbi}, \bibinfo{person}{Andr{\'e}
  Henn}, \bibinfo{person}{Gerhard Gr{\"o}ger}, \bibinfo{person}{Viktor Stroh},
  {and} \bibinfo{person}{Lutz Pl{\"u}mer}.} \bibinfo{year}{2021}\natexlab{}.
\newblock \showarticletitle{Robust and fast reconstruction of complex roofs
  with active sampling from 3D point clouds}.
\newblock \bibinfo{journal}{\emph{Transactions in GIS}} \bibinfo{volume}{25},
  \bibinfo{number}{1} (\bibinfo{year}{2021}), \bibinfo{pages}{112--133}.
\newblock


\bibitem[\protect\citeauthoryear{Demir, Aliaga, and Benes}{Demir
  et~al\mbox{.}}{2015}]%
        {demir2015procedural}
\bibfield{author}{\bibinfo{person}{Ilke Demir}, \bibinfo{person}{Daniel~G
  Aliaga}, {and} \bibinfo{person}{Bedrich Benes}.}
  \bibinfo{year}{2015}\natexlab{}.
\newblock \showarticletitle{Procedural editing of 3d building point clouds}. In
  \bibinfo{booktitle}{\emph{Proceedings of the IEEE International Conference on
  Computer Vision (ICCV)}}. \bibinfo{pages}{2147--2155}.
\newblock


\bibitem[\protect\citeauthoryear{Dinh, Krueger, and Bengio}{Dinh
  et~al\mbox{.}}{2015}]%
        {DBLP:journals/corr/DinhKB14}
\bibfield{author}{\bibinfo{person}{Laurent Dinh}, \bibinfo{person}{David
  Krueger}, {and} \bibinfo{person}{Yoshua Bengio}.}
  \bibinfo{year}{2015}\natexlab{}.
\newblock \showarticletitle{{NICE:} Non-linear Independent Components
  Estimation}. In \bibinfo{booktitle}{\emph{International Conference on
  Learning Representations (ICLR)}}, \bibfield{editor}{\bibinfo{person}{Yoshua
  Bengio} {and} \bibinfo{person}{Yann LeCun}} (Eds.).
\newblock


\bibitem[\protect\citeauthoryear{Dosovitskiy, Beyer, Kolesnikov, Weissenborn,
  Zhai, Unterthiner, Dehghani, Minderer, Heigold, Gelly, Uszkoreit, and
  Houlsby}{Dosovitskiy et~al\mbox{.}}{2021}]%
        {dosovitskiy2021an}
\bibfield{author}{\bibinfo{person}{Alexey Dosovitskiy}, \bibinfo{person}{Lucas
  Beyer}, \bibinfo{person}{Alexander Kolesnikov}, \bibinfo{person}{Dirk
  Weissenborn}, \bibinfo{person}{Xiaohua Zhai}, \bibinfo{person}{Thomas
  Unterthiner}, \bibinfo{person}{Mostafa Dehghani}, \bibinfo{person}{Matthias
  Minderer}, \bibinfo{person}{Georg Heigold}, \bibinfo{person}{Sylvain Gelly},
  \bibinfo{person}{Jakob Uszkoreit}, {and} \bibinfo{person}{Neil Houlsby}.}
  \bibinfo{year}{2021}\natexlab{}.
\newblock \showarticletitle{An Image is Worth 16x16 Words: Transformers for
  Image Recognition at Scale}. In \bibinfo{booktitle}{\emph{International
  Conference on Learning Representations (ICLR)}}.
\newblock


\bibitem[\protect\citeauthoryear{Eder and Held}{Eder and Held}{2018}]%
        {eder2018computing}
\bibfield{author}{\bibinfo{person}{G{\"u}nther Eder} {and}
  \bibinfo{person}{Martin Held}.} \bibinfo{year}{2018}\natexlab{}.
\newblock \showarticletitle{Computing positively weighted straight skeletons of
  simple polygons based on a bisector arrangement}.
\newblock \bibinfo{journal}{\emph{Inform. Process. Lett.}}
  \bibinfo{volume}{132} (\bibinfo{year}{2018}), \bibinfo{pages}{28--32}.
\newblock


\bibitem[\protect\citeauthoryear{Eppstein and Erickson}{Eppstein and
  Erickson}{1999}]%
        {eppstein1999raising}
\bibfield{author}{\bibinfo{person}{David Eppstein} {and} \bibinfo{person}{Jeff
  Erickson}.} \bibinfo{year}{1999}\natexlab{}.
\newblock \showarticletitle{Raising roofs, crashing cycles, and playing pool:
  Applications of a data structure for finding pairwise interactions}.
\newblock \bibinfo{journal}{\emph{Discrete \& Computational Geometry}}
  \bibinfo{volume}{22}, \bibinfo{number}{4} (\bibinfo{year}{1999}),
  \bibinfo{pages}{569--592}.
\newblock


\bibitem[\protect\citeauthoryear{Felkel and Obdrzalek}{Felkel and
  Obdrzalek}{1998}]%
        {felkel1998straight}
\bibfield{author}{\bibinfo{person}{Petr Felkel} {and} \bibinfo{person}{Stepan
  Obdrzalek}.} \bibinfo{year}{1998}\natexlab{}.
\newblock \showarticletitle{Straight skeleton implementation}. In
  \bibinfo{booktitle}{\emph{Proceedings of Spring Conference on Computer
  Graphics}}. Citeseer.
\newblock


\bibitem[\protect\citeauthoryear{Fisher, Savva, Li, Hanrahan, and
  Nie{\ss}ner}{Fisher et~al\mbox{.}}{2015}]%
        {fisher2015actsynth}
\bibfield{author}{\bibinfo{person}{Matthew Fisher}, \bibinfo{person}{Manolis
  Savva}, \bibinfo{person}{Yangyan Li}, \bibinfo{person}{Pat Hanrahan}, {and}
  \bibinfo{person}{Matthias Nie{\ss}ner}.} \bibinfo{year}{2015}\natexlab{}.
\newblock \showarticletitle{Activity-centric Scene Synthesis for Functional 3D
  Scene Modeling}.
\newblock \bibinfo{journal}{\emph{ACM Transactions on Graphics (TOG)}}
  \bibinfo{volume}{34}, \bibinfo{number}{6} (\bibinfo{year}{2015}).
\newblock


\bibitem[\protect\citeauthoryear{Gao, Yang, Wu, Yuan, Fu, Lai, and Zhang}{Gao
  et~al\mbox{.}}{2019}]%
        {gaosdmnet2019}
\bibfield{author}{\bibinfo{person}{Lin Gao}, \bibinfo{person}{Jie Yang},
  \bibinfo{person}{Tong Wu}, \bibinfo{person}{Yu-Jie Yuan},
  \bibinfo{person}{Hongbo Fu}, \bibinfo{person}{Yu-Kun Lai}, {and}
  \bibinfo{person}{Hao(Richard) Zhang}.} \bibinfo{year}{2019}\natexlab{}.
\newblock \showarticletitle{{SDM-NET}: Deep Generative Network for Structured
  Deformable Mesh}.
\newblock \bibinfo{journal}{\emph{ACM Transactions on Graphics (TOG)}}
  \bibinfo{volume}{38}, \bibinfo{number}{6} (\bibinfo{year}{2019}),
  \bibinfo{pages}{243:1--243:15}.
\newblock


\bibitem[\protect\citeauthoryear{Goodfellow, Pouget-Abadie, Mirza, Xu,
  Warde-Farley, Ozair, Courville, and Bengio}{Goodfellow et~al\mbox{.}}{2014}]%
        {goodfellow2014generative}
\bibfield{author}{\bibinfo{person}{Ian Goodfellow}, \bibinfo{person}{Jean
  Pouget-Abadie}, \bibinfo{person}{Mehdi Mirza}, \bibinfo{person}{Bing Xu},
  \bibinfo{person}{David Warde-Farley}, \bibinfo{person}{Sherjil Ozair},
  \bibinfo{person}{Aaron Courville}, {and} \bibinfo{person}{Yoshua Bengio}.}
  \bibinfo{year}{2014}\natexlab{}.
\newblock \showarticletitle{Generative adversarial nets}.
\newblock \bibinfo{journal}{\emph{Advances in Neural Information Processing
  Systems}}  \bibinfo{volume}{27} (\bibinfo{year}{2014}),
  \bibinfo{pages}{2672--2680}.
\newblock


\bibitem[\protect\citeauthoryear{Habbecke and Kobbelt}{Habbecke and
  Kobbelt}{2012}]%
        {habbecke2012linear}
\bibfield{author}{\bibinfo{person}{Martin Habbecke} {and} \bibinfo{person}{Leif
  Kobbelt}.} \bibinfo{year}{2012}\natexlab{}.
\newblock \showarticletitle{Linear analysis of nonlinear constraints for
  interactive geometric modeling}. In \bibinfo{booktitle}{\emph{Computer
  Graphics Forum}}, Vol.~\bibinfo{volume}{31}. Wiley Online Library,
  \bibinfo{pages}{641--650}.
\newblock


\bibitem[\protect\citeauthoryear{Held and Palfrader}{Held and
  Palfrader}{2017}]%
        {held2017straight}
\bibfield{author}{\bibinfo{person}{Martin Held} {and} \bibinfo{person}{Peter
  Palfrader}.} \bibinfo{year}{2017}\natexlab{}.
\newblock \showarticletitle{Straight skeletons with additive and multiplicative
  weights and their application to the algorithmic generation of roofs and
  terrains}.
\newblock \bibinfo{journal}{\emph{Computer-Aided Design}}  \bibinfo{volume}{92}
  (\bibinfo{year}{2017}), \bibinfo{pages}{33--41}.
\newblock


\bibitem[\protect\citeauthoryear{Holtzman, Buys, Du, Forbes, and Choi}{Holtzman
  et~al\mbox{.}}{2020}]%
        {DBLP:conf/iclr/HoltzmanBDFC20}
\bibfield{author}{\bibinfo{person}{Ari Holtzman}, \bibinfo{person}{Jan Buys},
  \bibinfo{person}{Li Du}, \bibinfo{person}{Maxwell Forbes}, {and}
  \bibinfo{person}{Yejin Choi}.} \bibinfo{year}{2020}\natexlab{}.
\newblock \showarticletitle{The Curious Case of Neural Text Degeneration}. In
  \bibinfo{booktitle}{\emph{International Conference on Learning
  Representations (ICLR)}}. \bibinfo{publisher}{OpenReview.net}.
\newblock


\bibitem[\protect\citeauthoryear{Hu, Huang, Tang, van Kaick, Zhang, and
  Huang}{Hu et~al\mbox{.}}{2020}]%
        {hu2020graph2plan}
\bibfield{author}{\bibinfo{person}{Ruizhen Hu}, \bibinfo{person}{Zeyu Huang},
  \bibinfo{person}{Yuhan Tang}, \bibinfo{person}{Oliver van Kaick},
  \bibinfo{person}{Hao Zhang}, {and} \bibinfo{person}{Hui Huang}.}
  \bibinfo{year}{2020}\natexlab{}.
\newblock \showarticletitle{Graph2Plan: Learning Floorplan Generation from
  Layout Graphs}.
\newblock \bibinfo{journal}{\emph{arXiv preprint arXiv:2004.13204}}
  (\bibinfo{year}{2020}).
\newblock


\bibitem[\protect\citeauthoryear{Jiang, Tang, Vaxman, Wonka, and
  Pottmann}{Jiang et~al\mbox{.}}{2015}]%
        {jiang2015polyhedral}
\bibfield{author}{\bibinfo{person}{Caigui Jiang}, \bibinfo{person}{Chengcheng
  Tang}, \bibinfo{person}{Amir Vaxman}, \bibinfo{person}{Peter Wonka}, {and}
  \bibinfo{person}{Helmut Pottmann}.} \bibinfo{year}{2015}\natexlab{}.
\newblock \showarticletitle{Polyhedral Patterns}.
\newblock \bibinfo{journal}{\emph{ACM Transactions On Graphics (TOG)}}
  \bibinfo{volume}{34}, \bibinfo{number}{6}, Article \bibinfo{articleno}{172}
  (\bibinfo{date}{Oct.} \bibinfo{year}{2015}), \bibinfo{numpages}{12}~pages.
\newblock
\showISSN{0730-0301}
\urldef\tempurl%
\url{https://doi.org/10.1145/2816795.2818077}
\showDOI{\tempurl}


\bibitem[\protect\citeauthoryear{Kelly, Femiani, Wonka, and Mitra}{Kelly
  et~al\mbox{.}}{2017}]%
        {kelly2017bigsur}
\bibfield{author}{\bibinfo{person}{Tom Kelly}, \bibinfo{person}{John Femiani},
  \bibinfo{person}{Peter Wonka}, {and} \bibinfo{person}{Niloy~J Mitra}.}
  \bibinfo{year}{2017}\natexlab{}.
\newblock \showarticletitle{BigSUR: large-scale structured urban
  reconstruction}.
\newblock \bibinfo{journal}{\emph{ACM Transactions On Graphics (TOG)}}
  \bibinfo{volume}{36}, \bibinfo{number}{6} (\bibinfo{year}{2017}).
\newblock


\bibitem[\protect\citeauthoryear{Kelly, Guerrero, Steed, Wonka, and
  Mitra}{Kelly et~al\mbox{.}}{2018}]%
        {kelly2018frankengan}
\bibfield{author}{\bibinfo{person}{Tom Kelly}, \bibinfo{person}{Paul Guerrero},
  \bibinfo{person}{Anthony Steed}, \bibinfo{person}{Peter Wonka}, {and}
  \bibinfo{person}{Niloy~J Mitra}.} \bibinfo{year}{2018}\natexlab{}.
\newblock \showarticletitle{FrankenGAN: Guided detail synthesis for building
  mass models using style-Synchonized Gans}.
\newblock \bibinfo{journal}{\emph{ACM Transactions On Graphics (TOG)}}
  \bibinfo{volume}{37}, \bibinfo{number}{6} (\bibinfo{year}{2018}),
  \bibinfo{pages}{1--14}.
\newblock


\bibitem[\protect\citeauthoryear{Kelly and Wonka}{Kelly and Wonka}{2011}]%
        {kelly2011interactive}
\bibfield{author}{\bibinfo{person}{Tom Kelly} {and} \bibinfo{person}{Peter
  Wonka}.} \bibinfo{year}{2011}\natexlab{}.
\newblock \showarticletitle{Interactive architectural modeling with procedural
  extrusions}.
\newblock \bibinfo{journal}{\emph{ACM Transactions on Graphics (TOG)}}
  \bibinfo{volume}{30}, \bibinfo{number}{2} (\bibinfo{year}{2011}),
  \bibinfo{pages}{1--15}.
\newblock


\bibitem[\protect\citeauthoryear{Kim, Lee, Kang, Lee, and Kim}{Kim
  et~al\mbox{.}}{2020}]%
        {kim2020softflow}
\bibfield{author}{\bibinfo{person}{Hyeongju Kim}, \bibinfo{person}{Hyeonseung
  Lee}, \bibinfo{person}{Woo~Hyun Kang}, \bibinfo{person}{Joun~Yeop Lee}, {and}
  \bibinfo{person}{Nam~Soo Kim}.} \bibinfo{year}{2020}\natexlab{}.
\newblock \showarticletitle{SoftFlow: Probabilistic Framework for Normalizing
  Flow on Manifolds}.
\newblock \bibinfo{journal}{\emph{Advances in Neural Information Processing
  Systems}}  \bibinfo{volume}{33} (\bibinfo{year}{2020}).
\newblock


\bibitem[\protect\citeauthoryear{Kingma and Ba}{Kingma and Ba}{2014}]%
        {kingma2014adam}
\bibfield{author}{\bibinfo{person}{Diederik~P Kingma} {and}
  \bibinfo{person}{Jimmy Ba}.} \bibinfo{year}{2014}\natexlab{}.
\newblock \showarticletitle{Adam: A method for stochastic optimization}.
\newblock \bibinfo{journal}{\emph{arXiv preprint arXiv:1412.6980}}
  (\bibinfo{year}{2014}).
\newblock


\bibitem[\protect\citeauthoryear{Kingma and Welling}{Kingma and
  Welling}{2014}]%
        {DBLP:journals/corr/KingmaW13}
\bibfield{author}{\bibinfo{person}{Diederik~P. Kingma} {and}
  \bibinfo{person}{Max Welling}.} \bibinfo{year}{2014}\natexlab{}.
\newblock \showarticletitle{Auto-Encoding Variational Bayes}. In
  \bibinfo{booktitle}{\emph{International Conference on Learning
  Representations (ICLR)}}, \bibfield{editor}{\bibinfo{person}{Yoshua Bengio}
  {and} \bibinfo{person}{Yann LeCun}} (Eds.).
\newblock


\bibitem[\protect\citeauthoryear{Kipf and Welling}{Kipf and Welling}{2016}]%
        {kipf2016semi}
\bibfield{author}{\bibinfo{person}{Thomas~N Kipf} {and} \bibinfo{person}{Max
  Welling}.} \bibinfo{year}{2016}\natexlab{}.
\newblock \showarticletitle{Semi-supervised classification with graph
  convolutional networks}.
\newblock \bibinfo{journal}{\emph{arXiv preprint arXiv:1609.02907}}
  (\bibinfo{year}{2016}).
\newblock


\bibitem[\protect\citeauthoryear{Larive and Gaildrat}{Larive and
  Gaildrat}{2006}]%
        {larive2006wall}
\bibfield{author}{\bibinfo{person}{Mathieu Larive} {and}
  \bibinfo{person}{Veronique Gaildrat}.} \bibinfo{year}{2006}\natexlab{}.
\newblock \showarticletitle{Wall grammar for building generation}. In
  \bibinfo{booktitle}{\emph{Proceedings of the 4th international conference on
  Computer graphics and interactive techniques in Australasia and Southeast
  Asia}}. \bibinfo{pages}{429--437}.
\newblock


\bibitem[\protect\citeauthoryear{Laycock and Day}{Laycock and Day}{2003}]%
        {laycock2003automatically}
\bibfield{author}{\bibinfo{person}{Robert~G Laycock} {and} \bibinfo{person}{AM
  Day}.} \bibinfo{year}{2003}\natexlab{}.
\newblock \showarticletitle{Automatically generating roof models from building
  footprints}.
\newblock  (\bibinfo{year}{2003}).
\newblock


\bibitem[\protect\citeauthoryear{Lin, Gao, Zhou, Lu, Ye, Zhang, Liu, and
  Yang}{Lin et~al\mbox{.}}{2013}]%
        {lin2013semantic}
\bibfield{author}{\bibinfo{person}{Hui Lin}, \bibinfo{person}{Jizhou Gao},
  \bibinfo{person}{Yu Zhou}, \bibinfo{person}{Guiliang Lu},
  \bibinfo{person}{Mao Ye}, \bibinfo{person}{Chenxi Zhang},
  \bibinfo{person}{Ligang Liu}, {and} \bibinfo{person}{Ruigang Yang}.}
  \bibinfo{year}{2013}\natexlab{}.
\newblock \showarticletitle{Semantic decomposition and reconstruction of
  residential scenes from LiDAR data}.
\newblock \bibinfo{journal}{\emph{ACM Transactions on Graphics (TOG)}}
  \bibinfo{volume}{32}, \bibinfo{number}{4} (\bibinfo{year}{2013}),
  \bibinfo{pages}{1--10}.
\newblock


\bibitem[\protect\citeauthoryear{Liu, Yang, Ceylan, Yumer, and Furukawa}{Liu
  et~al\mbox{.}}{2018}]%
        {liu2018planenet}
\bibfield{author}{\bibinfo{person}{Chen Liu}, \bibinfo{person}{Jimei Yang},
  \bibinfo{person}{Duygu Ceylan}, \bibinfo{person}{Ersin Yumer}, {and}
  \bibinfo{person}{Yasutaka Furukawa}.} \bibinfo{year}{2018}\natexlab{}.
\newblock \showarticletitle{Planenet: Piece-wise planar reconstruction from a
  single rgb image}. In \bibinfo{booktitle}{\emph{Proceedings of the IEEE
  Conference on Computer Vision and Pattern Recognition (CVPR)}}.
  \bibinfo{pages}{2579--2588}.
\newblock


\bibitem[\protect\citeauthoryear{Liu, Pottmann, Wallner, Yang, and Wang}{Liu
  et~al\mbox{.}}{2006}]%
        {liu2006geometric}
\bibfield{author}{\bibinfo{person}{Yang Liu}, \bibinfo{person}{Helmut
  Pottmann}, \bibinfo{person}{Johannes Wallner}, \bibinfo{person}{Yong-Liang
  Yang}, {and} \bibinfo{person}{Wenping Wang}.}
  \bibinfo{year}{2006}\natexlab{}.
\newblock \showarticletitle{Geometric modeling with conical meshes and
  developable surfaces}.
\newblock In \bibinfo{booktitle}{\emph{ACM Transactions On Graphics (TOG)}}.
  \bibinfo{pages}{681--689}.
\newblock


\bibitem[\protect\citeauthoryear{Merrell, Schkufza, and Koltun}{Merrell
  et~al\mbox{.}}{2010}]%
        {merrell2010computer}
\bibfield{author}{\bibinfo{person}{Paul Merrell}, \bibinfo{person}{Eric
  Schkufza}, {and} \bibinfo{person}{Vladlen Koltun}.}
  \bibinfo{year}{2010}\natexlab{}.
\newblock \showarticletitle{Computer-generated residential building layouts}.
\newblock In \bibinfo{booktitle}{\emph{ACM Transactions On Graphics (TOG)}}.
  \bibinfo{pages}{1--12}.
\newblock


\bibitem[\protect\citeauthoryear{Mo, Guerrero, Yi, Su, Wonka, Mitra, and
  Guibas}{Mo et~al\mbox{.}}{2019}]%
        {mo2019structurenet}
\bibfield{author}{\bibinfo{person}{Kaichun Mo}, \bibinfo{person}{Paul
  Guerrero}, \bibinfo{person}{Li Yi}, \bibinfo{person}{Hao Su},
  \bibinfo{person}{Peter Wonka}, \bibinfo{person}{Niloy Mitra}, {and}
  \bibinfo{person}{Leonidas Guibas}.} \bibinfo{year}{2019}\natexlab{}.
\newblock \showarticletitle{StructureNet: Hierarchical Graph Networks for 3D
  Shape Generation}.
\newblock \bibinfo{journal}{\emph{ACM Transactions on Graphics (TOG)}}
  \bibinfo{volume}{38}, \bibinfo{number}{6} (\bibinfo{year}{2019}),
  \bibinfo{pages}{Article 242}.
\newblock


\bibitem[\protect\citeauthoryear{M{\"u}ller, Wonka, Haegler, Ulmer, and
  Van~Gool}{M{\"u}ller et~al\mbox{.}}{2006}]%
        {muller2006procedural}
\bibfield{author}{\bibinfo{person}{Pascal M{\"u}ller}, \bibinfo{person}{Peter
  Wonka}, \bibinfo{person}{Simon Haegler}, \bibinfo{person}{Andreas Ulmer},
  {and} \bibinfo{person}{Luc Van~Gool}.} \bibinfo{year}{2006}\natexlab{}.
\newblock \showarticletitle{Procedural modeling of buildings}.
\newblock In \bibinfo{booktitle}{\emph{ACM Transactions On Graphics (TOG)}}.
  \bibinfo{pages}{614--623}.
\newblock


\bibitem[\protect\citeauthoryear{Musialski, Wonka, Aliaga, Wimmer, Van~Gool,
  and Purgathofer}{Musialski et~al\mbox{.}}{2013}]%
        {musialski2013survey}
\bibfield{author}{\bibinfo{person}{Przemyslaw Musialski},
  \bibinfo{person}{Peter Wonka}, \bibinfo{person}{Daniel~G Aliaga},
  \bibinfo{person}{Michael Wimmer}, \bibinfo{person}{Luc Van~Gool}, {and}
  \bibinfo{person}{Werner Purgathofer}.} \bibinfo{year}{2013}\natexlab{}.
\newblock \showarticletitle{A survey of urban reconstruction}. In
  \bibinfo{booktitle}{\emph{Computer Graphics Forum}},
  Vol.~\bibinfo{volume}{32}. Wiley Online Library, \bibinfo{pages}{146--177}.
\newblock


\bibitem[\protect\citeauthoryear{Nan and Wonka}{Nan and Wonka}{2017}]%
        {nan2017polyfit}
\bibfield{author}{\bibinfo{person}{Liangliang Nan} {and} \bibinfo{person}{Peter
  Wonka}.} \bibinfo{year}{2017}\natexlab{}.
\newblock \showarticletitle{Polyfit: Polygonal surface reconstruction from
  point clouds}. In \bibinfo{booktitle}{\emph{Proceedings of the IEEE
  International Conference on Computer Vision (ICCV)}}.
  \bibinfo{pages}{2353--2361}.
\newblock


\bibitem[\protect\citeauthoryear{Nash, Ganin, Eslami, and Battaglia}{Nash
  et~al\mbox{.}}{2020}]%
        {DBLP:conf/icml/NashGEB20}
\bibfield{author}{\bibinfo{person}{Charlie Nash}, \bibinfo{person}{Yaroslav
  Ganin}, \bibinfo{person}{S.~M.~Ali Eslami}, {and} \bibinfo{person}{Peter~W.
  Battaglia}.} \bibinfo{year}{2020}\natexlab{}.
\newblock \showarticletitle{PolyGen: An Autoregressive Generative Model of 3D
  Meshes}. In \bibinfo{booktitle}{\emph{Proceedings of the 37th International
  Conference on Machine Learning (ICML)}} \emph{(\bibinfo{series}{Proceedings
  of Machine Learning Research})}, Vol.~\bibinfo{volume}{119}.
  \bibinfo{publisher}{{PMLR}}, \bibinfo{pages}{7220--7229}.
\newblock


\bibitem[\protect\citeauthoryear{Para, Guerrero, Kelly, Guibas, and Wonka}{Para
  et~al\mbox{.}}{2020}]%
        {DBLP:journals/corr/abs-2011-13417}
\bibfield{author}{\bibinfo{person}{Wamiq~Reyaz Para}, \bibinfo{person}{Paul
  Guerrero}, \bibinfo{person}{Tom Kelly}, \bibinfo{person}{Leonidas~J. Guibas},
  {and} \bibinfo{person}{Peter Wonka}.} \bibinfo{year}{2020}\natexlab{}.
\newblock \showarticletitle{Generative Layout Modeling using Constraint
  Graphs}.
\newblock \bibinfo{journal}{\emph{CoRR}}  \bibinfo{volume}{abs/2011.13417}
  (\bibinfo{year}{2020}).
\newblock


\bibitem[\protect\citeauthoryear{Paszke, Gross, Massa, Lerer, Bradbury, Chanan,
  Killeen, Lin, Gimelshein, Antiga, Desmaison, Kopf, Yang, DeVito, Raison,
  Tejani, Chilamkurthy, Steiner, Fang, Bai, and Chintala}{Paszke
  et~al\mbox{.}}{2019}]%
        {NEURIPS2019_9015}
\bibfield{author}{\bibinfo{person}{Adam Paszke}, \bibinfo{person}{Sam Gross},
  \bibinfo{person}{Francisco Massa}, \bibinfo{person}{Adam Lerer},
  \bibinfo{person}{James Bradbury}, \bibinfo{person}{Gregory Chanan},
  \bibinfo{person}{Trevor Killeen}, \bibinfo{person}{Zeming Lin},
  \bibinfo{person}{Natalia Gimelshein}, \bibinfo{person}{Luca Antiga},
  \bibinfo{person}{Alban Desmaison}, \bibinfo{person}{Andreas Kopf},
  \bibinfo{person}{Edward Yang}, \bibinfo{person}{Zachary DeVito},
  \bibinfo{person}{Martin Raison}, \bibinfo{person}{Alykhan Tejani},
  \bibinfo{person}{Sasank Chilamkurthy}, \bibinfo{person}{Benoit Steiner},
  \bibinfo{person}{Lu Fang}, \bibinfo{person}{Junjie Bai}, {and}
  \bibinfo{person}{Soumith Chintala}.} \bibinfo{year}{2019}\natexlab{}.
\newblock \showarticletitle{PyTorch: An Imperative Style, High-Performance Deep
  Learning Library}.
\newblock In \bibinfo{booktitle}{\emph{Advances in Neural Information
  Processing Systems}}, \bibfield{editor}{\bibinfo{person}{H.~Wallach},
  \bibinfo{person}{H.~Larochelle}, \bibinfo{person}{A.~Beygelzimer},
  \bibinfo{person}{F.~d\textquotesingle Alch\'{e}-Buc},
  \bibinfo{person}{E.~Fox}, {and} \bibinfo{person}{R.~Garnett}} (Eds.).
  \bibinfo{publisher}{Curran Associates, Inc.}, \bibinfo{pages}{8024--8035}.
\newblock


\bibitem[\protect\citeauthoryear{Pottmann, Liu, Wallner, Bobenko, and
  Wang}{Pottmann et~al\mbox{.}}{2007}]%
        {pottmann2007geometry}
\bibfield{author}{\bibinfo{person}{Helmut Pottmann}, \bibinfo{person}{Yang
  Liu}, \bibinfo{person}{Johannes Wallner}, \bibinfo{person}{Alexander
  Bobenko}, {and} \bibinfo{person}{Wenping Wang}.}
  \bibinfo{year}{2007}\natexlab{}.
\newblock \showarticletitle{Geometry of multi-layer freeform structures for
  architecture}.
\newblock In \bibinfo{booktitle}{\emph{ACM Transactions On Graphics (TOG)}}.
  \bibinfo{pages}{65--es}.
\newblock


\bibitem[\protect\citeauthoryear{Pottmann, Schiftner, Bo, Schmiedhofer, Wang,
  Baldassini, and Wallner}{Pottmann et~al\mbox{.}}{2008}]%
        {pottmann2008freeform}
\bibfield{author}{\bibinfo{person}{Helmut Pottmann}, \bibinfo{person}{Alexander
  Schiftner}, \bibinfo{person}{Pengbo Bo}, \bibinfo{person}{Heinz
  Schmiedhofer}, \bibinfo{person}{Wenping Wang}, \bibinfo{person}{Niccolo
  Baldassini}, {and} \bibinfo{person}{Johannes Wallner}.}
  \bibinfo{year}{2008}\natexlab{}.
\newblock \showarticletitle{Freeform surfaces from single curved panels}.
\newblock \bibinfo{journal}{\emph{ACM Transactions on Graphics (TOG)}}
  \bibinfo{volume}{27}, \bibinfo{number}{3} (\bibinfo{year}{2008}),
  \bibinfo{pages}{1--10}.
\newblock


\bibitem[\protect\citeauthoryear{Radford, Wu, Child, Luan, Amodei, and
  Sutskever}{Radford et~al\mbox{.}}{2019}]%
        {radford2019language}
\bibfield{author}{\bibinfo{person}{Alec Radford}, \bibinfo{person}{Jeffrey Wu},
  \bibinfo{person}{Rewon Child}, \bibinfo{person}{David Luan},
  \bibinfo{person}{Dario Amodei}, {and} \bibinfo{person}{Ilya Sutskever}.}
  \bibinfo{year}{2019}\natexlab{}.
\newblock \showarticletitle{Language models are unsupervised multitask
  learners}.
\newblock  (\bibinfo{year}{2019}).
\newblock


\bibitem[\protect\citeauthoryear{Ranjan, Bolkart, Sanyal, and Black}{Ranjan
  et~al\mbox{.}}{2018}]%
        {ranjan2018generating}
\bibfield{author}{\bibinfo{person}{Anurag Ranjan}, \bibinfo{person}{Timo
  Bolkart}, \bibinfo{person}{Soubhik Sanyal}, {and} \bibinfo{person}{Michael~J
  Black}.} \bibinfo{year}{2018}\natexlab{}.
\newblock \showarticletitle{Generating 3D faces using convolutional mesh
  autoencoders}. In \bibinfo{booktitle}{\emph{Proceedings of the European
  Conference on Computer Vision (ECCV)}}. \bibinfo{pages}{704--720}.
\newblock


\bibitem[\protect\citeauthoryear{Razavi, van~den Oord, and Vinyals}{Razavi
  et~al\mbox{.}}{2019}]%
        {razavi2019generating}
\bibfield{author}{\bibinfo{person}{Ali Razavi}, \bibinfo{person}{Aaron van~den
  Oord}, {and} \bibinfo{person}{Oriol Vinyals}.}
  \bibinfo{year}{2019}\natexlab{}.
\newblock \showarticletitle{Generating diverse high-fidelity images with
  vq-vae-2}. In \bibinfo{booktitle}{\emph{Advances in Neural Information
  Processing Systems}}. \bibinfo{pages}{14866--14876}.
\newblock


\bibitem[\protect\citeauthoryear{Ren, Schneider, Ovsjanikov, and Wonka}{Ren
  et~al\mbox{.}}{2018}]%
        {ren2017graph}
\bibfield{author}{\bibinfo{person}{Jing Ren}, \bibinfo{person}{Jens Schneider},
  \bibinfo{person}{Maks Ovsjanikov}, {and} \bibinfo{person}{Peter Wonka}.}
  \bibinfo{year}{2018}\natexlab{}.
\newblock \showarticletitle{Joint Graph Layouts for Visualizing Collections of
  Segmented Meshes}.
\newblock \bibinfo{journal}{\emph{IEEE Transactions on Visualization and
  Computer Graphics}} \bibinfo{volume}{24}, \bibinfo{number}{9}
  (\bibinfo{year}{2018}), \bibinfo{pages}{2546--2558}.
\newblock


\bibitem[\protect\citeauthoryear{Rezende and Mohamed}{Rezende and
  Mohamed}{2015}]%
        {rezende2015variational}
\bibfield{author}{\bibinfo{person}{Danilo Rezende} {and}
  \bibinfo{person}{Shakir Mohamed}.} \bibinfo{year}{2015}\natexlab{}.
\newblock \showarticletitle{Variational Inference with Normalizing Flows}. In
  \bibinfo{booktitle}{\emph{International Conference on Machine Learning}}.
  \bibinfo{pages}{1530--1538}.
\newblock


\bibitem[\protect\citeauthoryear{Salimans, Karpathy, Chen, and Kingma}{Salimans
  et~al\mbox{.}}{2017}]%
        {DBLP:conf/iclr/SalimansK0K17}
\bibfield{author}{\bibinfo{person}{Tim Salimans}, \bibinfo{person}{Andrej
  Karpathy}, \bibinfo{person}{Xi Chen}, {and} \bibinfo{person}{Diederik~P.
  Kingma}.} \bibinfo{year}{2017}\natexlab{}.
\newblock \showarticletitle{PixelCNN++: Improving the PixelCNN with Discretized
  Logistic Mixture Likelihood and Other Modifications}. In
  \bibinfo{booktitle}{\emph{International Conference on Learning
  Representations (ICLR)}}. \bibinfo{publisher}{OpenReview.net}.
\newblock


\bibitem[\protect\citeauthoryear{Salinas, Lafarge, and Alliez}{Salinas
  et~al\mbox{.}}{2015}]%
        {salinas2015structure}
\bibfield{author}{\bibinfo{person}{David Salinas}, \bibinfo{person}{Florent
  Lafarge}, {and} \bibinfo{person}{Pierre Alliez}.}
  \bibinfo{year}{2015}\natexlab{}.
\newblock \showarticletitle{Structure-aware mesh decimation}. In
  \bibinfo{booktitle}{\emph{Computer Graphics Forum}},
  Vol.~\bibinfo{volume}{34}. Wiley Online Library, \bibinfo{pages}{211--227}.
\newblock


\bibitem[\protect\citeauthoryear{Stypu{\l}kowski, Zamorski, Zi{\k{e}}ba, and
  Chorowski}{Stypu{\l}kowski et~al\mbox{.}}{2019}]%
        {stypulkowski2019conditional}
\bibfield{author}{\bibinfo{person}{Micha{\l} Stypu{\l}kowski},
  \bibinfo{person}{Maciej Zamorski}, \bibinfo{person}{Maciej Zi{\k{e}}ba},
  {and} \bibinfo{person}{Jan Chorowski}.} \bibinfo{year}{2019}\natexlab{}.
\newblock \showarticletitle{Conditional invertible flow for point cloud
  generation}.
\newblock \bibinfo{journal}{\emph{arXiv preprint arXiv:1910.07344}}
  (\bibinfo{year}{2019}).
\newblock


\bibitem[\protect\citeauthoryear{Sugihara}{Sugihara}{2013}]%
        {sugihara2013straight}
\bibfield{author}{\bibinfo{person}{Kenichi Sugihara}.}
  \bibinfo{year}{2013}\natexlab{}.
\newblock \showarticletitle{Straight skeleton for automatic generation of 3-D
  building models with general shaped roofs}.
\newblock  (\bibinfo{year}{2013}).
\newblock


\bibitem[\protect\citeauthoryear{Sugihara}{Sugihara}{2019}]%
        {sugihara2019straight}
\bibfield{author}{\bibinfo{person}{Kenichi Sugihara}.}
  \bibinfo{year}{2019}\natexlab{}.
\newblock \showarticletitle{Straight Skeleton Computation Optimized for Roof
  Model Generation}. In \bibinfo{booktitle}{\emph{WSCG}},
  Vol.~\bibinfo{volume}{27}. \bibinfo{pages}{101--109}.
\newblock


\bibitem[\protect\citeauthoryear{Tan, Gao, Lai, and Xia}{Tan
  et~al\mbox{.}}{2018}]%
        {tan2018variational}
\bibfield{author}{\bibinfo{person}{Qingyang Tan}, \bibinfo{person}{Lin Gao},
  \bibinfo{person}{Yu-Kun Lai}, {and} \bibinfo{person}{Shihong Xia}.}
  \bibinfo{year}{2018}\natexlab{}.
\newblock \showarticletitle{Variational autoencoders for deforming 3d mesh
  models}. In \bibinfo{booktitle}{\emph{Proceedings of the IEEE Conference on
  Computer Vision and Pattern Recognition (CVPR)}}.
  \bibinfo{pages}{5841--5850}.
\newblock


\bibitem[\protect\citeauthoryear{Van~den Oord, Kalchbrenner, Espeholt, Vinyals,
  Graves, et~al\mbox{.}}{Van~den Oord et~al\mbox{.}}{2016}]%
        {van2016conditional}
\bibfield{author}{\bibinfo{person}{Aaron Van~den Oord}, \bibinfo{person}{Nal
  Kalchbrenner}, \bibinfo{person}{Lasse Espeholt}, \bibinfo{person}{Oriol
  Vinyals}, \bibinfo{person}{Alex Graves}, {et~al\mbox{.}}}
  \bibinfo{year}{2016}\natexlab{}.
\newblock \showarticletitle{Conditional image generation with pixelcnn
  decoders}.
\newblock \bibinfo{journal}{\emph{Advances in Neural Information Processing
  Systems}}  \bibinfo{volume}{29} (\bibinfo{year}{2016}),
  \bibinfo{pages}{4790--4798}.
\newblock


\bibitem[\protect\citeauthoryear{Van Den~Oord, Vinyals, et~al\mbox{.}}{Van
  Den~Oord et~al\mbox{.}}{2017}]%
        {van2017neural}
\bibfield{author}{\bibinfo{person}{Aaron Van Den~Oord}, \bibinfo{person}{Oriol
  Vinyals}, {et~al\mbox{.}}} \bibinfo{year}{2017}\natexlab{}.
\newblock \showarticletitle{Neural discrete representation learning}. In
  \bibinfo{booktitle}{\emph{Advances in Neural Information Processing
  Systems}}. \bibinfo{pages}{6306--6315}.
\newblock


\bibitem[\protect\citeauthoryear{Van~Oord, Kalchbrenner, and
  Kavukcuoglu}{Van~Oord et~al\mbox{.}}{2016}]%
        {van2016pixel}
\bibfield{author}{\bibinfo{person}{Aaron Van~Oord}, \bibinfo{person}{Nal
  Kalchbrenner}, {and} \bibinfo{person}{Koray Kavukcuoglu}.}
  \bibinfo{year}{2016}\natexlab{}.
\newblock \showarticletitle{Pixel Recurrent Neural Networks}. In
  \bibinfo{booktitle}{\emph{International Conference on Machine Learning}}.
  \bibinfo{pages}{1747--1756}.
\newblock


\bibitem[\protect\citeauthoryear{Vaswani, Shazeer, Parmar, Uszkoreit, Jones,
  Gomez, Kaiser, and Polosukhin}{Vaswani et~al\mbox{.}}{2017}]%
        {vaswani2017attention}
\bibfield{author}{\bibinfo{person}{Ashish Vaswani}, \bibinfo{person}{Noam
  Shazeer}, \bibinfo{person}{Niki Parmar}, \bibinfo{person}{Jakob Uszkoreit},
  \bibinfo{person}{Llion Jones}, \bibinfo{person}{Aidan~N Gomez},
  \bibinfo{person}{{\L}ukasz Kaiser}, {and} \bibinfo{person}{Illia
  Polosukhin}.} \bibinfo{year}{2017}\natexlab{}.
\newblock \showarticletitle{Attention is all you need}. In
  \bibinfo{booktitle}{\emph{Advances in Neural Information Processing
  Systems}}. \bibinfo{pages}{5998--6008}.
\newblock


\bibitem[\protect\citeauthoryear{Verdie, Lafarge, and Alliez}{Verdie
  et~al\mbox{.}}{2015}]%
        {verdie2015lod}
\bibfield{author}{\bibinfo{person}{Yannick Verdie}, \bibinfo{person}{Florent
  Lafarge}, {and} \bibinfo{person}{Pierre Alliez}.}
  \bibinfo{year}{2015}\natexlab{}.
\newblock \showarticletitle{LOD generation for urban scenes}.
\newblock \bibinfo{journal}{\emph{ACM Transactions On Graphics (TOG)}}
  \bibinfo{volume}{34}, \bibinfo{number}{ARTICLE} (\bibinfo{year}{2015}),
  \bibinfo{pages}{30}.
\newblock


\bibitem[\protect\citeauthoryear{Wang, Yeshwanth, and Nießner}{Wang
  et~al\mbox{.}}{2020}]%
        {wang2020sceneformer}
\bibfield{author}{\bibinfo{person}{Xinpeng Wang}, \bibinfo{person}{Chandan
  Yeshwanth}, {and} \bibinfo{person}{Matthias Nießner}.}
  \bibinfo{year}{2020}\natexlab{}.
\newblock \bibinfo{title}{SceneFormer: Indoor Scene Generation with
  Transformers}.
\newblock
\newblock
\showeprint[arxiv]{cs.CV/2012.09793}


\bibitem[\protect\citeauthoryear{Wang, Sun, Liu, Sarma, Bronstein, and
  Solomon}{Wang et~al\mbox{.}}{2019}]%
        {wang2019dynamic}
\bibfield{author}{\bibinfo{person}{Yue Wang}, \bibinfo{person}{Yongbin Sun},
  \bibinfo{person}{Ziwei Liu}, \bibinfo{person}{Sanjay~E Sarma},
  \bibinfo{person}{Michael~M Bronstein}, {and} \bibinfo{person}{Justin~M
  Solomon}.} \bibinfo{year}{2019}\natexlab{}.
\newblock \showarticletitle{Dynamic graph cnn for learning on point clouds}.
\newblock \bibinfo{journal}{\emph{ACM Transactions on Graphics (TOG)}}
  \bibinfo{volume}{38}, \bibinfo{number}{5} (\bibinfo{year}{2019}),
  \bibinfo{pages}{1--12}.
\newblock


\bibitem[\protect\citeauthoryear{Wu, Zhang, Xue, Freeman, and Tenenbaum}{Wu
  et~al\mbox{.}}{2016}]%
        {wu2016learning}
\bibfield{author}{\bibinfo{person}{Jiajun Wu}, \bibinfo{person}{Chengkai
  Zhang}, \bibinfo{person}{Tianfan Xue}, \bibinfo{person}{William~T Freeman},
  {and} \bibinfo{person}{Joshua~B Tenenbaum}.} \bibinfo{year}{2016}\natexlab{}.
\newblock \showarticletitle{Learning a probabilistic latent space of object
  shapes via 3D generative-adversarial modeling}. In
  \bibinfo{booktitle}{\emph{Proceedings of the 30th International Conference on
  Neural Information Processing Systems}}. \bibinfo{pages}{82--90}.
\newblock


\bibitem[\protect\citeauthoryear{Yang, Huang, Hao, Liu, Belongie, and
  Hariharan}{Yang et~al\mbox{.}}{2019}]%
        {yang2019pointflow}
\bibfield{author}{\bibinfo{person}{Guandao Yang}, \bibinfo{person}{Xun Huang},
  \bibinfo{person}{Zekun Hao}, \bibinfo{person}{Ming-Yu Liu},
  \bibinfo{person}{Serge Belongie}, {and} \bibinfo{person}{Bharath Hariharan}.}
  \bibinfo{year}{2019}\natexlab{}.
\newblock \showarticletitle{Pointflow: 3d point cloud generation with
  continuous normalizing flows}. In \bibinfo{booktitle}{\emph{Proceedings of
  the IEEE International Conference on Computer Vision (ICCV)}}.
  \bibinfo{pages}{4541--4550}.
\newblock


\bibitem[\protect\citeauthoryear{Yang, Mo, Lai, Guibas, and Gao}{Yang
  et~al\mbox{.}}{2020}]%
        {yang2020dsmnet}
\bibfield{author}{\bibinfo{person}{Jie Yang}, \bibinfo{person}{Kaichun Mo},
  \bibinfo{person}{Yu-Kun Lai}, \bibinfo{person}{Leonidas~J. Guibas}, {and}
  \bibinfo{person}{Lin Gao}.} \bibinfo{year}{2020}\natexlab{}.
\newblock \bibinfo{title}{DSM-Net: Disentangled Structured Mesh Net for
  Controllable Generation of Fine Geometry}.
\newblock
\newblock
\showeprint[arxiv]{cs.GR/2008.05440}


\bibitem[\protect\citeauthoryear{Yu, Ji, Liu, and Wei}{Yu
  et~al\mbox{.}}{2021}]%
        {yu2021automatic}
\bibfield{author}{\bibinfo{person}{Dawen Yu}, \bibinfo{person}{Shunping Ji},
  \bibinfo{person}{Jin Liu}, {and} \bibinfo{person}{Shiqing Wei}.}
  \bibinfo{year}{2021}\natexlab{}.
\newblock \showarticletitle{Automatic 3D building reconstruction from
  multi-view aerial images with deep learning}.
\newblock \bibinfo{journal}{\emph{ISPRS Journal of Photogrammetry and Remote
  Sensing}}  \bibinfo{volume}{171} (\bibinfo{year}{2021}),
  \bibinfo{pages}{155--170}.
\newblock


\bibitem[\protect\citeauthoryear{Yu, Yeung, Tang, Terzopoulos, Chan, and
  Osher}{Yu et~al\mbox{.}}{2011}]%
        {yu2011make}
\bibfield{author}{\bibinfo{person}{Lap~Fai Yu}, \bibinfo{person}{Sai~Kit
  Yeung}, \bibinfo{person}{Chi~Keung Tang}, \bibinfo{person}{Demetri
  Terzopoulos}, \bibinfo{person}{Tony~F Chan}, {and} \bibinfo{person}{Stanley~J
  Osher}.} \bibinfo{year}{2011}\natexlab{}.
\newblock \showarticletitle{Make it home: automatic optimization of furniture
  arrangement}.
\newblock \bibinfo{journal}{\emph{ACM Transactions on Graphics (TOG)}}
  \bibinfo{volume}{30}, \bibinfo{number}{4} (\bibinfo{year}{2011}).
\newblock


\bibitem[\protect\citeauthoryear{Zeng, Wu, and Furukawa}{Zeng
  et~al\mbox{.}}{2018}]%
        {zeng2018neural}
\bibfield{author}{\bibinfo{person}{Huayi Zeng}, \bibinfo{person}{Jiaye Wu},
  {and} \bibinfo{person}{Yasutaka Furukawa}.} \bibinfo{year}{2018}\natexlab{}.
\newblock \showarticletitle{Neural procedural reconstruction for residential
  buildings}. In \bibinfo{booktitle}{\emph{Proceedings of the European
  Conference on Computer Vision (ECCV)}}. \bibinfo{pages}{737--753}.
\newblock


\bibitem[\protect\citeauthoryear{Zhang, Nauata, and Furukawa}{Zhang
  et~al\mbox{.}}{2020}]%
        {zhang2020conv}
\bibfield{author}{\bibinfo{person}{Fuyang Zhang}, \bibinfo{person}{Nelson
  Nauata}, {and} \bibinfo{person}{Yasutaka Furukawa}.}
  \bibinfo{year}{2020}\natexlab{}.
\newblock \showarticletitle{Conv-mpn: Convolutional message passing neural
  network for structured outdoor architecture reconstruction}. In
  \bibinfo{booktitle}{\emph{Proceedings of the IEEE Conference on Computer
  Vision and Pattern Recognition (CVPR)}}. \bibinfo{pages}{2798--2807}.
\newblock


\bibitem[\protect\citeauthoryear{Zhou and Neumann}{Zhou and Neumann}{2008}]%
        {zhou2008fast}
\bibfield{author}{\bibinfo{person}{Qian-Yi Zhou} {and} \bibinfo{person}{Ulrich
  Neumann}.} \bibinfo{year}{2008}\natexlab{}.
\newblock \showarticletitle{Fast and extensible building modeling from airborne
  LiDAR data}. In \bibinfo{booktitle}{\emph{Proceedings of the 16th ACM
  SIGSPATIAL international conference on Advances in geographic information
  systems}}. \bibinfo{pages}{1--8}.
\newblock


\bibitem[\protect\citeauthoryear{Zhou and Neumann}{Zhou and Neumann}{2010}]%
        {zhou20102}
\bibfield{author}{\bibinfo{person}{Qian-Yi Zhou} {and} \bibinfo{person}{Ulrich
  Neumann}.} \bibinfo{year}{2010}\natexlab{}.
\newblock \showarticletitle{2.5 d dual contouring: A robust approach to
  creating building models from aerial lidar point clouds}. In
  \bibinfo{booktitle}{\emph{Proceedings of the European Conference on Computer
  Vision (ECCV)}}. Springer, \bibinfo{pages}{115--128}.
\newblock


\bibitem[\protect\citeauthoryear{Zhou and Neumann}{Zhou and Neumann}{2011}]%
        {zhou20112}
\bibfield{author}{\bibinfo{person}{Qian-Yi Zhou} {and} \bibinfo{person}{Ulrich
  Neumann}.} \bibinfo{year}{2011}\natexlab{}.
\newblock \showarticletitle{2.5 D building modeling with topology control}. In
  \bibinfo{booktitle}{\emph{Proceedings of the IEEE Conference on Computer
  Vision and Pattern Recognition (CVPR)}}. IEEE, \bibinfo{pages}{2489--2496}.
\newblock


\bibitem[\protect\citeauthoryear{Zhu, Shen, Gao, and Hu}{Zhu
  et~al\mbox{.}}{2018}]%
        {zhu2018large}
\bibfield{author}{\bibinfo{person}{Lingjie Zhu}, \bibinfo{person}{Shuhan Shen},
  \bibinfo{person}{Xiang Gao}, {and} \bibinfo{person}{Zhanyi Hu}.}
  \bibinfo{year}{2018}\natexlab{}.
\newblock \showarticletitle{Large scale urban scene modeling from MVS meshes}.
  In \bibinfo{booktitle}{\emph{Proceedings of the European Conference on
  Computer Vision (ECCV)}}. \bibinfo{pages}{614--629}.
\newblock


\end{thebibliography}

\appendix


\begin{figure}[!t]
    \centering
    \input{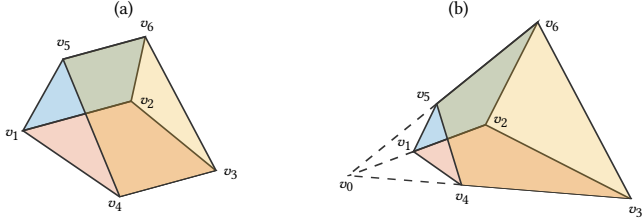}
    \vspace{-24pt}
    \caption{Illustration of Remark~\ref{rmk:valid_2D}}
    \label{fig:remark_proof}\vspace{-5pt}
\end{figure}

\section{Proof of Remark~\ref{rmk:valid_2D}}\label{appendix:remark_proof}
Recall Remark~\ref{rmk:valid_2D}: The intersecting line of two adjacent 3D planar faces with fixed outline edges, is either parallel to both outline edges, or intersecting at the same point with the two outline edges. To prove this, we only need to discuss two settings: (1) the outline edges of the two adjacent 3D faces are parallel to each other (see case (a) in Fig.~\ref{fig:remark_proof}); (2) the outline edges of the two adjacent faces intersect with each other (see case (b) in Fig.~\ref{fig:remark_proof}). 

For both cases in Fig.~\ref{fig:remark_proof}, we have two 3D planar faces $f_{1,2,6,5}$ and $f_{3,4,5,6}$, where $f_{1,2,6,5}$ has an outline edge $(v_1,v_2)$ and $f_{3,4,5,6}$ has an outline edge $(v_3,v_4)$. We know that $(v_1, v_2)$ and $(v_3, v_4)$ are two outline edges that belong to the roof outline. Therefore, these two edges are co-planar and belong to the plane $f_{1,2,3,4}$. In case (a), we have the outline edge $(v_1, v_2)$ being parallel to $(v_3, v_4)$. In case (b), we have the outline edge $(v_1, v_2)$ intersect with $(v_3, v_4)$ at point $v_0$. We are supposed to show that, in case (a), the intersecting line $(v_5, v_6) = f_{1,2,6,5}\cap f_{3,4,5,6}$ is parallel to $(v_1, v_2)$ and $(v_3, v_4)$; and show that in case (b), the intersecting line $(v_5, v_6)$ intersects with $(v_1, v_2)$ and $(v_3, v_4)$ at point $v_0$. We will give the simple proof as follows.

We prove that $(v_5, v_6)\,\parallelsum\, (v_1, v_2)$ in case (a) by contradiction. Assume $(v_5, v_6)$ is not parallel to $(v_1, v_2)$, i.e., $(v_5, v_6)$ intersect with $(v_1, v_2)$ at some point $x$. We then have $x\in(v_1, v_2)$ and $x\in(v_5,v_6)\in f_{3,4,5,6}$. Therefore, $x = (v_1, v_2) \cap f_{3,4,5,6}$, this is contradict to the fact that $(v_1, v_2)\,\parallelsum\, f_{3,4,5,6}$ since $(v_1, v_2) \,\parallelsum\,(v_3, v_4)\in f_{3,4,5,6}$. Therefore, our assumption does not hold. We then have $(v_5, v_6)\,\parallelsum\, (v_1, v_2)\,\parallelsum\,(v_3, v_4)$.

We then prove that in case (b) we have $v_0\in(v_5, v_6)$. We already know that $v_0 = (v_1, v_2)\cap(v_3, v_4)$. Therefore, $v_0\in (v_1, v_2)\in f_{1,2,6,5}$ and $v_0\in(v_3, v_4)\in f_{3,4,5,6}$. We then have $v_0 \in f_{1,2,6,5}\cap f_{3,4,5,6} = (v_5, v_6)$. This shows that the intersecting point $v_0$ belongs to the edge $(v_5, v_6)$. I.e., the intersecting edge intersects with the two outline edges at the same point. Note that the case (b) has a special case that two outline edges $(v_1, v_2)$ and $(v_1, v_3)$ intersect with each other at the same endpoint $v_1$.

The sufficient condition can be similarly proved by contradiction.
$\blacksquare$

\begin{figure}[!t]
    \centering
    \begin{overpic}[trim=0cm 12cm 17cm 0cm,clip,width=1\linewidth,grid=false]{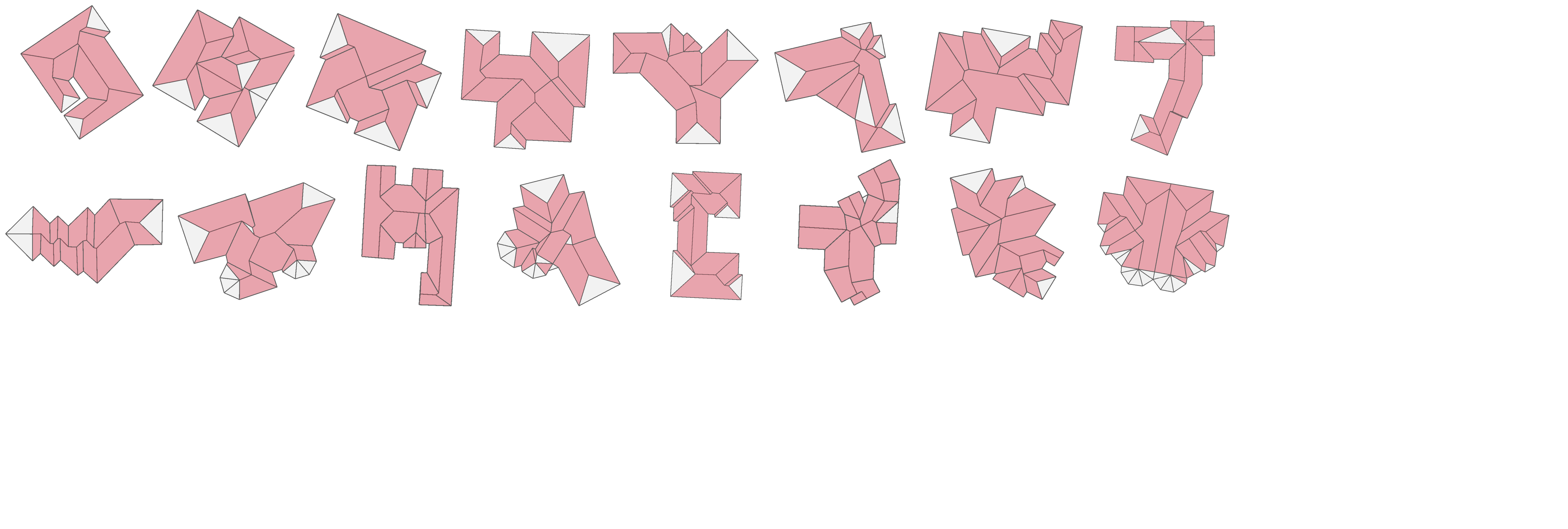}
    \end{overpic}\vspace{-6pt}
    \caption{\textbf{Roofs constructed in 3ds Max.} We highlight the roof faces that are \emph{not planar} in red.}
    \label{fig:res:3dmax}
\end{figure}

\begin{figure}[!t]
    \centering
    \begin{overpic}[trim=0cm 12cm 17cm 0cm,clip,width=1\linewidth,grid=false]{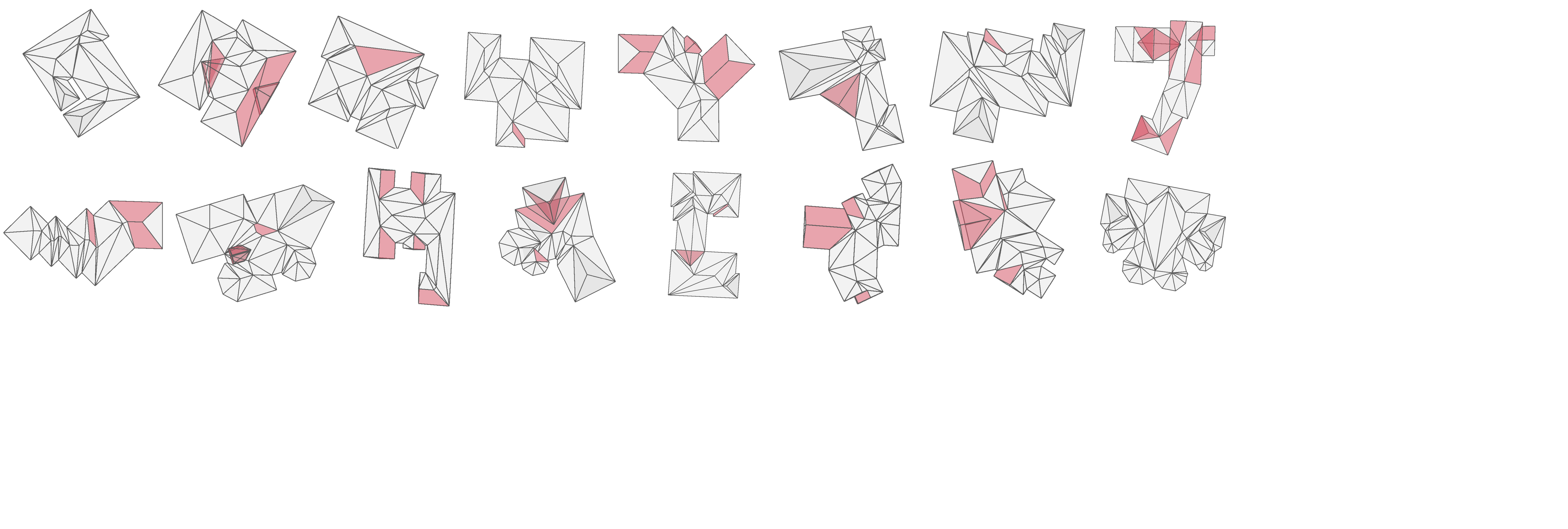}
    \end{overpic}\vspace{-6pt}
    \caption{\textbf{Roofs constructed in SketchUp.} We highlight the roof faces that are \emph{polygonal} in red.}
    \label{fig:res:su}
\end{figure}

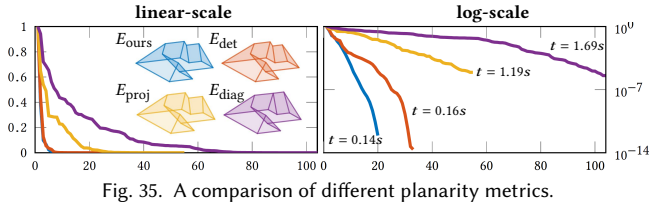
\begin{figure}[!t]
    \centering
    \begin{overpic}[trim=0cm 9cm 0cm 1.8cm,clip,width=1\linewidth,grid=false]{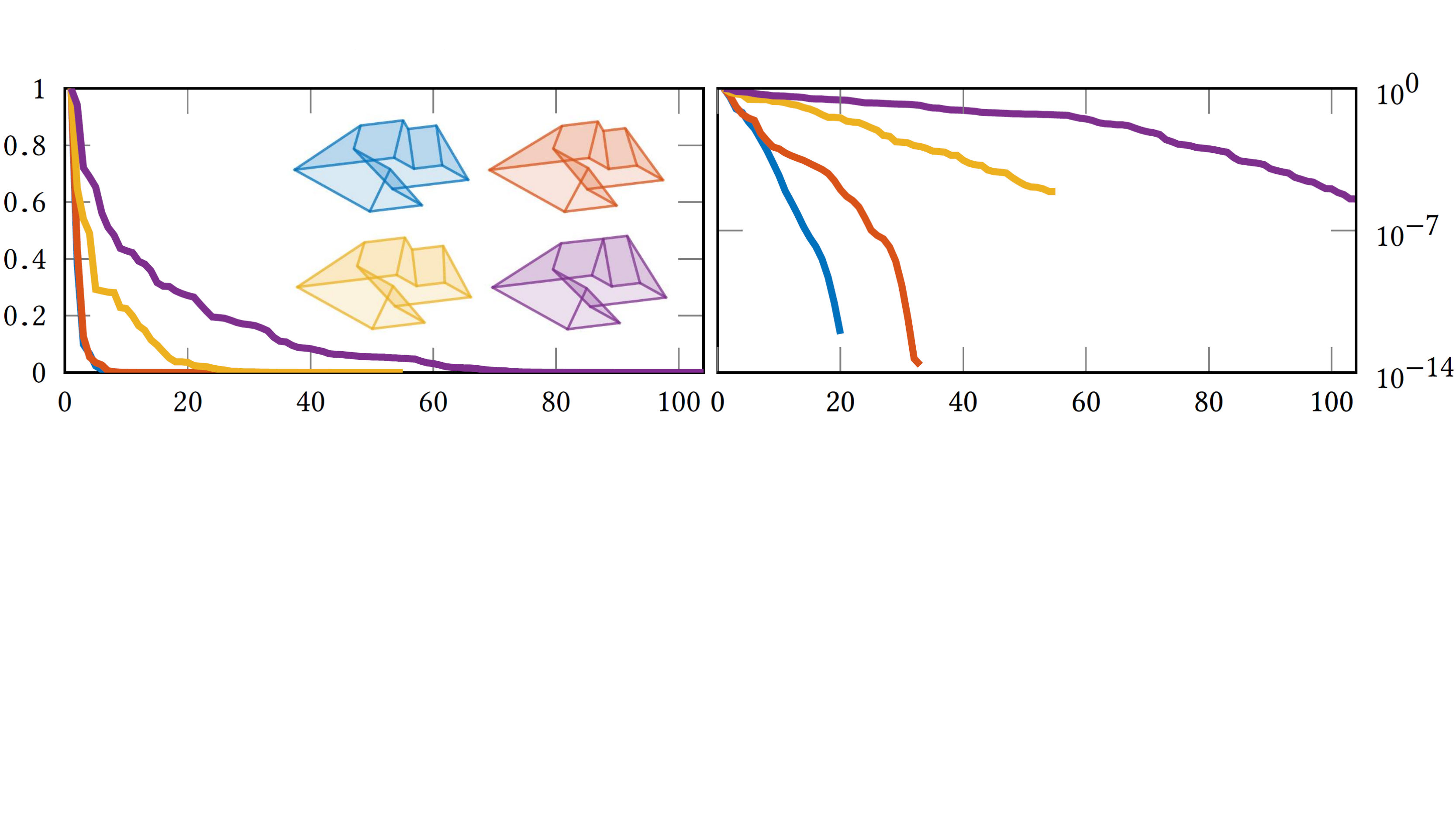}
    \put(20,25){\footnotesize\bfseries linear-scale}
    \put(70,25){\footnotesize\bfseries log-scale}
    \put(17,21){\scriptsize $E_{\text{ours}}$}
    \put(17,13){\scriptsize $E_{\text{proj}}$}
    \put(31.5,21){\scriptsize $E_{\text{det}}$}
    \put(31.5,13){\scriptsize $E_{\text{diag}}$}
    \put(50,5.5){\tiny $t=0.14s$}
    \put(63,10){\tiny $t=0.16s$}        
    \put(73,16){\tiny $t=1.19s$}
    \put(84.5,20){\tiny $t=1.69s$}
    \end{overpic}\vspace{-12pt}
     \caption{A comparison of different planarity metrics.\label{fig:ablation:planarity}}\vspace{-12pt}
\end{figure}

\revised{
\section{Alternative Planarity Metrics}\label{sec:diff_planarity}
In our method, we propose to use the smallest eigenvalue of the covariance matrix of the vertices in each face as planarity metric (see Eq.~\eqref{eq:energy:planarity}). There are other planarity metrics as well that could be considered: (1) one simple alternative is to measure the determinant of the covariance instead of the smallest eigenvalue; (2) to measure the planarity of a a set of 3D points, we can first sample 3 points to form a plane, and then measure the distance from the points to the plane; (3) another commonly used planarity metric on quad meshes~\cite{jiang2015polyhedral} is to measure the distance between two diagonal lines in a quad. We can generalize this metric to general polygon meshes as well. 
Specifically:
%
\begin{subequations}\label{eq:mtd:diff_planarity_metric}
\begin{gather}
\Scale[0.9]{E_{\text{det}}=\sum\limits_{i=1}^{n_f} \text{det}\Big( \text{Cov} \big( X_{f_i}\big) \Big)},   \quad  \Scale[0.9]{E_{\text{proj}} = \sum\limits_{i=1}^{n_f} \sum\limits_{x\in f_i} \text{dist}\big(x, P_{f_i}\big)},    \tag{\theequation a,b} \\
\Scale[0.9]{E_{\text{diag}}} \quad\Scale[0.9]{ = \sum\limits_{i: f_i = (x_{i_1},\cdots,x_{i_p})}\sum\limits_{j=1}^{p-3} \text{dist}\big( l_{x_{i_j}, x_{i_{j+2}}}, l_{x_{i_{j+1}}, x_{i_{j+3}}} \big)}    \tag{\theequation c}
\end{gather}
\end{subequations}
where in Eq.~(\ref{eq:mtd:diff_planarity_metric}b), $P_{f_i}$ is a plane formed by three sampled points on face $f_i$, and $\text{dist}(x, P)$
measures the projection distance from the point $x$ to the plane $P$; $\text{dist}(l_1, l_2)$ in Eq.~(\ref{eq:mtd:diff_planarity_metric}c) measures the distance between two 3D lines $l_1$ and $l_2$, and in our case, they are two diagonal lines connecting the vertices on the face. 

In Fig.~\ref{fig:ablation:planarity}, we compare our planarity metric (Eq.~\eqref{eq:mtd:planarity}) to the three alternatives discussed above:
we start with the same initial embedding (spectral initialization), use the same optimization solver (Quasi-Newton), and terminate w.r.t. the same criteria. We report the planarity error for each metric at different iterations in linear-scale (on the left) and log-scale (on the right). We also visualize the optimized roof on the left and report the runtime on the right. We can see that, all these four planarity metrics are valid and lead to planar 3D roofs at the end. Our choice of using the smallest eigenvalue of the covariance matrix has a much simpler form and can converge faster measured by both the running time and the number of iterations. 
}

\section{Optimize for a valid 2D embedding}\label{sec:opti_2D}
We can construct a 3D roof by optimizing for a valid 2D embedding according to Remark~\ref{rmk:valid_2D}, then lifting up the valid 2D embedding to obtain a valid 3D roof. 
Specifically, we can obtain a valid 2D embedding $\widebar{X}_{\R}$ by:
\begin{equation}\label{eq:energy:validity}
    \Scale[0.9]{\min\limits_{\widebar{X}_{\R}}\sum_{e\in\mathcal{E}_{\text{roof}}} \text{vad}(e)}
\end{equation}
where $\mathcal{E}_{\text{roof}}$ is the set of roof edges, and $\text{vad}(e)$ is a validity measurement defined on the edge $e = (x_1, x_2)$ based on Remark~\ref{rmk:valid_2D}. Specifically, assume the adjacent faces of the edge $e$ have the outline edges $e_1$ and $e_2$ respectively. If $e_1$ is parallel to $e_2$, then the edge $e$ is valid if $e$ is parallel to $e_1$ as well. In this case, we have $\text{vad}(e) = 1 -  \langle\vec{e}, \vec{e}_1\rangle^2$. If $e_1$ is not parallel to $e_2$, and the two edges intersect at the point $x$, then this point $x$ should be on the edge $e$ as well, i.e., $x_1 - x$ is parallel to $x_2 - x$. In this case, $\Scale[0.9]{\text{vad}(e) = 1 - \big\langle \frac{x_1 - x}{\Vert x_1 - x \Vert}, \frac{x_2 - x}{\Vert x_2 - x\Vert} \big\rangle ^2}$ (here $x_1, x_2, x$ are corresponding 2D positions). 


\section{Interactive Roof Editing \& Optimization}\label{sec:interactive}
Our roof modeling method can be user for interactive roof reconstruction: (1) modify/edit a valid roof graph $G$ or its dual $G^{\mathcal{D}}$ (2) starting from the modified roof graph, run our optimization method to obtain a \emph{valid} roof graph. We then go back to step (1) and edit again. In this way, we can edit the roof graph until we get a desirable one or the final valid roof graph is consistent with the input image. In the following, we first discuss some commonly used editing operations that are supported in step (1).
We then discuss how to efficiently solve step (2) by only optimizing the position of the vertices in the \emph{smallest affected region}.

\paragraph{\textbf{Editing Operations}}
Our primal-dual roof graph formulation allows us to design editing operations to modify the roof graph or its dual directly: 

\begin{itemize}[leftmargin=*]
    \item \textbf{Move a vertex}. We can modify the position of the selected vertex by an input 3D translation vector.
    \item \textbf{Move an edge}. We can modify the positions of the endpoint vertices of the selected edge by an input 3D translation vector.
    \item \textbf{Snap an edge}. We can snap an edge by merging its two endpoints into a single vertex. 
    \item \textbf{Merge two faces}. For two faces that are adjacent to each other, we can merge them into a single face by removing the shared edge and reordering the vertices in the two faces.
    \item \textbf{Split a face}. We can also split a face into two by adding an extra edge to connect two non-adjacent vertices in the selected face. 
    \item \textbf{Force two faces to be adjacent}. For two non-adjacent faces that are connected by an edge, we can force the two faces be adjacent.
\end{itemize}

\begin{figure}[!t]
    \centering
    \vspace{-8pt}
    \input{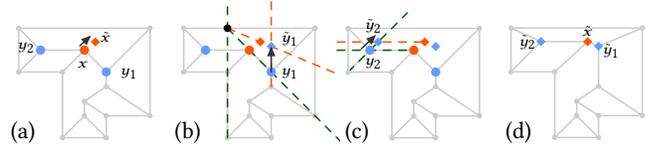}\vspace{-20pt}
    \caption{ \textbf{The smallest affected region}.
    (a) For a valid embedding of a roof graph, we change the position of the red vertex from $x$ to $\widebar{x}$. 
    (b) We then need to modify the position of the right blue vertex from $y_1$ to $\widebar{y}_1$ such that the line $(\widebar{x}, \widebar{y}_1)$ intersecting with the two outline edges at the same point. 
    (c) Similarly, we need to modify the position of the left blue vertex such that the intersecting line is parallel to the corresponding outline edges.
    (d) We then get a valid 2D embedding again after the change of $x$. Therefore, the smallest affected region of $x$ is $\{y_1, y_2\}$.}
    \label{fig:mtd:affected_region}
\end{figure}

\paragraph{\textbf{Smallest affected region}}
After applying some editing operations, we get the modified roof graph $G^{\text{mod}}$ with updated roof embedding $X^{\text{mod}}$, which is no longer valid. We need to run our roof optimization to enforce planarity constraint. Instead of rerunning our algorithm to update \emph{all} vertex positions, we only need to update a \emph{small set} of vertices (called "smallest affected region") to make the embedding valid again. 
Specifically, if we edit the position of some vertices in a valid embedding, we can detect the smallest group of vertices in the roof graph that need to be updated to satisfy the planarity constraint again by updating this group of vertices only. We therefore call this group of vertices the smallest affected region $P(x)$ of the modified vertex $x$. 
Fig.~\ref{fig:mtd:affected_region} illustrates how to detect the smallest affected region by investigating the validity of each roof vertex in the 2D embedding using Remark~\ref{rmk:valid_2D}.

Then, we only need to minimize the planarity energy in this restricted region $P(x)$:
\begin{equation}
    \Scale[0.9]{\min_{x_{\R}\in{P(x)}} \quad \sum_{i=1}^{n_f}  \sigma_1\Big(\text{Cov}\big(X_{f_i}\big)\Big)}
\end{equation}

The adoption of the smallest affected region has two main advantages for interactive editing: 1) less runtime of updating the positions of a smaller set of vertices instead of the complete vertex set. 2) more coherent embedding after the modification, which is more friendly for the users.

\clearpage

\pagebreak

\setcounter{equation}{0}
\setcounter{figure}{0}
\setcounter{table}{0}
\setcounter{page}{1}
\setcounter{section}{0}
\makeatletter
\renewcommand{\theequation}{S\arabic{equation}}
\renewcommand{\thefigure}{S\arabic{figure}}
\renewcommand{\bibnumfmt}[1]{[S#1]}
\renewcommand{\citenumfont}[1]{S#1}
\twocolumn[{\Huge\sffamily
 Supplemental Materials: Intuitive and Efficient Roof Modeling for Reconstruction and Synthesis}
 \medbreak
 \par\vspace{0.3cm}]

\begin{figure}[!t]
    \centering
    \vspace{6pt}
    \input{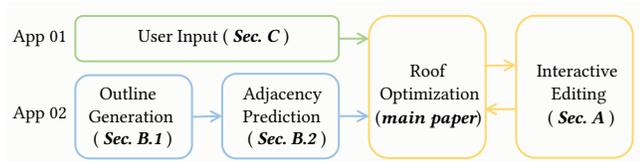}
    \vspace{-16pt}
    \caption{\textbf{Application overview}. The key components of roof reconstruction with interactive editing (App 01) and roof synthesis (App 02).}
    \label{fig:mtd:overview}\vspace{-6pt}
\end{figure}

\section{Interactive Roof Editing \& Optimization}\label{sec:interactive}

\begin{figure}[!t]
    \centering
    \vspace{18pt}
    \input{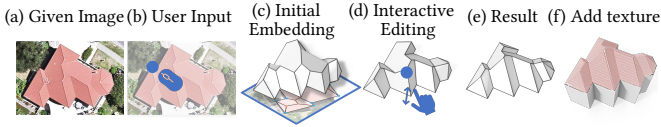}
    \vspace{-22pt}
    \caption{Overview of using our method to reconstruct a building from an input image. The user can first click on the image to specify the roof topology either by its primal or its dual as illustrated in (b). Then the roof graph is recovered and an initial embedding is computed in (c). We run our roof optimization algorithm and obtain a planar 3D roof as shown in (d). The user can edit the optimized 3D embedding (or the projected valid 2D embedding) to fix the input errors or change the vertex positions to make it more consistent with the input image. The roof embedding will get updated w.r.t. user edits as shown in (e). Finally, we can use the image as a texture for the roof we constructed as shown in (f).}
    \label{fig:app_overview:img_recon}
\end{figure}

Our roof modeling method can be directly used for interactive building reconstruction from aerial images. Specifically, for a given aerial image, one can draw the 2D outline of the roof and specify the roof structure (either using the roof graph or its dual). Then our method can construct a valid 3D (or 2D) embedding for the roof and the user is allowed to interactively modify the 3D (or 2D) embedding to improve the consistency with the input image or simply modify the roof as the user wants. Moreover, we can directly use the input image as a texture for the reconstructed building and obtain a realistic model. See Fig.~\ref{fig:app_overview:img_recon} for an overview.

This interactive roof reconstruction can be separated into two building blocks: (1) modify/edit the valid roof graph $G$ or its dual $G^{\mathcal{D}}$ (2) starting from the modified roof graph, run our optimization method to obtain a \emph{valid} roof graph. We then go back to step (1) and edit again. In this way, we can edit the roof graph until we get a desirable one or the final valid roof graph is consistent with the input image. In the following, we first discuss some commonly used editing operations that are supported in step (1).
We then discuss how to efficiently solve step (2) by only optimizing the position of the vertices in the \emph{smallest affected region}. 

\paragraph{\textbf{Editing Operations}}
Our primal-dual roof graph formulation allows us to design editing operations to modify the roof graph or its dual directly: 

\begin{itemize}[leftmargin=*]
    \item \textbf{Move a vertex}. We can modify the position of the selected vertex by an input 3D translation vector.
    \item \textbf{Move an edge}. We can modify the positions of the endpoint vertices of the selected edge by an input 3D translation vector.
    \item \textbf{Snap an edge}. We can snap an edge by merging its two endpoints into a single vertex. This can be done by modifying either the primal roof graph or the dual roof graph. For example, we would like to snap the edge $e_{p,q}$ that is shared by face $f_i$ and face $f_j$. We can remove the face adjacency of $f_i, f_j$ in the dual graph by setting $\Scale[0.9]{A^{\mathcal{D}}(i,j) = A^{\mathcal{D}}(j,i) = 0}$. After recovering the primal roof graph, $f_i$ is no longer adjacent to $f_j$ and the old selected edge is snapped. 
    \item \textbf{Merge two faces}. For two faces that are adjacent to each other, we can merge them into a single face by removing the shared edge and reordering the vertices in the two faces.
    \item \textbf{Split a face}. We can also split a face into two by adding an extra edge to connect two non-adjacent vertices in the selected face. 
    \item \textbf{Force two faces to be adjacent}. For two non-adjacent faces $f_i, f_j$ that are connected by an edge $e_{p,q}$, i.e., with one endpoint $v_p$ belongs to face $f_i$ and the other endpoint $v_q$ belongs to face $f_j$, we can force $f_i$ to be adjacent to $f_j$. Specifically, we can force the edge $e_{p,q}$ to be the shared edge by inserting the vertex $v_q$ to face $f_i$ and inserting the vertex $v_p$ to face $f_j$.
\end{itemize}

After applying some editing operations, we get the modified roof graph $G^{\text{mod}}$ with updated roof embedding $X^{\text{mod}}$. Note that the modified embedding $X^{\text{mod}}$ is no longer valid. We need to run our roof optimization to make it valid again, i.e., enforcing each roof face to be planar in 3D. A simple solution is to run our roof optimization algorithm to update all vertex positions initialized by the modified embedding $X^{\text{mod}}$. In practice, we only need to update a small set of vertices (determined by the modified vertices w.r.t. the editing operations, called "smallest affected region") to make the embedding valid again.
The adoption of the smallest affected region has two main advantages for interactive editing: 1) less runtime of updating the positions of a smaller set of vertices instead of the complete vertex set. 2) more coherent embedding after the modification, which is more friendly for the users. 


\begin{figure}[!t]
    \centering
    \input{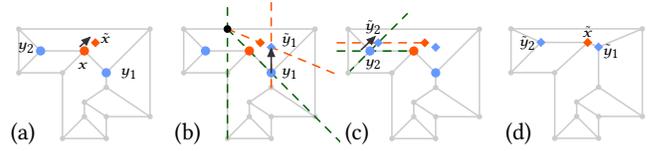}\vspace{-18pt}
    \caption{Illustration of detecting the smallest affected region. 
    (a) For a valid embedding of a roof graph, we change the position of the red vertex from $x$ to $\widebar{x}$. 
    (b) We then need to modify the position of the right blue vertex from $y_1$ to $\widebar{y}_1$ such that the line $(\widebar{x}, \widebar{y}_1)$ intersecting with the two outline edges at the same point. 
    (c) Similarly, we need to modify the position of the left blue vertex such that the intersecting line is parallel to the corresponding outline edges.
    (d) We then get a valid 2D embedding again after the change of $x$. Therefore, the smallest affected region of $x$ is $\{y_1, y_2\}$.}
    \label{fig:mtd:affected_region}
\end{figure}

\begin{figure}[!t]
    \centering
    \input{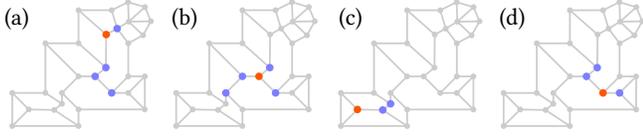}\vspace{-15pt}
    \caption{Example of the smallest affected region $P(x)$. We color the to-be-modified node $x$ in red, and then highlight its corresponding smallest affected region $P(x)$ in blue.}
    \label{fig:mtd:eg_affected_region}
\end{figure}

\paragraph{\textbf{Smallest affected region}}
If we edit the position of some vertices in a valid embedding (i.e., a planar roof), we can detect the smallest group of vertices in the roof graph that need to be updated to satisfy the planarity constraint again by updating this group of vertices only. We therefore call this group of vertices the smallest affected region $P(x)$ of the modified vertex $x$. 
Then, we only need to minimize the planarity energy in this restricted region $P(x)$:
\begin{equation}
    \Scale[0.9]{\min_{x_{\R}\in{P(x)}} \quad \sum_{i=1}^{n_f}  \sigma_1\Big(\text{Cov}\big(X_{f_i}\big)\Big)}
\end{equation}

Fig.~\ref{fig:mtd:affected_region} illustrates how to detect the smallest affected region by investigating the validity of each roof vertex in the 2D embedding using Remark 3.1. in the main paper.
Fig.~\ref{fig:mtd:eg_affected_region} includes multiple examples, where we show the smallest affected region $P(x)$ colored in blue for different vertices $x$ colored in red.

\section{Roof Synthesis from Scratch}\label{sec:mtd:roof_synthesis}
In this section, we explain how we can synthesize roofs \emph{from scratch}. Specifically, we develop a generative model for roof outline generation and face adjacency prediction. In Sec.~\ref{sec:mtd:outline_gen}, we design an auto-regressive generative model to generate roof outlines. In Sec.~\ref{sec:mtd:learn_adj}, we introduce a model to predict face adjacency for a given roof outline. 

\subsection{Outline generation} \label{sec:mtd:outline_gen}
\begin{figure}[!t]
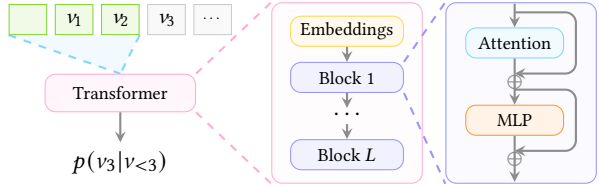

    \centering
    \includestandalone[mode=build]{figures_tex/transformer_supp}
    \vspace{-9pt}
    \caption{Our auto-regressive transformer with input of a flattened vertex sequence, and output of the probability distribution of the next token. The transformer model is composed of an embedding module and several other blocks, where each block contains a multihead-attention module and an MLP for feature projection.}
    \label{fig:transformer}
\end{figure}

Our goal is to model a distribution over $V_{\O}$ and $E_{\O}$. Most learning algorithms have better performance if the data is given in a canonical order. We therefore enforce a counter-clockwise order and encode an outline as a sequence of vertices $\left\{ v_1, v_2, \cdots, v_{n_{\O}} \right\}$. The vertex $v_1$ is the vertex closest to the lower left corner.


We flatten the coordinate matrix $\widebar{X}_{\O}\in\mathbf{R}^{n_{\O}\times 2}$ by concatenating each row in $\widebar{X}_{\O}$ and denote the flattened vertex sequence as $N^{seq} = \left\{\nu_1, \nu_2, \cdots, \nu_{2n_{\O}}\right\}$. The probability of $N^{seq}$ can be factorized into a chain of conditional probabilities,
\begin{equation}
    \Scale[0.9]{p\left(N^{seq};\phi\right) = \prod^{2n_{\O}}_{i=1}p\left( \nu_i|\nu_{<i};\phi \right)},
\end{equation}
where $\phi$ is the parameters of the model. The model is an auto-regressive network implemented with a transformer. The network outputs a probability $p$ at time step $i$ based on $$\nu_{<i}=\left\{ \nu_1, \nu_2, \cdots, \nu_{i-1} \right\}.$$ 
See Fig.~\ref{fig:transformer} for more details about the structure of our transformer. We train this model by minimizing the negative log-likelihood over all training sequences.

\paragraph{\textbf{Tokenization.}} We normalize the vertex values to the range $[0, 1]$ and quantize the vertex values to $b$-bits, which means any vertex value belongs to the set $\{1, 2, \cdots, 2^b\}$. We also append the sequence $N^{seq}$ with a stopping token $s$. Consequently, the sequence has the length of $2n_{\O}+1$ and each entry of the sequence has $2^b+1$ kinds of tokens.

\paragraph{\textbf{Learned embeddings.}} We convert the input tokens to embeddings. Specifically, we use three types of additive learned embeddings: 1) token embeddings $\mathbf{R}^{(2^b+1) \times d}$ which embed the input tokens, 2) position embeddings $\mathbf{R}^{(n_{\O}+1) \times d}$ which embed the positions of input tokens in the non-flattened sequence $V_O$, 3) coordinate embeddings $\mathbf{R}^{2 \times d}$ which embed the vertical/horizontal attribute of the input tokens. Here, $d$ is the dimension of the embeddings. We take a summation of the 3 embeddings as the inputs to the transformer (see Fig.~\ref{fig:tokens}). A similar embedding strategy can be found in PolyGen~\cite{DBLP:conf/icml/NashGEB20}.

\paragraph{\textbf{Transformer blocks}} The transformer is composed of a series of transformer blocks. Each block is as follows,
\begin{subequations}
\begin{align}
    \Scale[0.9]{h^{(l)}} & \Scale[0.9]{\leftarrow h^{(l-1)} + \mathrm{MultiheadAttention}\left(h^{(l-1)}\right)}, \\
    \Scale[0.9]{h^{(l)}} & \Scale[0.9]{\leftarrow \mathrm{LayerNorm}\left(h^{(l)}\right)}, \\
    \Scale[0.9]{h^{(l)}} & \Scale[0.9]{\leftarrow h^{(l)} + \mathrm{MLP}\left(h^{(l)}\right)}, \\
    \Scale[0.9]{h^{(l)}} & \Scale[0.9]{\leftarrow \mathrm{LayerNorm}\left(h^{(l)}\right)},
\end{align}
\end{subequations}
where $h^{(l)}$ represents the hidden representation of $N^{seq}$ in $l$-th block, $\mathrm{MultiheadAttention}(\cdot)$ is a (masked) multihead self-attention layer~\cite{vaswani2017attention} and $\mathrm{MLP}(\cdot)$ is a position-wise 2-layer fully-connected network.

\paragraph{\textbf{Architectures.}} We build the transformer with $6$ blocks. We use $384$ as the embedding dimension, $12$ heads in the self-attention modules and $1536$ as the hidden dimension in the MLPs. To make the model auto-regressive, we mask the sequences and allow the attention modules only attend previous tokens $\nu_{<i}$.

\begin{figure}[!t]
    \begin{center}
    \begin{tikzpicture}
    \node[fill=ffzzcc!8]{
        \begin{tabular}{rrrrrrrrr}
            \footnotesize{token:}& $\nu_1$ & $\nu_2$ & $\nu_3$ & $\nu_4$ & $\cdots$ & $\nu_{2n_{\O}-1}$ & $\nu_{2n_{\O}}$ & $s$ \\
            \footnotesize{position:}& $1$ & $1$ & $2$ & $2$ & $\cdots$ & $n_{\O}$ & $n_{\O}$ & $n_{\O}+1$ \\
            \footnotesize{coord:} & $1$ & $2$ & $1$ & $2$ & $\cdots$ & $1$ & $2$ & $1$
        \end{tabular}
    };
    \end{tikzpicture}
    \end{center}
    \vspace{-12pt}
    \caption{Sequence encoding. \emph{Top}: discretized vertex values which belong to $\{1, 2, 3, \cdots, 2^b+1\}$ and $s$ is the stopping token. \emph{Middle}: each entry shows the position of the token in the non-flattened vertex sequence. \emph{Bottom}: each entry represents if the token is a vertical or horizontal coordinate.}
    \label{fig:tokens}
\end{figure}

\begin{figure}[!t]
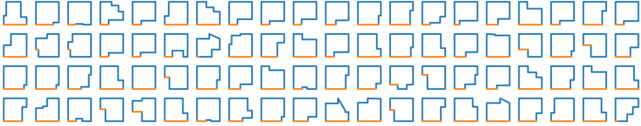

    \centering
    \begin{overpic}[trim=0cm 12.7cm 0cm 13cm,clip,width=1\linewidth,grid=false]{qual-trans.png}
    \end{overpic}\vspace{-10pt}
    \caption{Generated roof outlines with our auto-regressive model. We use our model to generate a sequence of 2D vertices and connect the tail vertex to the head by an orange line.}
    \label{fig:res-transformer-qual}
\end{figure}

\paragraph{\textbf{Inference.}} When making inference, at each time step, we apply Nucleus Sampling ~\cite{DBLP:conf/iclr/HoltzmanBDFC20}. The sampling strategy is commonly used in neural language processing.

\paragraph{\textbf{Training and test data.}} Our dataset contains 2300 training samples and 239 testing samples. We optionally skip samples with only four vertices to avoid simple roof outlines. After that, the numbers of training and testing samples are reduced to 2105 and 210, respectively.

\paragraph{\textbf{Learning Results.}} We train this model for 100 epochs. The optimizer is Adm~\cite{kingma2014adam} with a fixed learning rate $2e-4$.
We evaluate the model using the negative log-likelihood (NLL).
As a result, we achieve 1.051 NLL on the training dataset, and 0.964 NLL on the test dataset. 

\begin{figure}[!t]
    \centering
    \input{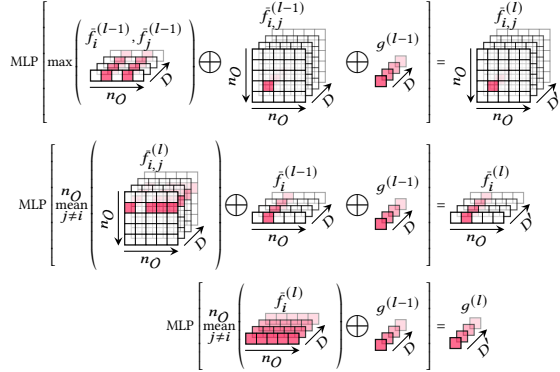}\vspace{-9pt}
    \caption{Face adjacency prediction building block. $D$ is the feature dimensions. $\bigoplus$ is the concatenation operator. From top to bottom, we show the adjacency model (Eq.~\eqref{eq:adj-model}), the edge model (Eq.~\eqref{eq:edge-model}) and the global model (Eq.~\eqref{eq:global-model})}
    \label{fig:adj-model}\vspace{-6pt}
\end{figure}

\subsection{Face adjacency prediction}\label{sec:mtd:learn_adj}
\subsubsection{Network Design}
In this step, the inputs are both $V_{\O}$ and $E_{\O}$. The output is the probability of face adjacency $(e_{i,i+1},e_{j,j+1})$ which is denoted as $p_{i,j}$.

We train a network which takes both $V_{\O}$ and $E_{\O}$ as input and outputs $p_{i,j}$ for all $1\leq i, j\leq n_{\O}$. The network is built by $L$ basic building blocks. The $l$-th block updates 3 types of representations: (1) an edge model updates the feature representation $\bar{f}_i^{(l)}$ for the edge $e_{i,i+1}$; (2) an adjacency model updates the feature representation $\bar{f}_{i,j}^{(l)}$ for the adjacency $(e_{i,i+1},e_{j,j+1})$; (3) a global model updates the global feature representation $g^{(l)}$.

\begin{figure}[!t]
    \centering
    \includegraphics[width=1\linewidth]{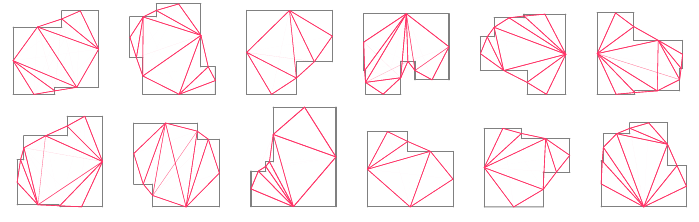}
    \vspace{-20pt}
    \caption{Predicted Adjacency. For an input outline, our trained transformer predicts the adjacency between the face $f_i$ and $f_j$ with probability $p_{ij}$. Here we visualize the probability $p_{ij}$ via opacity, i.e., the higher probability $p_{ij}$ is, the more red the corresponding edge is.}
    \label{fig:mtd:eg_learned_adj}
\end{figure}

\paragraph{\textbf{Building blocks.}} The designing of the building blocks is similar to the graph network block proposed by \cite{DBLP:journals/corr/abs-1806-01261}.
For the update of edge feature $\Scale[0.9]{\bar{f}_i^{(l)}}$, the process is,
\begin{subequations}\label{eq:edge-model}
\begin{align}
    \Scale[0.9]{\bar{f}_{i}^{(l)}} & \Scale[0.9]{\leftarrow \mathop{\mathrm{mean}}_{j\neq i} \left\{\bar{f}_{i,j}^{(l)} \right\}}, \label{eq:edge-adj}\\
    \Scale[0.9]{\bar{f}_i^{(l)}} & \Scale[0.9]{\leftarrow \mathrm{Concat}\left( \bar{f}_{i}^{(l)}, \bar{f}_i^{(l-1)}, g^{(l-1)} \right)}, \label{eq:edge-concat} \\
    \Scale[0.9]{\bar{f}_i^{(l)}} & \Scale[0.9]{\leftarrow \mathrm{MLP}\left(\bar{f}_i^{(l)}\right)}. \label{eq:edge-mlp}
\end{align}
\end{subequations}
In Eq.~\eqref{eq:edge-adj}, we aggregate all adjacency features related to $i$. Similar to the update of adjacency feature, we concatenate it with the edge feature from the previous layer $\Scale[0.9]{\bar{f}_{i}^{(l-1)}}$ and the global feature $\Scale[0.9]{g^{(l-1)}}$ (Eq.~\eqref{eq:edge-concat}) and project it with an MLP (Eq.~\eqref{eq:edge-mlp}).
The update of adjacency feature $\bar{f}_{i,j}^{(l)}$ is as follows,
\begin{subequations}\label{eq:adj-model}
\begin{align}
    \Scale[0.9]{\bar{f}_{i,j}^{(l)}} &\Scale[0.9]{\leftarrow \max\left\{\bar{f}_i^{(l-1)}, \bar{f}_j^{(l-1)}\right\}}, \label{eq:adj-max} \\
    \Scale[0.9]{\bar{f}_{i,j}^{(l)}} &\Scale[0.9]{\leftarrow \mathrm{Concat}\left(\bar{f}_{i,j}^{(l)}, \bar{f}_{i,j}^{(l-1)}, g^{(l-1)}\right)},\label{eq:adj-concat} \\ 
    \Scale[0.9]{\bar{f}_{i,j}^{(l)}} &\Scale[0.9]{\leftarrow \mathrm{MLP}\left(\bar{f}_{i,j}^{(l)}\right)}, \label{eq:adj-mlp}
\end{align}
\end{subequations}
where $\max(\cdot)$ is performed element-wisely, $\mathrm{Concat}(\cdot)$ is a concatenation operator and $\mathrm{MLP}(\cdot)$ is a multi-layered fully-connected network. In this update process, we first use a permutation-invariant operator ($\max(\cdot)$) to summarize the adjacency feature from both the edge features $\Scale[0.9]{\bar{f}_i^{(l-1)}}$ and $\Scale[0.9]{\bar{f}_j^{(l-1)}}$ (Eq.~\eqref{eq:adj-max}). Then we concatenate it with the adjacency feature from the previous layer $\Scale[0.9]{\bar{f}_{i,j}^{(l-1)}}$ and also the global feature $\Scale[0.9]{g^{(l-1)}}$ (Eq.~\eqref{eq:adj-concat}). Lastly an MLP is applied to project the adjacency feature to a new space (Eq.~\eqref{eq:adj-mlp}).

Analogously, the global feature representation is updated with a similar process. We concatenate the mean edge feature (Eq.~\eqref{eq:global-mean}) and the global feature from the previous layer $g^{(l-1)}$ (Eq.~\eqref{eq:global-concat}), and project the concatenated feature by an MLP (Eq.~\eqref{eq:global-mlp}).
\begin{subequations}\label{eq:global-model}
\begin{align}
    \Scale[0.9]{g^{(l)}} & \Scale[0.9]{\leftarrow \mathop{\mathrm{mean}}_i\left\{ \bar{f}_i^{(l)}\right\}}, \label{eq:global-mean} \\
    \Scale[0.9]{g^{(l)}} & \Scale[0.9]{\leftarrow \mathrm{Concat}\left(g^{(l)}, g^{(l-1)}\right)}, \label{eq:global-concat} \\
    \Scale[0.9]{g^{(l)}} & \Scale[0.9]{\leftarrow \mathrm{MLP}\left(g^{(l)}\right)}. \label{eq:global-mlp}
\end{align}
\end{subequations}

Each block has 3 MLPs (Eq.~\eqref{eq:adj-mlp}, Eq.~\eqref{eq:edge-mlp} and Eq.~\eqref{eq:global-mlp})  which have trainable parameters. We denote the hyper-parameters of a block with $ \big((\alpha_1, \alpha_2), (\epsilon_1, \epsilon_2), (\gamma_1, \gamma_2)\big),$ where $(\alpha_1, \alpha_2)$ means the MLP in the adjacency model have two fully-connected layers with output dimensions of $a1$ and $a2$. Similarly, $(\epsilon_1, \epsilon_2)$ and $(\gamma_1, \gamma_2)$ give the dimensions of the MLPs in the edge model and the global model, respectively. See Fig.~\ref{fig:adj-model} for an illustration. 

\paragraph{\textbf{Network inputs.}}
The inputs to the network are represented as $\Scale[0.9]{\bar{f}_{i,j}^{(0)}}$, $\Scale[0.9]{\bar{f}_{i}^{(0)}}$ and $\Scale[0.9]{g^{(0)}}$. While we do not have information for the initial adjacency feature $\Scale[0.9]{\bar{f}_{i,j}^{(0)}}$ and the initial global feature $\Scale[0.9]{g^{(0)}}$, we initialize them with zeros. For the edge feature $\Scale[0.9]{\bar{f}_{i}^{(0)}}$, we utilize the vertices $\Scale[0.9]{\widebar{X}_{\O}\in\mathbf{R}^{n_{\O}\times 2}}$. Specifically, $\Scale[0.9]{\bar{f}_{i}^{(0)}}$ is composed of the midpoint, the normal vector and the length of the edge segment, thus making it a $5$-dimensional vector.

\paragraph{\textbf{Loss function.}} An initial input is transformed by $L$ blocks and we have the final adjacency representation $\Scale[0.9]{\bar{f}_{i,j}^{(L)}}$. We project the representation to a scalar and apply the Sigmoid activation function to obtain the probability,
\begin{equation}
    \Scale[0.9]{p_{i,j} = \mathrm{Sigmoid}\left(\mathrm{FC}\left(\bar{f}_{i,j}^{(L)}\right)\right) \in [0,1]},
\end{equation}
where $\mathrm{FC}(\cdot)$ is a fully-connected layer which outputs a scalar.
The loss function is the binary cross entropy between the predicted probability $p_{i,j}$ and the ground-truth adjacency $A_F$.

\paragraph{\textbf{Architectures.}} In our experiment we designed a shallow network with 4 blocks with hyperparameters as follows,
\begin{align*}
    \Scale[0.9]{\big((\phantom{0}32, \phantom{0}32), (\phantom{0}64, \phantom{0}32), (\phantom{0}64, \phantom{0}32)\big)}, \\
    \Scale[0.9]{\big((\phantom{0}96, \phantom{0}64), (\phantom{0}96, \phantom{0}64), (\phantom{0}96, \phantom{0}64)\big)}, \\
    \Scale[0.9]{\big((192, 128), (192, 128), (192, 128)\big)}, \\
    \Scale[0.9]{\big((512, 256), (\phantom{000}, \phantom{000}), (\phantom{000}, \phantom{000})\big)}.
\end{align*}
Note that in the last ($4$-th) block, we do not need the edge model and the global model since we already have the final adjacency representation $\bar{f}_{i,j}^{(L)}$ from the adjacency model. Therefore, we leave them blank. 

\paragraph{\textbf{Training and test data.}} The training and test splits are the same as in Sec.~\ref{sec:mtd:outline_gen}. All vertex coordinates are normalized to the range $[-1, 1]$. We apply random rotation and random scaling with a factor between $[0.8, 1.2]$ as data augmentation.

\paragraph{\textbf{Learning Results.}} For each input outline with $n_{\O}$ vertices, we need to predict the probability of $n_{\O} (n_{\O}-1) / 2$ adjacencies. The task bears a resemblance to image foreground/background segmentation. Thus we use intersection over union (IoU) in the field of image segmentation as our main metric for evaluation. 
As a result, we achieve 98.18\% IoU on the training dataset, and 97.30\% on the test dataset. 
See Fig.~\ref{fig:mtd:eg_learned_adj} for some qualitative examples, where we visualize the probability $p_{ij}$ via opacity.


\begin{figure}[!t]
    \centering
    \includegraphics[width=1\linewidth]{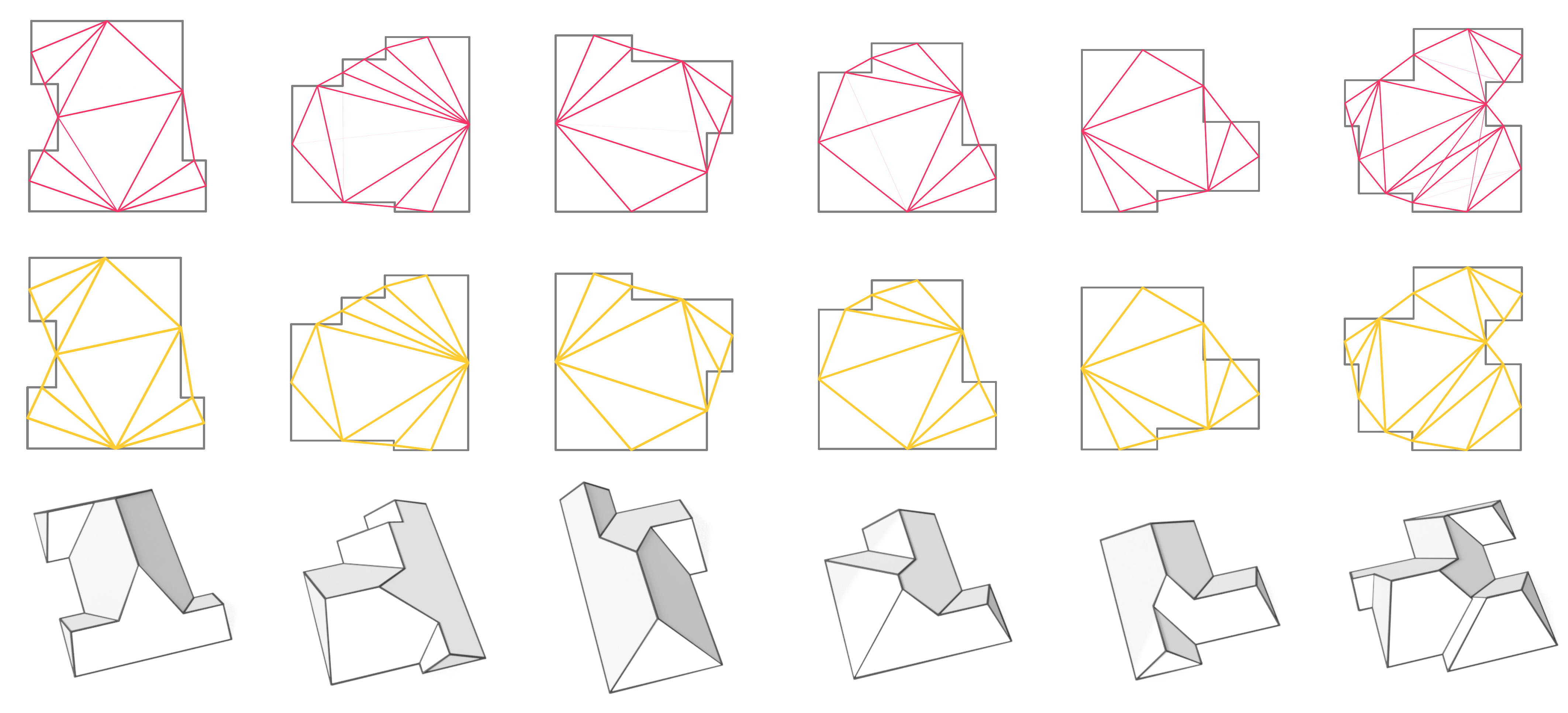}
    \vspace{-20pt}
    \caption{\emph{Top}: the predicted adjacency with probability using our transformer. \emph{Middle}: post-processed adjacency that forms a valid dual graph. \emph{Bottom}: the corresponding constructed 3D roofs using our method.}
    \label{fig:res:learned_adj}\vspace{-3pt}
\end{figure}

\subsubsection{Resolving ambiguities in learned adjacency}\label{appendix:fix_learned_adj}

As discussed above, we proposed a network to learn the adjacency probability $p_{ij}$ between a pair of faces $f_i$ and $f_j$ (see the top row of Fig.~\ref{fig:res:learned_adj}, where $p_{ij}$ is visualized via color opacity). To construct the dual graph of the roof, we need to know the exact (binary) adjacency $A_F(i,j)$ between the face $f_i$ and $f_j$ by discretizing the probability $p_{ij}$. To achieve this, we can use a simple hard thresholding that set $A_F(i,j) = 1$ if $p_{ij} > 0.5$, which is commonly used for classification. However, this is not sufficient for our case. Specifically, for the same roof outline, it is possible to have different roof styles with different sets of face adjacencies. Therefore, the learned face adjacency can contain ambiguities that multiple roof styles are mixed and cannot be realized at the same time. 

\begin{figure}[!t]
    \centering
    \vspace{6pt}
    \input{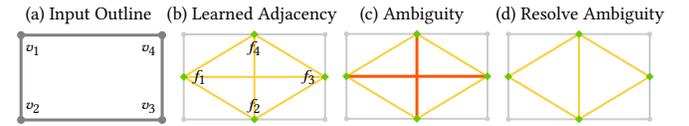}\vspace{-20pt}
    \caption{Ambiguity of the learned adjacency (type 01). Here we show an example outline with four edges in (a). In (b) we show the predicted adjacency between the face $f_1, f_2, f_3, f_4$ using our transformer. Note that, as highlighted in (c), $(f_2, f_4)$ and $(f_1, f_3)$ are both likely to be valid, however, they cannot be valid simultaneously. Therefore, once such an ambiguity occurs (i.e., two edges in the dual graph intersect with each other and the intersection is not an existing node), we keep the edge with a higher predicted probability as shown in (d). In this case, we can obtain a valid dual graph and recover the roof graph from it successfully.}
    \label{fig:learned_adj:ambiguity01}\vspace{-6pt}
\end{figure}

\begin{figure}[!t]
    \centering
    \vspace{3pt}
    \input{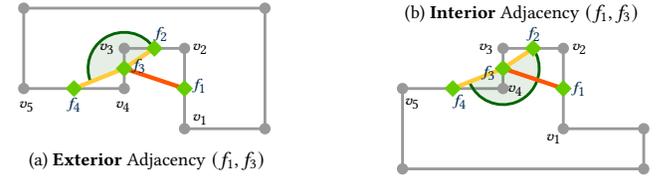}\vspace{-20pt}
    \caption{Ambiguity of the learned adjacency (type 02). Here we show two outlines with eight outline edges in (a) and (b), and we only focus on the part formed by $(v_1, v_2,\cdots, v_5)$ in both graphs. In both cases, our transformer predicts that $f_1$ is highly likely to be adjacent to $f_3$. 
    However, in reality, for the roof graph (a), it is \emph{unlikely} to have the face $f_1$ and $f_3$ adjacent to each other, otherwise, the two faces will intersect with each other \emph{outside} of the outline. The roof graph (b) shows the opposite. This ambiguity happens due to the fact that the transformer cannot tell the \emph{interior} region from the \emph{exterior} region of the roof. We can resolve this ambiguity be removing the edge in the dual graph (i.e., the face adjacency) that is in the exterior region as shown in (a).}
    \label{fig:learned_adj:ambiguity02}\vspace{-6pt}
\end{figure}

\paragraph{\textbf{Ambiguities of the Learned Adjacency}}
We show two typical ambiguities of our learned face adjacency in Fig.~\ref{fig:learned_adj:ambiguity01} and Fig.~\ref{fig:learned_adj:ambiguity02}. Specifically, for the first type of ambiguity (Fig.~\ref{fig:learned_adj:ambiguity01}), the transformer predicts that two pairs of faces are equally likely to be adjacent to each other, which cannot hold at the same time in reality. We can remove such ambiguities by detecting self-intersections in the dual graph. Once a self-intersection occurs, we only keep one of the edge and remove the other (note that an edge in the dual graph indicates the face adjacency). As for the second type of ambiguity shown in Fig.~\ref{fig:learned_adj:ambiguity02}, we can see that case (a) has complementary outline to the case (b), and our transformer gives the same prediction on the face pair $(f_1, f_3)$ that they should be adjacent to each other. However, it is unlikely to have face $f_1$ being adjacent to face $f_3$ in case (a) since otherwise they will intersect with each other outside of the roof region. On the contrary, $f_1$ is very likely to be adjacent to $f_3$ in case (b). To resolve this type of ambiguity, we can check if two faces will intersect in the interior region of the roof.  Specifically, if we assume the outline vertices are ordered in a clockwise order (e.g., $v_1, \cdots, v_5$ in case (a)), and we would like to check if $(f_i, f_j)$ is in the interior region or not. We know that, the angle forms by the two vectors $(f_i, f_{i-1})$ and $(f_i, f_{i+1})$ in counter-clockwise order lies inside the roof (i.e., the green arc region highlighted in Fig.~\ref{fig:learned_adj:ambiguity02}). We can then simply check if $(f_i, f_j)$ lies in this region or not.

Therefore, we can extract valid dual graphs from the learned adjacency by firstly resolving the second type of ambiguity that all the exterior adjacency are removed. We then resolve the first type of ambiguity in two ways:
\begin{itemize}[leftmargin=*]
    \item \textbf{Greedy strategy.} Once a self-intersection is detected, we always keep the edge with the largest probability and remove the other one.  In this case, we can extract the \emph{most likely} dual graph from the learned face adjacency (see Algorithm~\ref{alg:mtd:fix_learned_adj}).
    \item \textbf{Sampling strategy.} Once a self-intersection is detected, we bifurcate and obtain two different adjacency matrices with different choices of the edge to keep at the intersection. We then recursively check the intersections until we get a set of feasible dual graphs. In this case we can obtain \emph{a set of} valid 3D roofs with different style (different dual graph) from the same roof outline. This also justifies the advantage of our design choice of learning the adjacency in the dual graph.
\end{itemize}

\begin{algorithm}[!t]
    \DontPrintSemicolon
    \SetKwData{Left}{left}\SetKwData{This}{this}\SetKwData{Up}{up}
    \SetKwFunction{Union}{Union}\SetKwFunction{FindCompress}{FindCompress}
    \SetKwInOut{Input}{Input}\SetKwInOut{Output}{Output}
    \Input{User input 2D outline $\widebar{X}_{\O}$, Learned face adjacency $A_F$}
    \Output{Updated face adjacency $A_F$}
    (0) Make $\widebar{X}_{\O}$ in clockwise order.\;
    (1) Compute outline edge center $\vec{c}_i$ between $\widebar{X}_{\O}^{i}$ and $\widebar{X}_{\O}^{i+1}$ as the embedding for face $f_i$ in $G^{\D}$\;
    (2) Resolve ambiguity 01: for \emph{each pair} of adjacency, i.e., $(i_1, j_1)$ and $(i_2, j_2)$ where $A_F(i_1, j_1) = A_F(i_2, j_2) = 1$, check if the two line segments $l_1 = (\vec{c}_{i_1}, \vec{c}_{j_1})$ and $l_2 = (\vec{c}_{i_2}, \vec{c}_{j_2})$ intersect with each other. If so, set $A_F(i_1, j_1) = A_F(j_1, i_1) = 0$ if
    $p_{i_1, j_1} > p_{i_2, j_2}$ and set $A_F(i_2, j_2) = A_F(j_2, i_2) = 0$ otherwise.\;
    (3) Resolve ambiguity 02: for \emph{each} adjacency, i.e., $(i,j)$ where $A_F(i,j) = 1$, check if this edge is in the \emph{interior} region of the roof. If not, set $A_F(i,j) = A_F(j,i) = 0$\;
    (4) Return updated $A_F$
    \caption[caption]{Resolve Adjacency Ambiguities}
    \label{alg:mtd:fix_learned_adj}
    \end{algorithm}

\begin{figure*}[!t]
    \centering
    \includegraphics[width=1\linewidth]{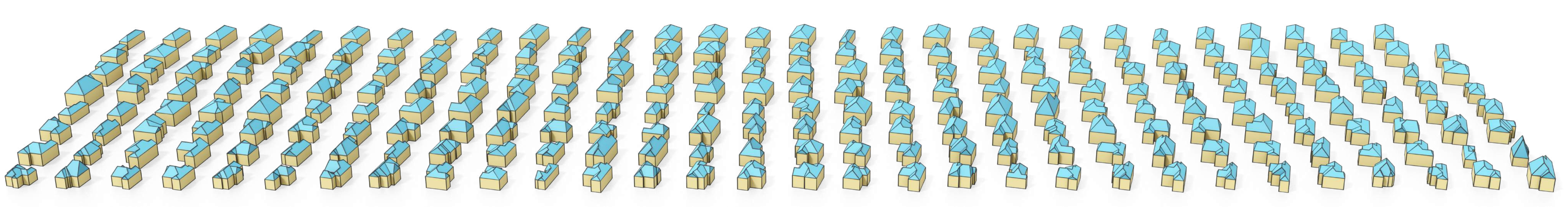}
    \vspace{-25pt}
    \caption{Synthesized buildings from learned adjacency. We use the trained network to predict the adjacency on 239 test outlines and show the corresponding reconstructed buildings.}
    \label{fig:res:learned_adj_on_test_dataset}\vspace{-6pt}
\end{figure*}

The second row of Fig.~\ref{fig:res:learned_adj} shows the extracted dual graph using the greedy strategy, and the bottom row shows the corresponding reconstructed roofs. We also use the predicted adjacency to reconstruct roofs on the 239 test outlines (see Fig.~\ref{fig:res:learned_adj_on_test_dataset}).

\subsection{Comparison to Variational Auto-encoder}\label{appendix:vae}
We justify our design choices for automatic roof synthesis by comparing to a Variational Auto-Encoder (VAE) based generative model \cite{DBLP:journals/corr/KingmaW13} for roof graph generation. The model consists of two modules, an encoder and a decoder. The encoder is a graph convolution network for latent variable encoding. The decoder has several MLPs for recovering a graph from a latent variable.
\paragraph{\textbf{Encoder}}
The encoder $\mathrm{Enc}(\cdot)$ maps an input roof graph $(V, E)$ to a latent variable $z$. Traditional convolutional networks do not work on graph data. Thus we use a graph convolution operator EdgeConv~\cite{wang2019dynamic} to build the encoder network. In our experiment, the encoder consists two EdgeConv layers which have $64$ and $128$ output channels, respectively.
\paragraph{\textbf{Decoder}}
The decoder $\mathrm{Dec}(\cdot)$ maps the latent $z$ back to a roof graph. Firstly, we decode the latent $z$ into a set of vertex features $\{z_i\}_{i=1}^{N^v}$, where $N^v$ is the maximum number of vertices and we fix it as $35$. This is done by a multi-layer perception ($\mathrm{MLP_{vf}}(\cdot)$). The MLP contains three fully-connected layers whose output channels are 512, 512, $N^v \times 32$. Secondly, we decode vertex feature $z_i$ to a 2D position and a probability which represents the probability that vertex $i$ exists. We design two MLPs for this step, $\mathrm{MLP}_{pos}(\cdot)$ for positions and $\mathrm{MLP}_{vx}(\cdot)$ for vertex existence. Finally, we have an MLP to decode pair-wise vertex features to edge existences, $\mathrm{MLP}_{ex}(\cdot, \cdot)$. For the rest MLPs, $\mathrm{MLP}_{pos}(\cdot)$, $\mathrm{MLP}_{vx}(\cdot)$ and $\mathrm{MLP}_{ex}(\cdot, \cdot)$, we use two-layered MLPs with hidden channels of $32$.
\paragraph{\textbf{Loss functions}}
The objective of the variational auto-encoder has two parts, a KL-divergence regularization term and a reconstruction loss term. The regularization term $\mathcal{L}_{KL}$ is the same as in vanilla variational auto-encoders. For the reconstruction loss, we aim to learn to predict a graph which is an approximation of the input graph. We calculate a linear assignment to match predicted vertices and input vertices, then minimize the mean squared error (MSE) $\mathcal{L}_{pos}$. For the vertex and edge existences, we minimize the cross-entropy loss $\mathcal{L}_{vx}$ and $\mathcal{L}_{ex}$.

\begin{figure}[!t]
    \centering
    \includegraphics[width=1\linewidth]{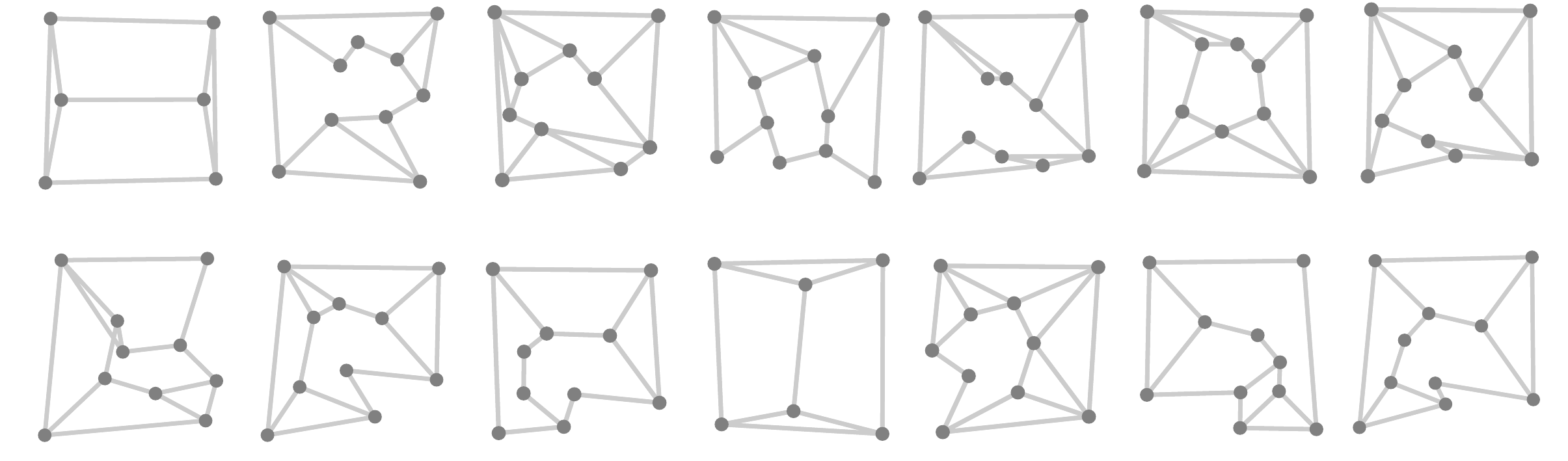}\vspace{-12pt}
    \caption{Synthesized roof graph via VAE. Synthesizing a valid roof directly can be hard since the model needs to take care of the discrete constraints and the continuous constraints at the same time.}
    \label{fig:res:vae_roofgraph}
\end{figure}

\paragraph{\textbf{Comparison.}}
Recall that we propose to tackle the roof synthesis problem by combining generative models for roof topology generation (dealing with \emph{discrete} constraints only) with roof optimization (dealing with \emph{continuous} constraints only). We believe that designing a generative model to synthesize a valid 2D or 3D roof directly is difficult since existing machine learning methods struggle with a mixture of continuous and discrete constraints. 
Take the VAE-based generative model discussed above as an example. We train a VAE on \emph{valid} 2D roof graphs consisting of the 2D positions of all the vertices and the edge connectivity, with the overall goal 
to synthesize valid 2D roof graphs directly. We synthesized 360 roof graphs, and only 119 of them are fully connected graphs while the remaining graphs have up to 19 disconnected components. We then only focus on fully connected cases for potentially valid roof graphs. Fig.~\ref{fig:res:vae_roofgraph} shows some example roof graphs synthesized by the VAE-based model. Even most of the fully connected roof graphs do not have a reasonable topology. Among the few synthesized roofs that do have a reasonable topology (e.g., the first and the last one) the geometry is not reasonable and violates aesthetic constraints. We therefore conclude that the task of constraint geometry generation is very difficult for a VAE. 
This shows that separating the continuous constraints from the discrete constraints can simplify the problem and make it easier for training a generative model to learn roof topology. At the same time, encoding the topology via the dual graph and recovering the primal graph in an algorithmic way is an easier task than predicting the primal graph directly with a potentially varying numbers of interior vertices.


\begin{figure}[!t]
    \centering
    \begin{overpic}[trim=0cm 0cm 0cm 0cm,clip,width=1\linewidth,grid=false]{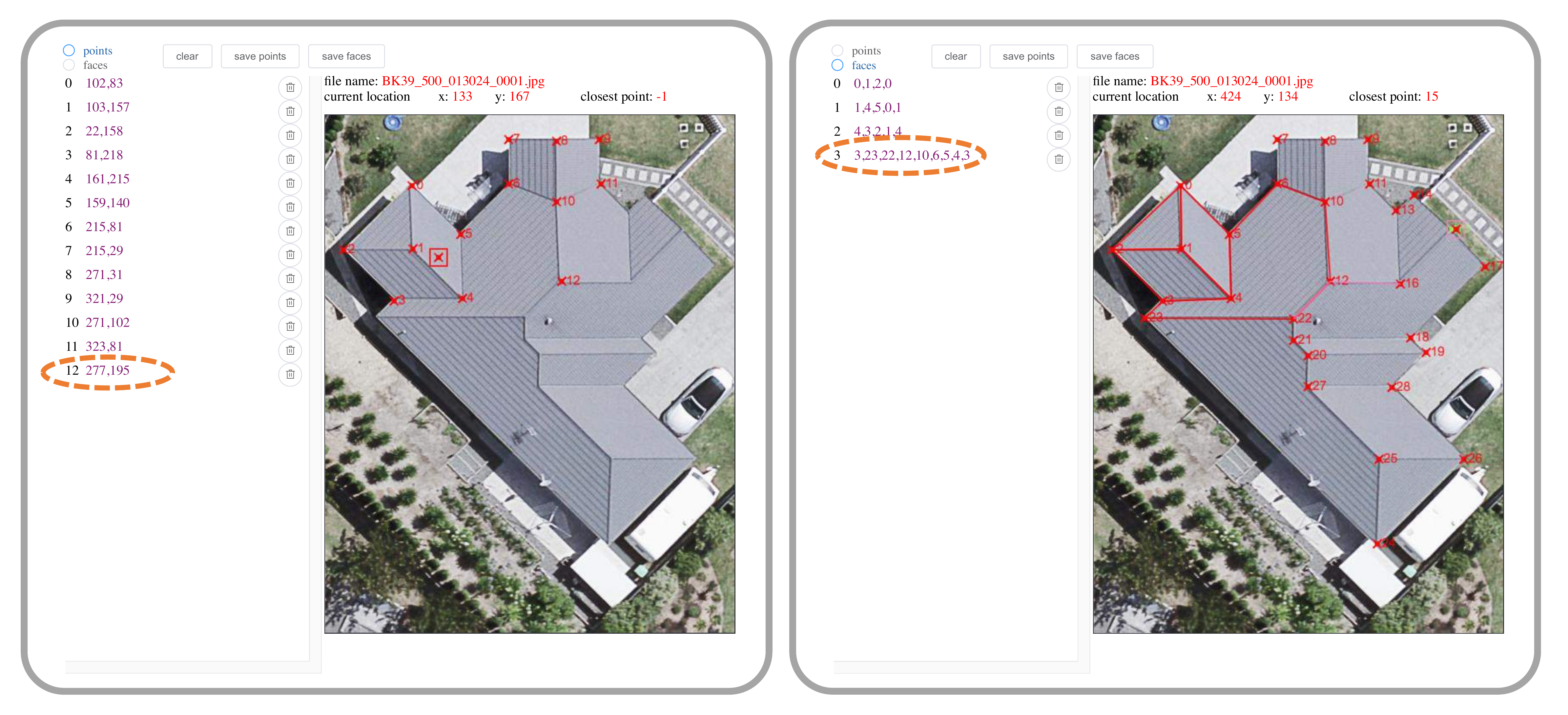}
    \put(11,46){\footnotesize (1) label the vertices}
    \put(64,46){\footnotesize (2) label the faces}
    \put(3,18){\tiny (save coordinates)}
    \put(53,32){\tiny (save vtxIDs)}
    \end{overpic}\vspace{-9pt}
    \caption{GUI mode01: label the \emph{roof graph} by specifying the vertices and the faces.}
    \label{fig:gui:mode1}
\end{figure}

\begin{figure}[!t]
    \centering
    \begin{overpic}[trim=0cm 0cm 0cm 0cm,clip,width=1\linewidth,grid=false]{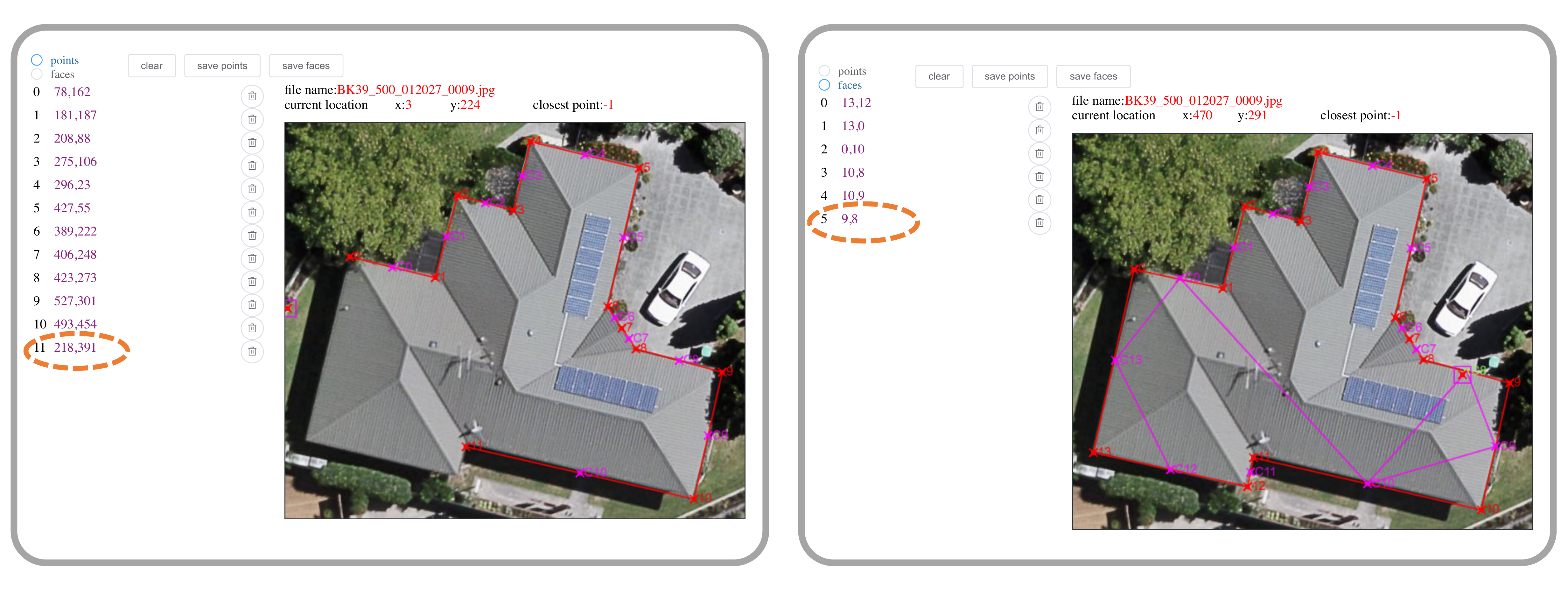}
    \put(6,38){\footnotesize (1) label the outline vertices}
    \put(58,38){\footnotesize (2) label the face adjacency}
    \put(1,12){\tiny (save coordinates)}
    \put(52,20){\tiny (save faceIDs)}
    \end{overpic}\vspace{-9pt}
    \caption{GUI mode02: label the \emph{dual graph} by specifying the outline vertices and the face adjacency.}
    \label{fig:gui:mode2}\vspace{-3pt}
\end{figure}

\section{User Interface for Roof Reconstruction}\label{sec:ui}
To construct a consistent 3D planar roof as shown in an image, 
we need to specify the 2D outlines and the topology/structure of 
the roof. We designed a GUI to collect these inputs. 
Specifically, we allow the users to specify the roof structure 
in two modes, either via the roof graph (see Fig.~\ref{fig:gui:mode1})
 or the dual graph (see Fig.~\ref{fig:gui:mode2}).




Specifically, for the first mode, as illustrated in Fig.~\ref{fig:gui:mode1}, the user can annotate the \emph{roof graph} by labeling all the vertices in the roof and then labeling each face by clicking the annotated vertices to form a polygon (i.e., in either clockwise or counter-clockwise order). For the second mode (see Fig.~\ref{fig:gui:mode2}), the user can label the \emph{dual graph} of the roof. Specifically, the user needs to first annotate the outline vertices in either the clockwise or counter-clockwise order to form an outline polygon. The \emph{center} of the each outline edge is automatically computed afterwards for selection in the next step. Then the user is asked to specify the face adjacency in the dual graph: if two faces are adjacent to each other, the user can simply click the two centers of the outline edges of the corresponding faces. 

The two labeling modes have their own advantages. For example, for a roof with $n_{\O}$ outline vertices and $n_{\R}$ roof vertices, assume it has $n_e$ roof edges. To label such a roof in the first mode, the user needs to click $n_{\O} + n_{\R}$ times to specify all the vertices, and then click $2n_{e} + n_{\O}$ times to specify the topology of all the faces. As a comparison, for the second mode, the user only needs to click $n_{\O}$ times to specify the outline vertices, and click $n_{e}$ times to specify the face adjacency. Therefore, the second mode need less user input for optimization.

However, the first mode can handle a larger group of images than the second mode, since for the second mode we assume that each of the roof face contains an outline edge. For the cases, where there exist interior roof faces that do not contain an outline edge, we can only use the first mode to specify the face topology. Another advantage of the first mode is that, since all the vertices in the roof are specified by the user, we already have an initial 2D embedding of the roof. Starting from this 2D embedding can guide us to obtain an optimized feasible 3D roof that is similar to the input image, and therefore we do not need heavy interactive edits to improve the optimized 3D roof.  In general, the user can pick the more convenient mode for labeling regarding different images.

\begin{figure*}[!t]
    \centering
    \vspace{-6pt}
    \includegraphics[width=0.85\linewidth]{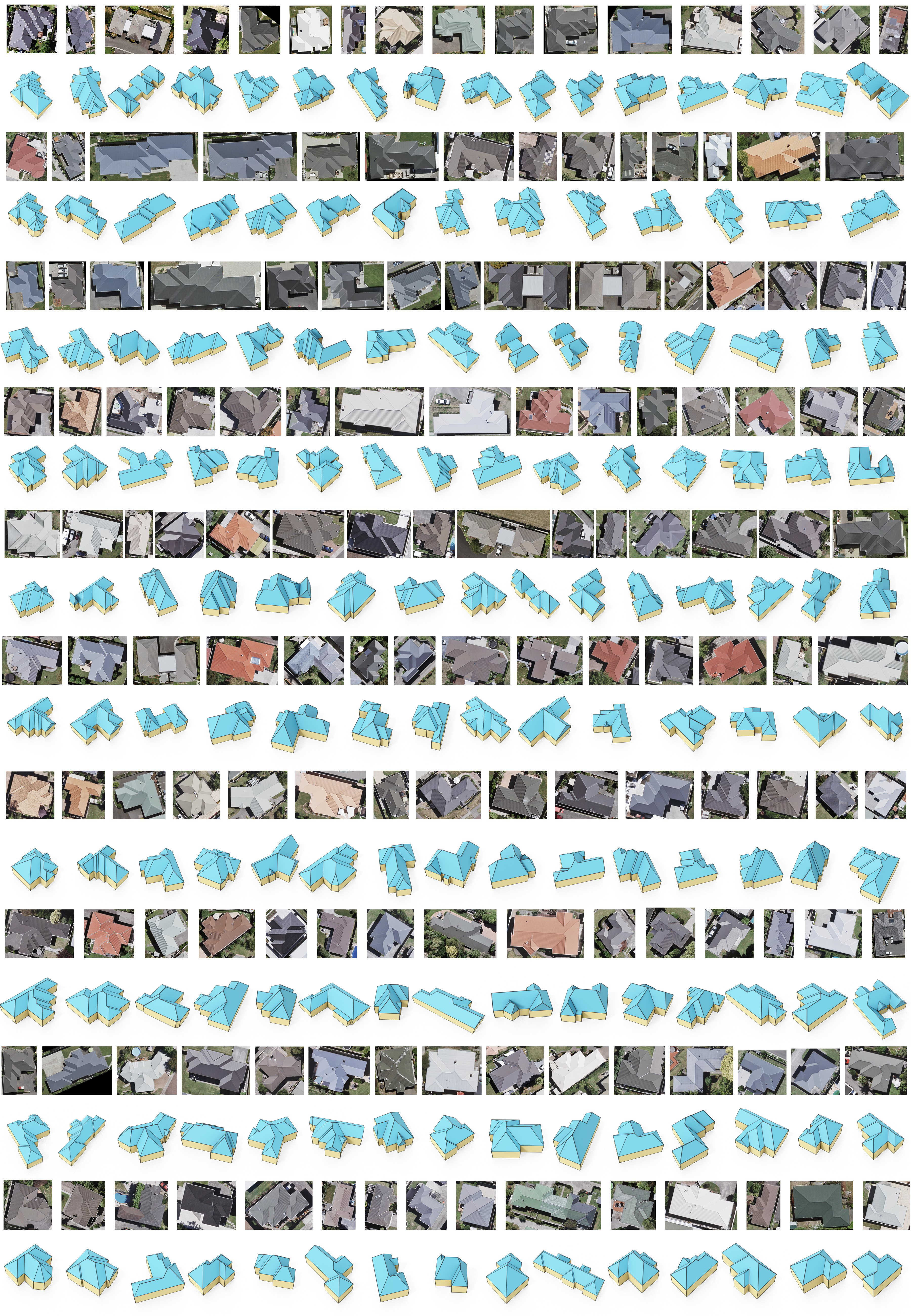}\vspace{-15pt}
    \caption{We show 150 example roofs with the corresponding aerial images from our dataset.}
    \label{fig:dataset}
\end{figure*}

\end{document}


\definecolor{wwccqq}{rgb}{0.4,0.8,0}
\definecolor{wwccff}{rgb}{0.4,0.8,1}
\definecolor{ffzzcc}{rgb}{1,0.6,0.8}
\definecolor{orange}{RGB}{255,204,51}
\definecolor{violet}{RGB}{125,125,255}
\definecolor{olive}{RGB}{255,127,0}
\begin{tikzpicture}[remember picture,]
    \tikzstyle{transformer} = [inner sep=0pt, rectangle, rounded corners, minimum width=2cm, minimum height=0.5cm, text centered, draw=ffzzcc!80, fill=ffzzcc!10]
    \tikzstyle{embedding} = [inner sep=0pt, rectangle, rounded corners, minimum width=1.5cm, minimum height=0.4cm, text centered, draw=orange!80, fill=orange!10]
    \tikzstyle{block} = [inner sep=0pt, rectangle, rounded corners, minimum width=1.5cm, minimum height=0.4cm, text centered, draw=violet!80, fill=violet!10]
    \tikzstyle{attention} = [inner sep=0pt, rectangle, rounded corners, minimum width=1.3cm, minimum height=0.4cm, text centered, draw=wwccff!80, fill=wwccff!10]
    \tikzstyle{mlp} = [inner sep=0pt, rectangle, rounded corners, minimum width=1.3cm, minimum height=0.4cm, text centered, draw=olive!80, fill=olive!10]
    \matrix[matrix of nodes, column sep=1mm, nodes={rectangle, minimum width=0.5cm, minimum height=0.4cm, text centered, draw=wwccqq!80, fill=wwccqq!10, anchor=center}] (token) {
     \small\phantom{0} & \small$\nu_1$ & \small$\nu_2$ & |[draw=gray!40, fill=gray!5]| \small$\nu_3$ & |[draw=gray!40, fill=gray!5]| \tiny$\cdots$ \\
    };
    \node[below = 0.4cm of token, transformer] (trans) {\footnotesize Transformer};
    \begin{scope}[on background layer]
    \draw[line width=0.8pt, dashed,draw=wwccff!80] (token-1-1.south west) -- (trans.north);
    \draw[line width=0.8pt, dashed,draw=wwccff!80] (token-1-3.south east) -- (trans.north);
    \fill [opacity=0.1,wwccff!80] (token-1-1.south west) -- (trans.north) -- (token-1-3.south east) -- cycle;
    \end{scope}
    
    \node[below = 0.4cm of trans] (prob) {$p(\nu_3|\nu_{<3})$};
    \draw[line width=0.8pt, -stealth, draw=gray!90] (trans) -- (prob);
    
    \node[right = 1.0cm of trans, transformer, minimum width=2.0cm, minimum height=2.4cm, fill=ffzzcc!3] (trans2) {
        \begin{tikzpicture}[remember picture]
            \node[embedding] (embeddings) {\footnotesize Embeddings};
            \node[below = 0.2cm of embeddings, block] (block1) {\footnotesize Block 1};
            \draw[line width=0.8pt, -stealth, draw=gray!90] (embeddings) -- (block1);
            \node[below = 0.2cm of block1, inner sep=0pt, minimum width=1cm, minimum height=0.2cm] (vdots) {$\,\cdots\,$};
            \draw[line width=0.8pt, -stealth, draw=gray!90] (block1) -- (vdots);
            \node[below = 0.2cm of vdots, block] (blockl) {\footnotesize Block $L$};
            \draw[line width=0.8pt, -stealth, draw=gray!90] (vdots) -- (blockl);
        \end{tikzpicture}
    };
    \draw[line width=0.8pt, -, dashed, draw=ffzzcc!80] (trans.north east) -- (trans2.north west);
    \draw[line width=0.8pt, -, dashed, draw=ffzzcc!80] (trans.south east) -- (trans2.south west);
    
    \node[right = 0.3cm of trans2, block, minimum width=2cm, minimum height=2.4cm, fill=violet!3] (blocki) {
        \begin{tikzpicture}[remember picture]
            \node[minimum width=0.2cm, minimum height=0.1cm, inner sep=0pt] (blockinput) {};
            \node[below = 0.2cm of blockinput, attention] (atten) {\footnotesize Attention};
            \node[below = 0.2cm of atten, minimum width=0.3cm, minimum height=0.2cm, inner sep=0pt] (res) {\textcolor{gray!90}{$\oplus$}};
            \node[below = 0.2cm of res, mlp] (mlp) {\footnotesize MLP};
            \node[below = 0.2cm of mlp, minimum width=0.3cm, minimum height=0.2cm, inner sep=0pt] (res2) {\textcolor{gray!90}{$\oplus$}};
            \node[below = 0.2cm of res2, minimum width=0.3cm, minimum height=0.01cm, inner sep=0pt] (blockoutput) {};
            
            \draw[line width=0.8pt, -stealth, draw=gray!90] (blockinput.north) -- (atten);
            \draw[line width=0.8pt, -, draw=gray!90] (atten) -- (res);
            \draw[line width=0.8pt, -stealth, draw=gray!90] (res.north) -- (mlp);
            \draw[line width=0.8pt, -, draw=gray!90] (mlp) -- (res2);
            \draw[line width=0.8pt, -stealth, draw=gray!90]  (blockinput.south) -- ++(6ex,0)coordinate (a) -- ($(a|-res.north)$) -- (res.north);
            \draw[line width=0.8pt, -stealth, draw=gray!90]  (res.south) -- ++(6ex,0)coordinate (a) -- ($(a|-res2.north)$) -- (res2.north);
            \draw[line width=0.8pt, -stealth, draw=gray!90] (res2.north) -- (blockoutput);
        \end{tikzpicture}
    };

    \draw[line width=0.8pt, -, dashed, draw=violet!80] (block1.north east) -- (blocki.north west);
    \draw[line width=0.8pt, -, dashed, draw=violet!80] (block1.south east) -- (blocki.south west);
\end{tikzpicture}